\documentclass[rmp,twocolumn,english,noeprint]{revtex4-2}

\usepackage{hyperref}

\hypersetup{
    unicode=true, 
    colorlinks=true, 
    linkcolor=blue, 
    citecolor=blue,
    urlcolor=blue
}

\usepackage{latexsym,amssymb,amsmath,amsthm,bm,bbold,amsfonts,amstext,graphicx,bbm,dsfont,enumerate,float,enumitem}

\usepackage{times}
\usepackage{bold-extra}
\usepackage[table]{xcolor}
\usepackage[utf8]{inputenc}
\usepackage[english]{babel}
\usepackage{tikz}
\usepackage{transparent}
\usepackage[normalem]{ulem}
\usepackage{setspace}
\usepackage{textcomp}
\usepackage[most,skins]{tcolorbox}
\usepackage{tabularx}
\definecolor{light_blue}{HTML}{f0f5ff}
\usepackage{color}

\newcommand{\etal}{\textit{et al.$\:$}}
\newcommand{\comma}{,{ }}

\newcommand{\tr}{\mathrm{Tr}}

\def\bra#1{\langle{#1}|}
\def\ket#1{|{#1}\rangle}
\def\braket#1{\langle{#1}\rangle}
\newcommand{\ketbra}[2]{\ket{#1}\!\bra{#2}}

\newcommand{\phdagger}{{\phantom{\dagger}}}
\newcommand{\MBL}{\textrm{MBL}}
\newcommand{\SYK}{\textrm{SYK}}

\let\realhbar\hbar
\renewcommand{\hbar}{}

\definecolor{boxcolor}{HTML}{e5e3fa}
\definecolor{lightgrey}{HTML}{e0e0e0}
\definecolor{basegrey}{HTML}{f1f2f2}
\definecolor{quantum_purple}{HTML}{8f05ff}
\definecolor{volt}{HTML}{f1ffcc}
\definecolor{dark_green}{HTML}{006340}
\definecolor{energy_orange}{HTML}{fc3503}
\definecolor{bright_orange}{HTML}{fff5e6}
\definecolor{energy_yellow}{HTML}{ff7300}
\definecolor{light_yellow}{HTML}{fff9e6}

\begin{document}
\sloppy 

\title{\emph{Colloquium:} Quantum Batteries}

\author{Francesco Campaioli}
\email{francesco.campaioli@rmit.edu.au}
\affiliation{Dipartimento di Fisica e Astronomia “G. Galilei” Università degli Studi di Padova\comma I-35131 Padua\comma Italy, Padua Quantum Technologies Research Center\comma Università degli Studi di Padova\comma I-35131 Padua\comma Italy, ARC Centre of Excellence in Exciton
Science\comma and School of Science\comma RMIT University\comma Melbourne 3000\comma Australia}

\author{Stefano Gherardini}
\email{stefano.gherardini@ino.cnr.it}
\affiliation{CNR-INO, Area Science Park\comma Basovizza\comma I-34149 Trieste\comma Italy, 
SISSA, via Bonomea 265\comma I-34136 Trieste\comma Italy,
European Laboratory for Non-linear Spectroscopy\comma Università di Firenze\comma I-50019 Sesto Fiorentino\comma Italy}

\author{James Q. Quach}
\email{james.quach@csiro.au}
\affiliation{CSIRO\comma Ian Wark Laboratory\comma Bayview Ave\comma Clayton\comma Victoria\comma 3168\comma Australia, The University of Adelaide\comma South Australia 5005\comma Australia}

\author{Marco Polini}
\email{marco.polini@unipi.it}
\affiliation{Dipartimento di Fisica dell'Universit\`a di Pisa\comma Largo Bruno Pontecorvo 3\comma I-56127 Pisa\comma Italy}
\affiliation{Planckian srl\comma Pisa\comma Italy}
\affiliation{Istituto Italiano di Tecnologia\comma Graphene Labs\comma Via Morego 30\comma  I-16163 Genova\comma Italy}
\affiliation{ICFO-Institut de Ci\`{e}ncies Fot\`{o}niques\comma The Barcelona Institute of Science and Technology\comma Av. Carl Friedrich Gauss 3\comma 08860 Castelldefels (Barcelona)\comma Spain}

\author{Gian Marcello Andolina}
\email{gian.andolina@icfo.eu}
\affiliation{ICFO-Institut de Ci\`{e}ncies Fot\`{o}niques\comma The Barcelona Institute of
Science and Technology\comma 08860 Castelldefels (Barcelona)\comma Spain}
\affiliation{JEIP\comma USR 3573 CNRS\comma Collège de France\comma PSL Research University\comma 11 Place Marcelin Berthelot\comma F-75321 Paris\comma France}

\date{\today}

\begin{abstract}
{Recent years have witnessed an explosion of interest in quantum devices for the production, storage, and transfer of energy. In this Colloquium we concentrate on the field of quantum energy storage by reviewing recent theoretical and experimental progress in quantum batteries. We first provide a theoretical background discussing the advantages that quantum batteries offer with respect to their classical analogues. We then review the existing quantum many-body battery models and present a thorough discussion of important issues related to their ``open nature''. We finally conclude by discussing promising experimental implementations, preliminary results available in the literature, and perspectives.}
\end{abstract}

\maketitle
\makeatletter

{\hypersetup{linkcolor=black}
\tableofcontents
}

\section{Introduction}
\label{s:introduction}
Developments in the field of quantum information have generated great expectations that quantum effects like entanglement could be exploited to perform certain tasks with sizable advantages over classical devices~\cite{Horodecki2009}. Theoretical examples for the existence of such advantages have led to considerable research and industry operations in the fields of computations~\cite{Ladd2010,fedorov2022quantum}, cryptography~\cite{Gisin2002_QC,Portman2022}, and sensing~\cite{Giovannetti2011,Degen2017}. The emergence of new quantum technologies, based on these effects, is expected to eventually lead to a disruptive technological revolution~\cite{Acin2018}.

In the past, other technological revolutions have been driven by the development of a new scientific theory. For example, two centuries ago, the success of the first industrial revolution was deeply intertwined with the development of thermodynamics~\cite{Carnot1824,Fermi1956}. As an empirical theory based on laws postulated from experience~\cite{Fermi1956}, thermodynamics has a universal character, offering predictions that are valid for both classical and quantum settings. For example, just like heat cannot flow form a cold to a hot bath, the efficiency of a heat engine based on a quantum system cannot surpass the Carnot limit. Analogously, entanglement cannot be used to extract more work from an energy reservoir~\cite{Hovhannisyan2013}. Thus, at first glance, there seems to be no place for a quantum advantage in thermodynamics. 

However, thermodynamics at equilibrium does not set bounds on how fast energy is transformed into heat and work. Therefore it seems natural to seek thermodynamic quantum advantages in quantum systems that are driven out of equilibrium~\cite{Vinjanampathy2016,Binder2018a}. Indeed, groundbreaking theoretical results in the field of quantum thermodynamics have shown that entanglement generation is linked to faster work extraction, when energy is stored in many-body quantum systems~\cite{Hovhannisyan2013}. These and other results have sparked the interest for quantum systems used as heat engines~\cite{Kieu2004,Uzdin2016} and energy storage devices. This led to the emergence of research on quantum batteries, first introduced in the seminal work by~\citet{Alicki2013}, and the search for quantum effects that improve their performance~\cite{Binder2015a,Campaioli2017}.

Like electrochemical batteries, quantum batteries are temporary energy storage systems. They have a finite energetic capacity and power density~\cite{Julia-Farre2020}, and can lose energy to the environment~\cite{Liu2019,Gherardini2020}.
However, quantum batteries can be charged (or expended) via operations that generate coherent superpositions between different states~\cite{Binder2015}. In quantum batteries composed of many sub-cells, these coherences can form entanglement and other non-classical correlations~\cite{Campaioli2017}, leading to a superextensive scaling of the charging power. This quantum effect, akin to the Heisenberg scaling in quantum metrology and that of Grover's search algorithm~\cite{Campaioli2017}, leads to an advantage over classical devices and has thus been one of the main driving forces of this field.

A major boost to this research field occurred when the first model of a quantum battery---dubbed ``Dicke quantum battery"---that could be engineered in a solid-state architecture, was proposed by~\citet{Ferraro2018}. Since then, many other concrete quantum battery models have been theoretically proposed, from one-dimensional spin-chains~\cite{Le2018} to strongly interacting (Sachdev-Ye-Kitaev) fermionic batteries~\cite{Rossini2020}. More recently, important but preliminary steps towards the experimental implementation of quantum batteries have also been made~\cite{Quach2022,Hu2022,Joshi2022}. 

Meanwhile, theoretical studies have clarified the role of quantum correlations and collective effects towards achieving superextensive power scaling~\cite{Campaioli2017,Andolina2018,Julia-Farre2020,Gyhm2022}, i.e., a charging power that grows faster than the number of sub-cells. Recent studies have also proposed protocols to maximize charging efficiency and precision~\cite{Friis2018,Santos2019a,Rosa2020}, and methods to prevent energy losses due to the environment~\cite{Liu2019,Gherardini2020,Quach2020,Mitchison2021,HernandezPRXQuantum2022}. However, many aspects of the physics of quantum batteries remain unexplored, such as the ultimate limits on energy density, absolute power, and lifetime of energy storage~\cite{Mohan2021}. Furthermore, experimental work on quantum batteries is still in its infancy, and a fully-operational proof of principle is yet to be demonstrated. 

This {\it Colloquium} aims to be a self-contained, pedagogical review of this rapidly developing field. In Sec.~\ref{s:fundamental_theory}, we introduce the theoretical framework to study quantum batteries and look at theorems and bounds of general (i.e.~architecture-independent) validity. We then examine the most prominent models of quantum batteries in Sec.~\ref{s:quantum_battery_models}, focusing on the superextensive scaling of the charging power. The effect of work fluctuations on the precision of charging and work extraction protocols are reviewed in Sec.~\ref{s:charging_precision}. We then discuss open quantum batteries in Sec.~\ref{s:open_quantum_batteries}, where we review approaches for charging and stabilization in the presence of decoherence and energy-loss processes. In Sec.~\ref{s:architectures}, we survey the most promising platforms for the experimental realization of quantum batteries, from exciton batteries based on organic semiconductors to superconducting architectures. With the aim of providing scope and momentum to this emerging research field, we finally conclude by presenting in Sec.~\ref{s:conclusions_and_outlooks} a forward-looking overview of urgent research questions.


\section{Theoretical background and methods}
\label{s:fundamental_theory}
In this Section we formally introduce quantum batteries and the mathematical framework to study their performance, focusing on theorems and results of general validity. Relying on these tools we show that entanglement generation leads to faster charging. We then elucidate the nature of the quantum advantage for the charging power, setting the scene for a survey of the prominent models of many-body quantum batteries.

\subsection{Unitary charging and work extraction}
\label{ss:reversible_work}

In their seminal work, \citet{Alicki2013} define a quantum battery as a $d$-dimensional system whose internal Hamiltonian $H_0$ (or bare Hamiltonian) has non-degenerate energy levels $\epsilon_k < \epsilon_{k+1}$,
\begin{equation}
    \label{eq:quantum_battery}
    H_0 = \sum_{k=1}^d \epsilon_k \ketbra{k}{k}~.
\end{equation}
However, this condition can be relaxed (as in this review) by allowing the eigenvalues to be partially degenerate $\epsilon_k \leq \epsilon_{k+1}$, as long as the Hamiltonian $H_0$ has a non-zero \textit{bandwidth}\footnote{The difference between the maximum and minimum eigenvalue of the Hamiltonian, corresponding to the maximal amount of energy that can be stored in the system.} $w[H_0]:=\epsilon_{\textrm{max}}-\epsilon_{\textrm{min}} > 0$, where $\epsilon_\mathrm{max}$ ($\epsilon_\mathrm{min}$) is the largest (smallest) eigenvalue.
Thus, energy can be stored in this system by preparing it in some excited state $\rho$, such that its energy $\tr[H_0\rho] > \epsilon_1$. Examples of quantum systems that can be used as quantum batteries therefore include (but are not limited to) spins in a magnetic field~\cite{Le2018}, semiconductor quantum dots~\cite{Wenniger2022}, superconducting qubits~\cite{Santos2019a,Dou2022a}, the electronic states of an organic molecule~\cite{Liu2018,Quach2022}, and the states of the electromagnetic field confined in a high-quality photonic cavity~\cite{Friis2018}.
In contrast with its classical counterpart, a quantum battery can be charged via unitary operations that may temporarily generate coherences between its eigenstates $\ket{k}$. Besides minimizing heat production, unitary charging can generate non-classical correlations in many-body quantum batteries, leading to the superextensive charging power scaling discussed in Secs.~\ref{sss:charging_power} and~\ref{s:quantum_battery_models}. While many of the charging protocols considered in this review are based on cyclic unitaries, work extraction and injection have also been extended to the case of non-unitary processes~\cite{Garcia-Pintos2020}, as discussed in Sec.~\ref{s:open_quantum_batteries}.

For the moment, let us focus on the amount of energy that can be reversibly injected (charging) or extracted (discharging) via a cyclic unitary process,
\begin{equation}
\label{eq:unitary_work_extraction}
    \dot{\rho}(t) = - i[H_0+H_1(t),\rho(t)]~,
\end{equation}
where $H_1(t)$ is a Hermitian time-dependent interaction that is turned on at time $t=0$ and off at time $t=\tau$, and where $\dot{\rho}(t)$ represents the time derivative of $\rho(t)$. Note that $\realhbar$ is set to $1$ unless specified otherwise. The energy $W$ deposited in such way, starting from some initial state $\rho_0 := \rho(0)$, is measured with respect to the internal Hamiltonian $H_0$,
\begin{equation}
    \label{eq:energy_deposited}
    W(\tau) = \tr[H_0\rho(\tau)] - \tr[H_0\rho_0]~,
\end{equation}
where $\rho(\tau) = U(\tau;0)\rho_0U^\dagger(\tau;0)$ is obtained from the solution of Eq.~\eqref{eq:unitary_work_extraction}, $U(t;0) = \mathcal{T}\{-i\int_0^t ds [H_0+H_1(s)]\}$ being the time-evolution operator expressed in terms of the time-ordering operator $\mathcal{T}$. Note that if we restrict ourselves to unitary evolution, work injection (charging) and extraction are effectively equivalent tasks. More precisely, the work extracted from the system $W_\mathrm{out}$ is simply related to the energy deposited $W_\mathrm{in}$ by the following relation $W_\mathrm{in} = - W_\mathrm{out}$, where $W_\mathrm{in}$ is obtained as in Eq.~\eqref{eq:energy_deposited}. These subscripts will be often omitted to simplify the notation.

\begin{figure}
    \centering    \includegraphics[width=0.48\textwidth]{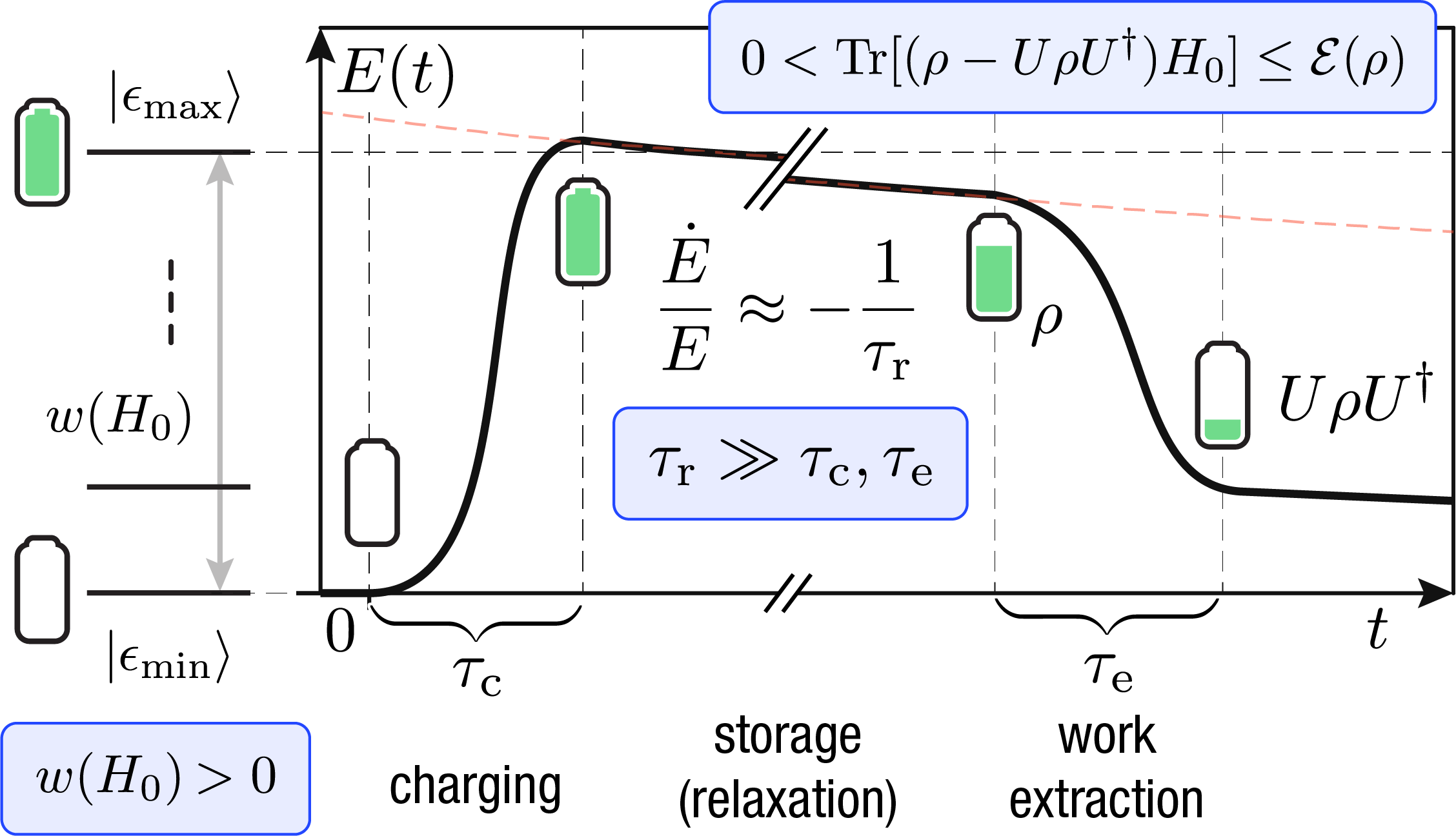}
    \caption{(Color online) A quantum battery is a quantum system with finite energy bandwidth $w(H_0) = \epsilon_\mathrm{\max}-\epsilon_\mathrm{\min}>0$. Charging and work extraction may be performed by means of cyclic unitary operations, as prescribed by Eq.~\eqref{eq:unitary_work_extraction}. 
    An active state $\rho$ and a unitary operation $U$ are needed (i.e., a non-passive state), so that some work $\tr[(\rho - U\rho U^\dagger)H_0] > 0$ can be extracted from the system. Optimally, $U$ extracts the ergotropy $\mathcal{E}(\rho)$, defined in Eq.~\eqref{eq:ergotropy_definition}.
    In practice, a quantum battery may lose energy at some relaxation rate $\tau^{-1}_\mathrm{r}$, due to the interaction with the environment, as discussed in Sec.~\ref{s:open_quantum_batteries}. Ideally, charging/extraction and relaxation time scales should be well separated, $\tau_\mathrm{r}\gg\tau_\mathrm{c},\tau_\mathrm{e}$.}
    \label{fig:battery_processes}
\end{figure}

The processes of reversible charging and work extraction are illustrated in Fig.~\ref{fig:battery_processes} in relation to those of energy loss (or leakage), discussed in Sec.~\ref{s:open_quantum_batteries}. We will now introduce some figures of merit and key concepts, like passive states and ergotropy. By presenting some key results obtained for the tasks of reversible work extraction, we gradually take the reader to the fundamental relation between charging power and the formation of quantum correlations, later reviewed in Secs.~\ref{ss:powerful charging}.

\subsubsection{Ergotropy and passive states}
\label{sss:ergotropy}
Crucially, restricting the dynamics to unitary cycles imposes a bound on the amount of energy that can be deposited or extracted via  Eq.~\eqref{eq:unitary_work_extraction}, which also depends on the initial state of the system $\rho$ and on the control interaction $H_1(t)$. This observation leads to the definition of \textit{ergotropy}~\cite{Allahverdyan2004}, denoted with $\mathcal{E}$ or $W_\mathrm{max}$, as the maximal amount of work that can be extracted from a state $\rho$ via unitary operations:
\begin{equation}
    \label{eq:ergotropy_definition}
    \mathcal{E}(\rho) := \tr[H_0\rho]-\min_{U\in\mathrm{SU}(d)}\big\{\tr[H_0 U\rho U^\dagger]\big\}~.
\end{equation}
Here, the optimization is performed with respect to unitary operators $U$ in the special unitary group $\mathrm{SU}(d)$. As such, the ergotropy is one of the key figures of merit for the performance of quantum batteries~\cite{Tirone2021}.

When no work can be extracted from some state, such state is called \textit{passive}. In other words, a state $\sigma$ is passive when $\tr[H_0\sigma]\leq \tr[H_0U\sigma U^\dagger]$ for all unitaries $U$. It turns out that $\sigma$ is passive if and only if it is diagonal in the basis of the Hamiltonian $H_0$, and its eigenvalues are non-increasing with the energy~\cite{Pusz1978,Lenard1978}:
\begin{equation}
    \label{eq:passive_state_theorem}
    \sigma = \sum_{k=1}^d s_k \ketbra{k}{k}, \;\;\; s_{k+1} \leq s_k~.
\end{equation}

Interestingly, for any state $\rho = \sum_k r_k \ketbra{k}{k}$, there exists a unique\footnote{If $H_0$ has a non-degenerate spectrum.} passive state $\sigma_\rho$ that minimizes the term $\tr[H_0U\rho U^\dagger]$ in Eq.~\eqref{eq:ergotropy_definition}. The state $\sigma_\rho$ is obtained via some unitary operation $U_\rho$ that sorts the eigenvalues of $\rho$ in non-increasing order, $\{r_k\}\to\{r'_k\}$, such that
\begin{equation}
    \label{eq:passive_of_rho}
    \sigma_\rho = U_\rho^\phdagger \rho U_\rho^\dagger = \sum_{k=1}^d r'_k\ketbra{k}{k}~.
\end{equation}
Accordingly, the ergotropy can be expressed in terms of such passive state: $\mathcal{E}(\rho) = \tr[H_0\rho]-\tr[H_0\sigma_\rho]$.

As one would expect from Eq.~\eqref{eq:passive_state_theorem}, all thermal states\footnote{In $G_\beta$, $\beta = 1/k_B T$ is the inverse temperature, or inverse thermal energy, $\mathcal{Z} = \tr\exp[-\beta H_0]$, and $k_B$ is the Boltzmann constant.} $G_\beta = \exp[-\beta H_0]/\mathcal{Z}$ are passive, since they commute with $H_0$ and their eigenvalues do not increase with energy. Less trivial is the fact that the ergotropy of some state $\rho$ is upper-bounded as
\begin{equation}
    \label{eq:ergotropy_lower_bound}
    \mathcal{E}(\rho) \leq \tr[H_0\rho] - \tr[H_0 G_{\overline{\beta}}]~,
\end{equation}
where $\overline{\beta}$ is such that $\rho$ and $G_{\overline{\beta}}$ have the same von Neumann entropy $S(\rho) = -\tr[\rho\log\rho] = S(G_\beta)$~\cite{Alicki2013}.

Since every state of a two-level system (TLS) can be seen as a thermal state, when $d=2$ all passive states are thermal, and Eq.~\eqref{eq:ergotropy_lower_bound} becomes a tight bound. In general, passive states may not be thermal, and finding $\sigma_\rho$ effectively becomes a sorting problem in the eigenbasis of the Hamiltonian $H_0$, whose computational complexity typically scales as $O(d\log d)$~\cite{Zutshi2021}.

\subsubsection{Completely passive states}
\label{sss:completely_passive_states}

A key question in quantum thermodynamics is to determine whether or not quantum phenomena like coherence and entanglement can be harnessed in some thermodynamics task~\cite{Binder2018a}. To address this question,~\citet{Alicki2013} consider the case of composite systems with $N$ constituents, and focus on states of the form $\otimes^{N}\rho:=\bigotimes_{j=1}^N\rho$, i.e.~states that are given by $N$ copies of the same state $\rho$. They consider a system whose Hamiltonian $H^{(N)}_0$ is given by $N$ local copies of $H_0$,
\begin{equation}
    \label{eq:local_hamiltonian}
    H^{(N)}_0 = \sum_{i=1}^N H_i~,
\end{equation}
where $H_i := \mathbb{1}_1\otimes\cdots\otimes\mathbb{1}_{i-1}\otimes H_0\otimes\mathbb{1}_{i+1}\cdots\otimes\mathbb{1}_N$. This system can be seen as a quantum battery given by $N$ non-interacting cells\footnote{Note that, in general, $H^{(N)}$ will have degenerate eigenvalues.}. Later in Secs.~\ref{s:quantum_battery_models} and~\ref{s:architectures} we will discuss possible experimental implementations of such Hamiltonian. 

Interestingly, $N$ copies of a passive state may not form a passive state with respect to $H^{(N)}_0$. It makes therefore sense to define a subclass of passive states, which are dubbed ``\textit{completely passive}''. These are $N$-copy states that are passive for any $N$. Notably, a state $\rho$ is completely passive if and only if it is thermal~\cite{Lenard1978}.

\subsubsection{Bounds on extractable and injectable work}
\label{ss:bounds_extractable_work}

Using the above result,~\citet{Alicki2013} show that the bound in Eq.~\eqref{eq:ergotropy_lower_bound} can be achieved asymptotically in the $N\to\infty$ limit for systems with $N$ constituents. Given a $N$-copy state $\otimes^N\rho$, the ergotropy-per-copy $\varepsilon(N)$ is defined as
\begin{equation}
    \label{eq:ergotorpy_per_copy}
    \varepsilon(N) := \frac{1}{N}\bigg(\tr[H_0^{(N)}\otimes^N\rho]-\tr[H_0^{(N)}\sigma_{\otimes^N\rho}]\bigg)~.
\end{equation}
In the $N\to\infty$ limit, $\varepsilon(N)$ is tightly bounded as in Eq.~\eqref{eq:ergotropy_lower_bound}, $ \lim_{N\to\infty} \varepsilon(N) = \tr[H_0\rho] - \tr[H_0 G_{\overline{\beta}}]$, which follows from the fact that the energy difference between the passive state of the copies $\sigma_{\otimes^N\rho}$ and $\otimes^N G_{\overline{\beta}}$ vanishes in the limit $N\to\infty$~\cite{Alicki2013}. 

These bounds can also be cast for the task of unitary \emph{work injection} by introducing the \textit{anti-ergotropy} and \textit{anti-passive} states\footnote{A state is anti-passive state if it commutes with internal Hamiltonian and its eigenvalues are non-decreasing with the energy.}~\cite{Salvia2021,Mazzoncini2022}, related by the transformation $\beta\to-\beta$.

\subsubsection{Entanglement generation and work extraction/injection}
\label{sss:entanglement_work_extraction}

Since the passive state of the copies $\sigma_{\otimes^N\rho}$ is diagonal in the eigenbasis of the local Hamiltonian $H^{(N)}_0$, it follows that it is also separable\footnote{A state $\rho$ is separable if it is a convex of combination of product states~\cite{Mintert2005}}. However, using local operations (here, permutations\footnote{Local permutations are permutations of the eigenvalues of each individual copy $\rho$.}) of the eigenvalues of $\otimes^{N}\rho$ we can only reach $\otimes^{N}\sigma_\rho$ at best. Therefore, in order to extract the ergotropy---or, equivalently, to inject the anti-ergotropy---, one must use \textit{entangling} operation, i.e.~operations that can generate entanglement between two or more subsystems. 

This observation led~\citet{Alicki2013} to suggest that entanglement generation is needed for optimal reversible work extraction.
However, it was later shown by~\citet{Hovhannisyan2013} that the ergotropy can in fact be extracted from $N$ copies of $\rho$ while keeping their state separable at all times, as long as at least 2-body operations are available. 
The Authors support their conclusion with an explicit protocol to achieve such task, showing that entanglement generation can always be suppressed by adding extra steps in the protocol (where each step is a unitary operation)
used to connect some state $\otimes^N\rho$ to its passive $\sigma_{\otimes^N\rho}$. In other words, there is always a longer path (here, a unitary trajectory) that connects a state to its passive without generating entanglement. The results of Ref.~\cite{Hovhannisyan2013} hint to a relation between entanglement generation and extraction power, answering the key questions of Sec.~\ref{sss:completely_passive_states}, and leading the way towards the quantitative study of such relation.

\subsection{Charging power}
\label{ss:powerful charging}

While so far we have focused on work extraction, the task of \textit{charging} a quantum battery can be treated equivalently, as long as we consider closed systems that evolve according to Eq.~\eqref{eq:unitary_work_extraction}. In this Section we review some notions of ``charging power'' and outline how time-energy uncertainty relations impose bounds on the timescale of charging and work extraction.

\subsubsection{Average and instantaneous power}
\label{sss:charging_power}

For the first time,~\citet{Binder2015a} shift the focus from work extraction to charging.
The Authors consider the \textit{average charging power} $\braket{P}_\tau$ of some unitary process as the ratio between the average deposited energy $W(\tau)$ and the time $\tau$ required to complete the procedure. Let $P(t):=\dot{W}(t)$ be the instantaneous power,
\begin{equation}
    \label{eq:instant_power}
    P(t) = \tr[H_0\dot{\rho}(t)]~,
\end{equation}
intended as time-local energy-gain with respect to the (time-independent) internal Hamiltonian of the battery. 
Then, the average power is given by
\begin{equation}
    \label{eq:average_power}
    \braket{P}_\tau = \frac{W(\tau)}{\tau}~,
\end{equation}
where $\braket{f}_\tau$ denotes the time-average of some function $f(t)$ in the time interval $t\in[0,\tau]$, i.e.~$\braket{f}_\tau:= \frac{1}{\tau}\int_0^\tau dt f(t)$. For simplicity, we will often omit the subscript $\tau$ in $\braket{P}_\tau$, and assume the implicit dependence of the average power on the time interval over which it is calculated.

Arguably, it is always desirable for charging to be as fast as possible. \citet{Binder2015a} therefore seek to maximize the average power $\braket{P}$. While this task has been the main focus of the research on quantum batteries, we mention in passing that~\citet{Binder2015a} also propose an alternative family of objective functions $\mathcal{F}_\alpha[P,W] : = \braket{P}^\alpha \braket{W}^{1-\alpha}$ with $0\leq \alpha\leq1$, which also puts emphasis on the amount of work done and not only on maximizing the power. These objective functions have been only partially explored in this field. 

\subsubsection{A bound on the maximal power}
\label{sss:bound_on_power}

To seek a bound on the average power $\braket{P}$,
\citet{Binder2015a} formulate the charging problem in terms of finding the minimal time required to
reach some target state $\rho^\star$ starting from some initial state $\rho_0$, by means of unitary evolution. This problem, known as quantum speed limit (QSL)~\cite{Giovannetti2003,Deffner2013}, has been extensively studied as an operational interpretation of time-energy uncertainty relations~\cite{Campaioli2020}. When considering pairs of pure states $\ket{\psi}$ and $\ket{\phi}$, the minimal time $\tau$ required to unitarily evolve between them is bounded as
\begin{equation}
    \label{eq:qsl}
    \tau \geq \tau_\mathrm{QSL} = \frac{\arccos|\braket{\psi|\phi}|}{\min\{\braket{E},\braket{\Delta E}\}}~,
\end{equation}
where $\braket{E}$ ad $\braket{\Delta E}$ are the time-averaged expectation value and standard deviation of the total Hamiltonian $H(t) = H_0+H_1(t)$. These are calculated from $E(t) = \tr[H(t)\rho(t)] - \omega_1(t)$, where $\omega_1(t)$ is the instantaneous ground-state energy of $H(t)$, and $\Delta E(t) = \sqrt{\tr[H^{2}(t)\rho(t)]-\tr[H(t)\rho(t)]^2}$~\cite{Deffner2013}. Originally derived as two different bounds by~\citet{Mandelstam1945} ($\braket{\Delta E}$), and~\citet{Margolus1998} ($\braket{E}$), they have been recently unified into the bound of Eq.~\eqref{eq:qsl} by~\citet{Levitin2009}.
The QSL for the evolution of pure states is tight, and the bound in Eq.~\eqref{eq:qsl} is attainable, in the absence of restrictions on the interaction Hamiltonian $H_1(t)$ of Eq~\eqref{eq:unitary_work_extraction}. In this case, a prescription for the interaction $H_1(t)$ to saturate the bound is
\begin{equation}
    \label{eq:pure_orthogonal_qsl_saturation}
    H_1(t) = \lambda(t)\Big[ -H_0 + \alpha\ketbra{\psi}{\phi}+\alpha^*\ketbra{\phi}{\psi} \Big]~,
\end{equation}
where $\lambda(t) = 1$ for $0< t \leq \tau$ and zero otherwise,
for orthogonal pairs of states\footnote{See~\cite{Campaioli2020} for the case of non-orthogonal pairs of pure states,~\cite{Campaioli2019} for arbitrary pairs of mixed states.}, where $\alpha$ is a non-zero complex number. By imposing the interaction Hamiltonian to have finite energy, e.g., via the operator norm $\lVert H(t) \rVert = E_\mathrm{max}$ for some $E_\mathrm{max}>0$, the minimal evolution time between orthogonal pure states becomes $\tau = \pi/(2E_\mathrm{max})$. From this result,~\citet{Binder2015a} conclude that the following inequality must hold true
\begin{equation}
    \label{eq:power_bound}
    \braket{P}\leq 2W E_\mathrm{max}/\pi~,
\end{equation}
when considering a charging process between orthogonal states, associated with an injection of energy $W$. The right-hand side of the bound in Eq.~\eqref{eq:power_bound} has clearly units of power, provided that $\realhbar$ is reintroduced.

Note that the QSL does not provide a prescription for the optimal driving interaction $H_1(t)$, but just a bound on the minimal time of evolution. The problem of optimizing $H_1(t)$ to minimize $\tau$, known as the \textit{quantum brachistochrone}~\cite{Borras2008}, is generally hard, and solutions\footnote{Solutions to the quantum brachistochrone problem are of central importance in quantum optimal control, and find application in quantum computing, nuclear magnetic resonance and other fields of quantum technology~\cite{Caneva2009}. 
} are known only for a handful of particular cases. For the case of composite systems, the relation between power and entanglement outlined by~\citet{Hovhannisyan2013} can be quantitatively framed in terms of the QSL and quantum brachistochrone, as reviewed in Sec.~\ref{ss:quantum_advantage}. 

\subsection{Quantum advantage}
\label{ss:quantum_advantage}

In this Section, following the steps of~\citet{Binder2015a} and~\citet{Campaioli2017}, we review how a quantum advantage can be achieved for the task of charging a quantum battery. To seek a formal definition we consider a figure of merit $\Gamma$, built in such a way that $\Gamma > 1$ when a quantum advantage is achieved. A first formulation appears in the work by~\citet{Campaioli2017}, where the Authors considered
\begin{equation}
    \label{eq:quantum_advantage}
    \Gamma := \frac{\braket{P_\mathrm{q}}}{\braket{P_\mathrm{c}}}~.
\end{equation}
Here, $\braket{P_\mathrm{q}}$ ($\braket{P_\mathrm{c}}$) represents the maximal charging power of a \textit{quantum} (\textit{classical}) charging protocol.

While the distinction between \textit{quantum} and \textit{classical} is here very loose, initial consensus on this definition was based on the idea that a charging protocol is defined to be quantum mechanical when it generates non-classical correlations~\cite{Campaioli2017,Le2018,Ferraro2018}, such as entanglement or quantum discord~\cite{Modi2011}. Nevertheless, the notion of an advantage that is genuinely quantum mechanical is still a matter of investigation~\cite{Andolina2019a,Rossini2020}, as we will see below in Sec.~\ref{sss:genuine_quatum_advantage}.

\subsubsection{Local and global charging}
\label{sss:global_charging}

Let us consider again the quantum battery given by the composite system with Hamiltonian $H_0^{(N)}$ defined in Eq.~\eqref{eq:local_hamiltonian}. When depositing energy onto such system by means of some unitary process, we may consider two different scenarios: i) Local charging, if each subsystem is charged independently, and ii) \textit{global}\footnote{Local and global are otherwise referred to as \textit{parallel} and \textit{collective} in some references. Here we use global for generality, in order to make a distinction between collective and quantum advantage as discussed in Sec.~\ref{sss:genuine_quatum_advantage}.} charging, if the control interaction couples different subsystems. In this scenario, it is insightful to consider a charging task $\ket{G}\to\ket{E}$, which deposits some energy $W$, from the $N$-copy passive state $\ket{G}:=\otimes^N \ket{g}$ (a \textit{dead} battery) to the active state $\ket{E}:=\otimes^N\ket{e}$ (a \textit{charged} battery), where $\ket{g}:=\ket{1}$ and $\ket{e}:=\ket{d}$ are the ground and excited states of the $d$-dimensional system defined in Eq.~\eqref{eq:quantum_battery}.

If we can only use local interactions, the best approach to achieve the $\ket{G}\to\ket{E}$ task is to drive each subsystem independently at the speed provided by the QSL. This can be done by using the local (i.e.~1-body interactions only) charging Hamiltonian
\begin{equation}
    \label{eq:local_charging_hamiltonian}
    H_\|^{(N)} = \sum_{i=1}^N V_i~,
\end{equation}
where each $V_i$ is a local copy of $-H_0 + \alpha\ketbra{e}{g}+\alpha^*\ketbra{e}{g}$ acting on the subsystem $i$, as discussed in Sec.~\ref{sss:bound_on_power}. If instead we can use arbitrary $N$-body interactions, the best approach is to use the global Hamiltonian~\cite{Binder2015a}
\begin{equation}
    \label{eq:global_charging_hamiltonian}
    H_\sharp^{(N)} = -H^{(N)}_0 + \alpha_N\ketbra{E}{G} + \alpha_N^*\ketbra{E}{G}~.
\end{equation}

In order to make a fair comparison between these two charging approaches, we impose that both Hamiltonians have the same operator norm, i.e.~that $\lVert H_\|^{(N)} \rVert = \Vert H_\sharp^{(N)} \rVert = E_\mathrm{max}$.
We then obtain $E_\mathrm{max} = |\alpha_N| = N |\alpha|$, which implies the minimal time of local charging is $\tau_\| = N \pi / (2 E_\mathrm{max})$, that is, $N$ times larger than that of global charging $\tau_\sharp = \pi / (2 E_\mathrm{max})$. We then arrive at an explicit expression for the quantum advantage of this charging task,
\begin{equation}
    \label{eq:quantum_advantage_explicit}
    \Gamma = \frac{\braket{P_\sharp}}{\braket{P_\|}} = \frac{\tau_\|}{\tau_\sharp} = N~.
\end{equation}
Since the global Hamiltonian generates entanglement between the subsystems during the charging process, this advantage was initially considered to be exclusive to quantum systems. In particular, \citet{Binder2015a} concluded that entanglement generation was responsible for the $N$-fold speed-up, and necessary to obtain some non-trivial advantage $\Gamma > 1$. 

However, as discussed in the Sec~\ref{sss:role_of_enanglement}, a quantum advantage can be achieved even without generating entanglement, at the price of reducing the amount of energy extracted or injected~\cite{Campaioli2017}. For this reason, the nature of the advantage $\Gamma > 1$ has been explored by many Authors~\cite{Andolina2018,Farina2019,Andolina2019a,Andolina2020thesis,Rossini2020,Julia-Farre2020,Fan2021,Carrasco2022,Abah2022,Mondal2022}, and is currently a matter of investigation. In the remainder of this Section we review the main results of these research efforts, aiming to clarify the role that quantum correlations, coherences, and many-body interactions have on $\Gamma$.

\subsubsection{Role of entanglement, correlations and coherence}
\label{sss:role_of_enanglement}
\begin{figure}
    \centering
    \includegraphics[width=0.48\textwidth]{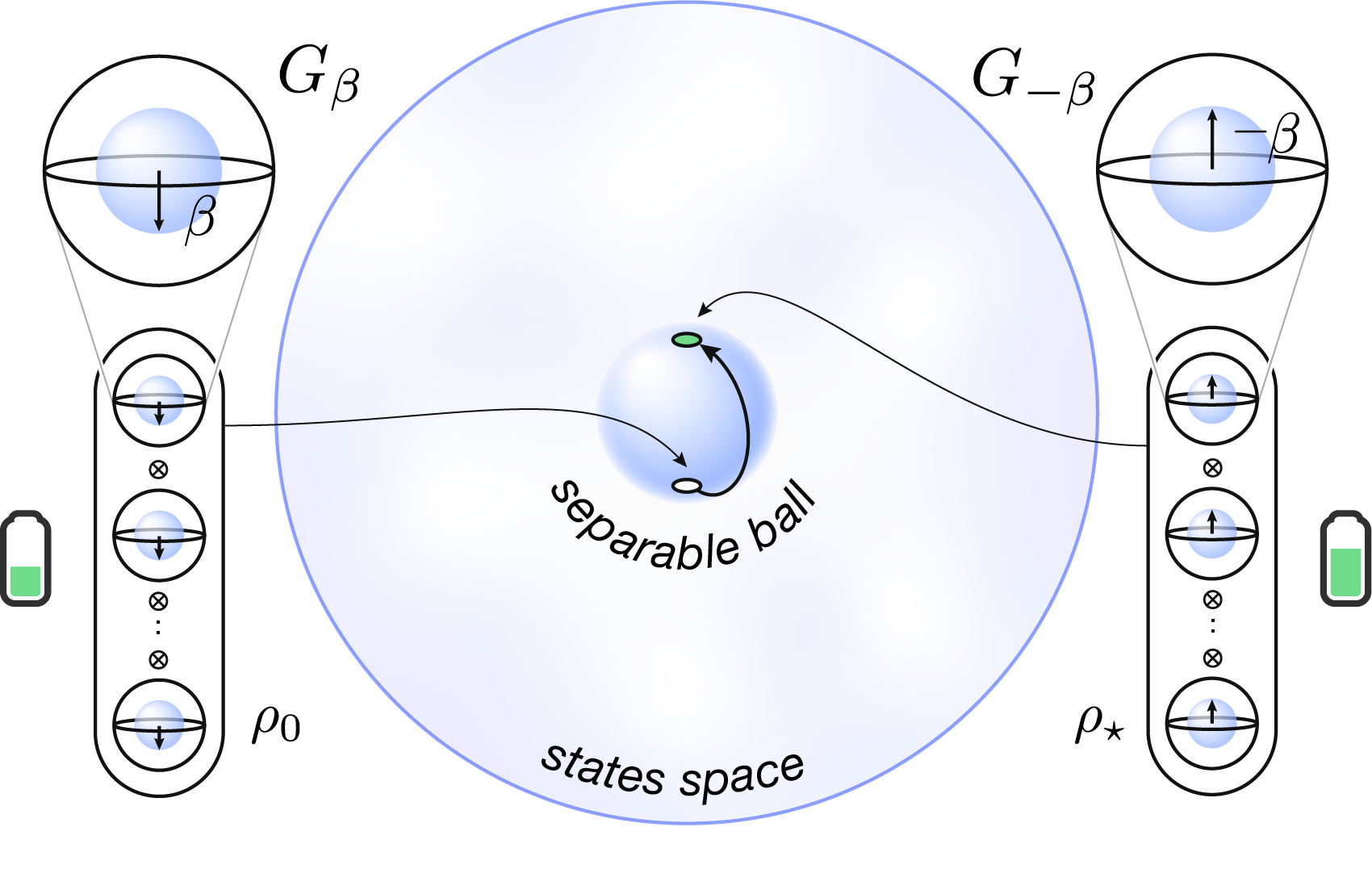}
    \caption{(Color online) A scalable power advantage can be obtained without generating entanglement, at the expense of reducing the total work exchanged $
    W = \tr[H_0(\rho_\star-\rho_0)]$, as discussed in Sec.~\ref{sss:role_of_enanglement}. The initial state $\rho_0=\otimes^{N}G_\beta$, given by $N$ copies of a TLS state $G_\beta$ with sufficiently small $\beta$, falls into the \textit{separable ball}, a open region of the states space with spherical symmetry~\cite{Gurvits2005}.  Any unitary charging procedure keeps the state in the separable ball and may generate \textit{quantum discord}. Adapted from Ref.~\cite{Campaioli2018c}.}
    \label{fig:discord_example}
\end{figure}

With an explicit example, illustrated in Fig.~\ref{fig:discord_example}, \citet{Campaioli2017} prove that it is possible to achieve an advantage that scales with a power-law of $N$, like that in Eq.~(\ref{eq:quantum_advantage_explicit}) even without generating entanglement, at the cost of dramatically reducing the amount of energy injected (or extracted). The example involves an initial completely passive $N$-copy state $\rho_0 = \otimes^N G_\beta$, for some inverse temperature $\beta$, and its corresponding active state $\rho^\star = \otimes^N G_{-\beta}$, both calculated with respect to the local internal Hamiltonian $H_0$. By choosing a sufficiently small $\beta > 0$, the state $\rho_0$ is within the region of the states space known as the \emph{separable ball}~\cite{Aubrun2006}. The latter is a spherically symmetric region of the space of states that is centered on the maximally mixed state, and that contains only separable state. Any unitary charging procedure that drives $\rho_0$ to $\rho^\star$ will keep $\rho(t)$ within the separable ball at all times. Nevertheless, the charging power of the time-optimal global Hamiltonian is $N$-fold\footnote{When using a constraint on the operator norm of the charging Hamiltonian, as described in Sec.~\ref{sss:global_charging}.} larger than the one of the local optimal Hamiltonian. However, in this case no entanglement is generated during the evolution. 

Two remarks are in order. First, while the power advantage is still scalable, i.e., yielding a power that scales with $N^2$, the total deposited energy and the absolute charging power reduce as $\beta$ decreases. In other words, the closer $\rho_0$ is to the maximally mixed state, the less energy will be extractable from $\rho^\star$~\cite{Campaioli2017}. Since the separable ball is a comparatively small region of the space of states~\cite{Gurvits2005}, achieving quantum advantage without entanglement generation comes at the cost of a considerably reduction in the exchanged work $W=\tr[H_0(\rho_\star-\rho_0)]$. This example shows the importance of entanglement, which is necessary to jointly maximize the injected work and the charging power. Second, while no entanglement is generated during the evolution in this example, this type of Hamiltonian is known to produce \textit{quantum discord}~\cite{Modi2011}, which need not to vanish for separable states.

The role of coherence on the charging power has also been explored independently from that of quantum correlations in Ref.~\cite{Caravelli2020}. There, the Authors bound the extractable work and power of a charging protocol using the Hilbert-Schmidt coherence of the density matrix in the basis of the driving Hamiltonian. The bounds express the intricate interplay between the coherences of the state, the internal Hamiltonian and the interaction. These and other results~\cite{Shi2022} depict the nontrivial role of coherences in quantum batteries, which is still object of investigation. For example, coherences in the battery subspace are usually beneficial, and can increase the ergotropy. On the other hand, as discussed in Sec.~\ref{s:open_quantum_batteries}, coherence between the battery and a charger subsystem can be detrimental when they lead to a mixed reduced state of the quantum battery. Recently, a study on the capacity of energy-storing quantum systems has further investigated the relation between extractable work~\cite{Yang2023}, entanglement and coherence, which should be investigated for some of the quantum battery models discussed in Sec.~\ref{s:quantum_battery_models}.

\subsubsection{Role of interaction order}
\label{sss:role_of_interaction}
\begin{figure}
    \centering
    \includegraphics[width=0.40\textwidth]{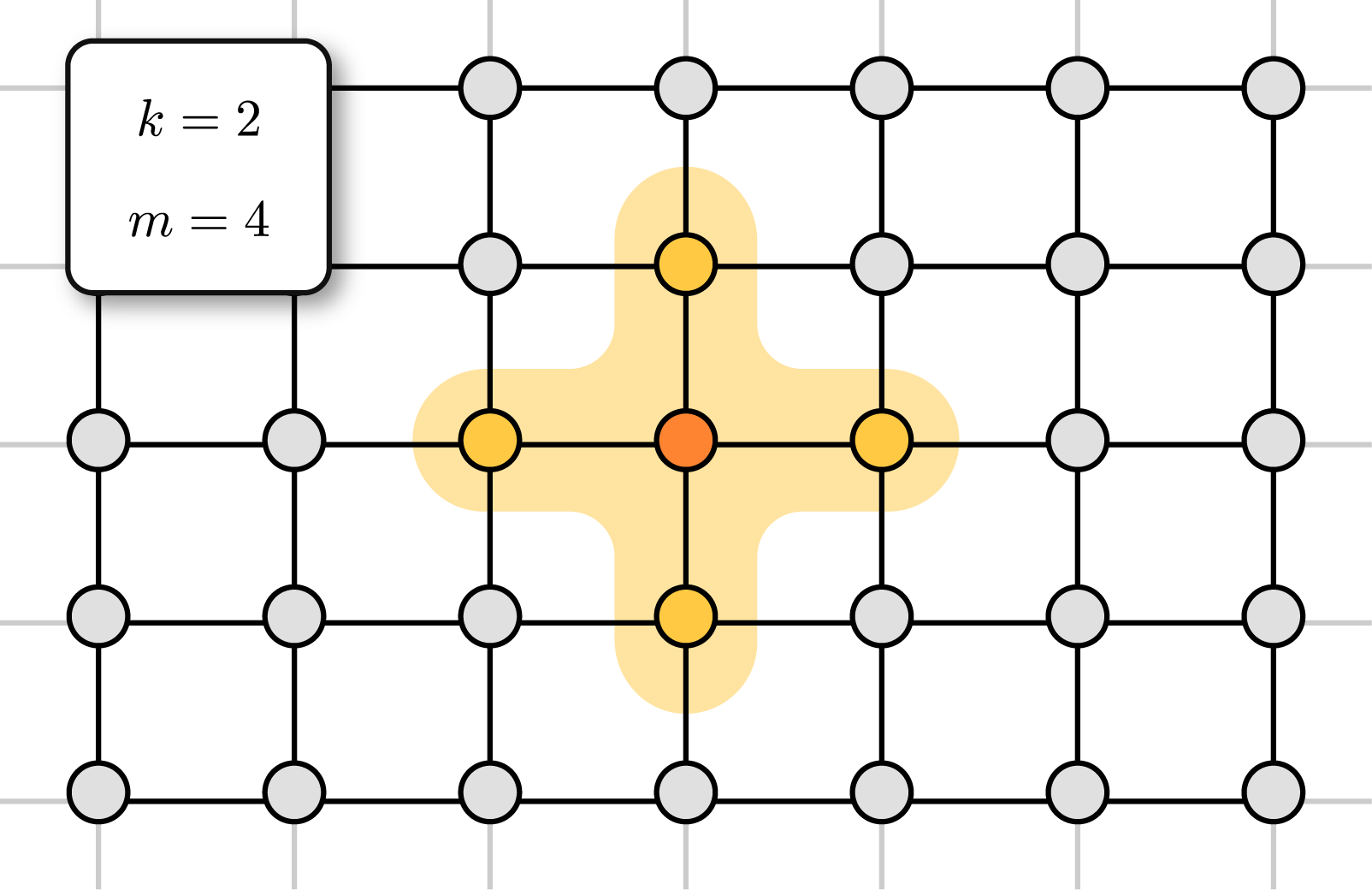}
    \caption{(Color online) The bound on the quantum advantage of Eq.~\eqref{eq:quantum_advantage_order_participation} depends on the \textit{interaction order} $k$ (the number of subsystems directly coupled by some interaction) and the \textit{participation number} $m$ (the number of subsystems that are coupled with any given subsystem). In the 2d lattice with nearest-neighbor interaction shown here, each subsystem directly interacts via pairwise interactions ($k=2$) at most with four other neighbors ($m=4$). Adapted from Ref.~\cite{Campaioli2020}.}
    \label{fig:lattice}
\end{figure}

The examples considered in the previous sections to achieve $\Gamma = N$ require the use of $N$-body interactions, i.e., interactions that directly couple $N$ subsystems. However, the vast majority of physical systems displays at most 2-body interactions, with a few exceptions such as the 3-body (Efimov) interaction in nuclear~\cite{Naidon2017} and atomic systems~\cite{Roy2013}. It is therefore important to determine whether a quantum advantage can be achieved with such limitations on the Hamiltonian.
To address this question,~\citet{Campaioli2017} examined the scaling of $\Gamma$ for $N$-body systems when the charging Hamiltonian is limited to $k$-body interactions, where $k$ is sometimes referred to as the \textit{interaction order}. The Authors show that, for composite systems with local internal Hamiltonians $H_0^{(N)}$, the quantum advantage is bounded as
\begin{equation}
\label{eq:quantum_advantage_order_participation}
    \Gamma < \gamma[k^2(m-1)+k]~,
\end{equation}
where $\gamma$ is a constant that does not depend on $N$, and where the participation number $m$ is the number of subsystems that are coupled with a given one. The meaning of $k$ and $m$ with respect to $N$ is illustrated in Fig.~\ref{fig:lattice}.
The Authors also prove that for the case of a circuit model with at most $k$-body gates, illustrated in Ref.~\cite{Campaioli2020}, the quantum advantage is limited to $\Gamma < \gamma k$, and it is conjectured that the same holds for an arbitrary $k$-body Hamiltonian. 
This conclusion, in line with the findings of~\citet{Hovhannisyan2013}, has sparked work to determine the properties of the charging Hamiltonian that is required to achieve a scalable power advantage~\cite{Le2018,Ferraro2018,Andolina2018,Farina2019,Andolina2019}.

Recently,~\citet{Gyhm2022} have indeed demonstrated that a scalable quantum advantage for the charging power cannot be achieved without global operations. The Authors bound the instantaneous power $P(t)$ using the norm of the commutator that generates the evolution, $|P(t)|\leq\lVert[H_0,H_1(t)]\rVert$, for composite systems of batteries driven by Hamiltonians with at most $k$-body interactions. In this way, they obtain a tight bound for the quantum advantage, $\Gamma \leq \gamma k$, where $\gamma$ does not depend on $N$ and $k$. This important result has cemented the role of the interaction order $k$, which can be seen as a resource necessary to achieve a speed-up in the charging and work extraction tasks. In Secs.~\ref{sss:genuine_quatum_advantage} and~\ref{sss:syk} we will discuss some proposed approaches to achieve a quantum advantage that scales with a power-low of $N$, such as quantum batteries based on the Sachdev-Ye-Kitaev (SYK) model~\citet{Rossini2020,Kim2022}. We will also discuss how super-extensive charging or extraction power can be achieved via many-body interactions that do not require the generation of genuine quantum correlations.

\subsubsection{Genuine quantum advantage}
\label{sss:genuine_quatum_advantage}
The results on the role that entanglement and other quantum correlations have on charging and extraction power have provoked a discussion around the definition of quantum advantage $\Gamma$.~\citet{Andolina2018} present a new figure of merit to fairly compare quantum protocols against classical ones, different from that of Eq.~\eqref{eq:quantum_advantage}. \citet{Andolina2018} aim is to quantitatively distinguish charging speed-ups emerging from many-body interactions (which may indeed have a  fundamentally classical nature) from those that stem from the quantum-mechanical nature of the system. They propose that an unbiased comparison can only be made when a quantum system defined by some Hamiltonian $H$ admits a classical analogue $\mathcal{H}$. 

If the dynamics of the quantum system is governed by Eq.~\eqref{eq:unitary_work_extraction}, that of the classical system is governed by Hamilton's equations of motion,
\begin{equation}
    \label{eq:hamilton}
    \dot{q}_i = \partial_{p_i} \mathcal{H}, \;\; \dot{p}_i = -\partial_{q_i} \mathcal{H}~.
\end{equation}
Here, $\vec{q}$ and $\vec{p}$ are a set of canonically conjugated variables (coordinates and momenta, respectively) such that $\{q_i,p_i\} = \delta_{ij}$ where $\{\cdot,\cdot\}$ denotes the Poisson brackets. Thus, \citet{Andolina2018} suggest to measure the advantage stemming from genuine quantum mechanical effects by comparing the power scaling of the quantum and classical Hamiltonians, $H$ and $\mathcal{H}$ respectively, i.e.,
\begin{equation}\label{eq:genuine_quantum_advantage}
    R := \frac{\Gamma_\mathrm{q}}{\Gamma_\mathrm{c}}~. 
\end{equation}
Charging protocols that yield $R> 1$ are said to be characterized by a \emph{genuine quantum advantage}. Here, $\Gamma_\mathrm{q} = \braket{P_\sharp}/\braket{P_\|}$ represents the advantage of using global operation over local ones, as in Eq.~\eqref{eq:quantum_advantage_explicit}, and $\Gamma_\mathrm{c}$ is the same quantity but calculated for the corresponding  classical model.

The ratio $R$ has been studied for a variety of charging models~\cite{Andolina2018,Farina2019,Andolina2019,Andolina2020thesis,Fan2021,Carrasco2022,Abah2022,Mondal2022}. Interestingly, the genuine quantum advantage $R$ is model dependent and, even for the same model, it depends on the microscopic parameters of the charging Hamiltonian. Note that $R$ is well defined only for quantum battery models which admit a classical analogue. There are plenty of models of quantum batteries that do not have a classical analogue such as, for example,  the Sachdev-Ye-Kitaev (SYK) model considered by~\citet{Rossini2020} and discussed in Sec.~\ref{sss:syk}. In these cases, it is not possible to evaluate $R$ and any advantage $\Gamma>1$ is expected to be stemming from the underlying quantum many-body dynamics~\cite{Rossini2020}.
\begin{figure}
    \centering
    \includegraphics[width=0.48\textwidth]{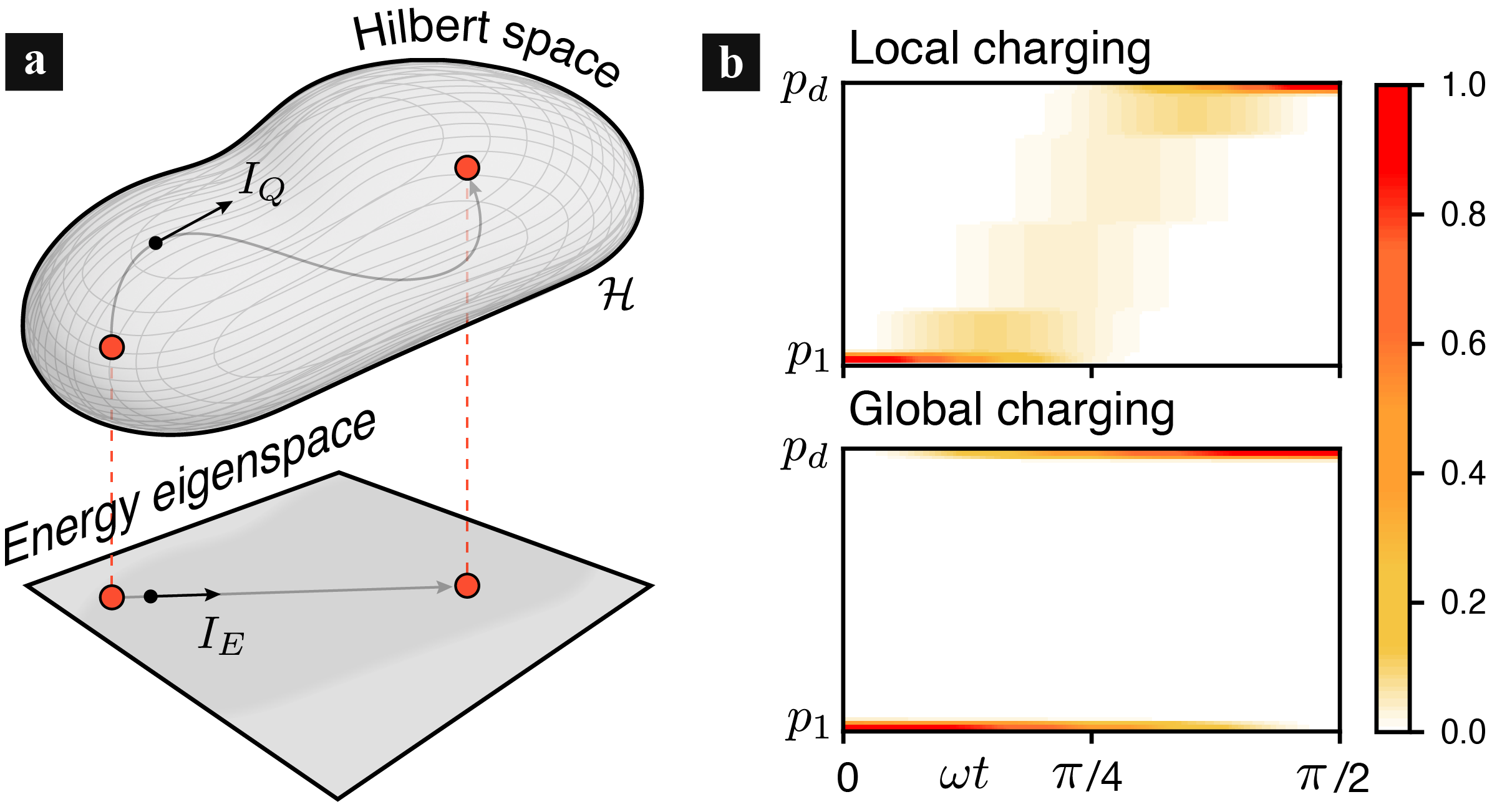}
    \caption{(Color online) (a) The speed $I_Q$ of quantum dynamics in the Hilbert space $\mathcal{H}$ and the speed $I_E\leq I_Q$ in the energy eigenspace of the battery Hamiltonian are related to the quantum Fisher information. The instantaneous charging power is bounded by $I_E$ as in Eq.~\eqref{eq:insta_power_bound}. (b) Dynamics of the energy levels $p_k$ associated to the internal Hamiltonian $H^{(N)}$ of Eq.~\eqref{eq:local_hamiltonian} during local ($H^{(N)}_1$) and global ($H^{(N)}_N$) charging, where $\omega$ is the associated charging frequency. The global charging scheme takes a \textit{shortcut} in the Hilbert space by driving the battery through the maximally entangled state $\ket{\psi}\propto\ket{G}+\ket{E}$. Adapted from Ref.~\cite{Julia-Farre2020}.}
    \label{fig:gqa}
\end{figure}

A crucial step in disentangling the contribution of collective interactions from that of quantum correlations was made by~\citet{Julia-Farre2020}. These Authors separate the two contributions by using a geometric approach, where the instantaneous charging power $P(t)$ is studied. First, they notice that the quantum Fisher information~\cite{Barndorff-Nielsen2000} is related to the speed in the Hilbert space ($I_Q$) and the speed in the energy eigenspace ($I_E$), while the variance of the battery Hamiltonian $\Delta E_\mathrm{B}^2 = \tr[H_\mathrm{B}^2\rho_\mathrm{B}]-\tr[H_\mathrm{B}\rho_\mathrm{B}]^2$ encodes non-local correlations between subsystems. Interestingly, if the battery Hamiltonian $H_\mathrm{B}$ is made of a sum of local terms, $H_\mathrm{B}=\sum_{i=1}^N h_i$,
it is possible to write $\Delta E_\mathrm{B}^2 = \Delta E_\mathrm{B}^2|_{\rm Loc}+\Delta E_\mathrm{B}^2|_{\rm Ent}$,
where
\begin{align}
  \label{eq:BoundHb}
   \Delta E_\mathrm{B}^2|_{\rm Loc} & \equiv  \sum_i \Big[ \tr[h_i^2 \rho_\mathrm{B}] - \tr[h_i \rho_\mathrm{B}]^2 \Big]~,  \\
  \Delta E_\mathrm{B}^2|_{\rm Ent}~ & \equiv 
  \sum_{i\neq j} \Big[ \tr[ h_i  h_j \rho_\mathrm{B}] - \tr[ h_i \rho_\mathrm{B}]\tr[  h_j \rho_\mathrm{B}] \Big]~. 
\end{align}
The first quantity $\Delta E_\mathrm{B}^2|_{\rm Loc}$, being a sum of local terms, scales linearly with $N$ by construction.
On the other hand, $\Delta E_\mathrm{B}^2|_{\rm Ent}$, whose explicit form can be immediately linked to correlations between sites $i$ and $j$, may display a super-linear scaling with $N$ if correlations between different battery units are developed. With this approach, \citet{Julia-Farre2020} obtain a bound on the instantaneous power $P(t)$,
\begin{equation}
    \label{eq:insta_power_bound}
    P(t) \leq \sqrt{\Delta E_\mathrm{B}^2(t) I_E(t)}~,
\end{equation}
whose geometric interpretation is illustrated in Fig.~\ref{fig:gqa}, as well as similar bound on the average power $\braket{P}_\tau \leq \sqrt{\braket{\Delta E_\mathrm{B}^2}_\tau \braket{I_E}_\tau}$~\cite{Julia-Farre2020}.

As detailed in Sec.~\ref{sss:originofadvatageTD}, \citet{Julia-Farre2020} use this method to confirm that the advantage discussed in~\cite{Binder2015a} and~\cite{Campaioli2017} is \textit{quantum}, since it stems from non-local correlations, while that of the Dicke battery (Sec.~\ref{ss:dicke_quantum_battery}) is \textit{collective}, since it stems from a larger speed in the energy eigenspace, characterized by a larger Fisher information $I_E$. The importance of the bound on power of Ref.~\cite{Julia-Farre2020} is striking, since it enables the discrimination between a \textit{collective} advantage in power, emerging from the Fisher information, and a genuine \textit{quantum} advantage, due to quantum correlations between battery cells. It is also worth noticing that their approach does not require the definition of an analogue classical Hamiltonian $\mathcal{H}_\mathrm{B}$.


\section{Models of many-body batteries}
\label{s:quantum_battery_models}

At the microscopic level, matter is granular as it can be described in terms of a collection of $N$ elementary units, such as atoms or molecules.  In many cases, the behavior of a macroscopic system can be accurately described by assuming that these units behave independently, leading to {\it intensive} quantities (such as pressure) that do not depend on the size of the system, or {\it extensive quantities} (such as energy or volume) that scale linearly with the number of constituent units $N$.
 However, in some cases, interactions between the elementary units can give rise to collective effects that cannot be explained by the properties of a single unit. 
 These collective effects can result in macroscopic quantities showing a super-extensive (i.e.~$N^\alpha$ with $\alpha>1$) scaling in the number of units $N$. 
 An example of paramount importance is provide by the Dicke model~\cite{Dicke54}, where an ensemble of $N$ atoms collectively radiates with a superextensive intensity that scales as $N^2$, i.e.~enhanced by a factor $N$ with respect to ordinary fluorescence. In the latter case, atoms emit independently. In the former, synchronization of the electrical dipoles of the atoms occurs, leading to an enhanced emission which has been dubbed ``superradiance''~\cite{RevModPhys.85.1083,gross_pr_1982}.

Superradiant emission has been measured in a plethora of different systems, such as Rydberg atoms in a cavity \cite{Kaluzny83} or color centers in diamonds \cite{Angerer2018}.
The concept of super-absorption \cite{Higgins14}, where collective effects are used to speed up energy absorption, has also been proposed based on the time-reversal symmetry of the dynamics of an ensemble of emitters interacting with electromagnetic radiation~\cite{Yang2021}.

In 2018,~\citet{Ferraro2018} proposed a model of a battery that could be engineered in a solid-state device. The quantum many-body model comprised $N$ two-level systems (TLSs) coupled to the very same mode of an electromagnetic field. For these reasons, the battery was termed ``Dicke quantum battery". Three were the reasons for proposing such a model: i) the fact that the $N$ TLSs were coupled to the same cavity mode effectively provided a way to couple all the TLSs together during the non-equilibrium dynamics; ii) the collective effects displayed by the Dicke model discussed above were thought to be useful in determining an advantage in the charging process; iii) as we will see below in Sec.~\ref{s:architectures}, Dicke models can be realized experimentally in a variety of ways. A first step towards the realization of a Dicke battery has been experimentally realized in an excitonic system~\cite{Quach2022}, where collective effects in the charging process of a superabsorber have been observed.

In the remainder of this Section we review the charging properties of a variety of many-body battery models.

\subsection{Charging protocols and figure of merits}

Before looking at many-body model of quantum batteries, let us briefly present here a general framework for describing the charging process of a quantum battery.

The battery is a quantum system $\rm B$, described by a Hamiltonian $H^{(0)}_{\rm B}$, consisting of $N$ identical units. It is often assumed that the battery Hamiltonian is the sum of $N$ local Hamiltonians, $H^{(0)}_{\rm B}=\sum_{i=1}^N h_i^{({\rm B})}$. The battery is initially prepared in the ground state of the battery Hamiltonian, and energy is injected into it through a charging protocol with a time duration $\tau$. The specifics of different protocols will be discussed later.
After the protocol, the energy stored in the battery $W(\tau)$ is given by Eq.~\eqref{eq:energy_deposited}, i.e.~$W (\tau)= \tr[{H}^{(0)}_{\rm B} \rho_{\rm B}(\tau)]$ where $\rho_{\rm B}(\tau)$ is the state of the battery at time $\tau$ and the second term in Eq.~\eqref{eq:energy_deposited} vanishes as the ground-state energy is fixed to be zero.

The aim is to find charging processes that maximize the battery energy $W(\tau)$, while at the same time minimizing $\tau$ (i.e.~displaying fast-charging). A figure of merit that weighs energy and time is given by the average charging power, $\langle P\rangle_\tau={ W(\tau)}/{\tau}$, which was introduced in Eq.~\eqref{eq:average_power}. 
A \textit{good} charging process should provide a sufficient amount of energy to the battery in a short time, thus displaying a high charging power. In what follows we describe two different ways of modeling the charging phase of a battery.

\subsubsection{Direct charging protocol}

In a {\it direct charging protocol}, the Hamiltonian of the battery is externally changed by suddenly switching on a suitable interaction Hamiltonian $H^{(1)}_{\rm B}$ for a finite amount of time $\tau$, while switching off $H^{(0)}_{\rm B}$~\cite{Binder2015a,Campaioli2017,Rossini19,Rossini2020,Le2018,Rosa2020}, as depicted in Fig.~\ref{fig:charging_protocols}~{(a)}.
The most general charging protocol without a charger has already been presented in Eq.~\eqref{eq:unitary_work_extraction}, which describes energy being supplied to a quantum system through an arbitrary unitary operation. Here, we focus on a more practical scenario, in which only a single fixed operator, $H^{(1)}_{\rm B}$, can be modulated over time.
The battery dynamics is therefore dictated by the following time-dependent Hamiltonian
\begin{equation}
  \label{eq:protocol}
 { H}_{\rm B}(t) = H^{(0)}_{\rm B} + \lambda(t) \big( H^{(1)}_{\rm B} - H^{(0)}_{\rm B}\big)~,
\end{equation}
where $\lambda(t)$ is a classical parameter representing an external control. 
The term $H^{(1)}_{\rm B}$ is referred to as the {\it charging} Hamiltonian, and $H^{(0)}_{\rm B}$ is the {\it battery} Hamiltonian. In this case, all energy is injected in the battery by the external classical control.
For simplicity, we have assumed that the control is a step function that is equal to $1$ during the charging time $t\in[0,\tau]$ and zero otherwise. This sudden quench can be realized, for instance, in superconducting circuits, where Hamiltonians can be rapidly modulated by adjusting parameters like external magnetic flux or gate voltage. These adjustments enable precise control over energy levels and interactions within the system on timescales significantly shorter than the system's dynamics \cite{Koch2007}.
 
In general, more complex controls $\lambda(t)$ can also be considered. 
It is worth noting that this study falls within the field of quantum control \cite{Dalessandro2007}. In quantum control, external fields or parameters, such as electromagnetic fields or gate voltages, are meticulously adjusted to steer the evolution of quantum states towards desired outcomes with the highest possible accuracy given the resources at disposal.

\subsubsection{Charger-mediated protocol}

\begin{figure}
    \centering
    \includegraphics[width=0.45\textwidth]{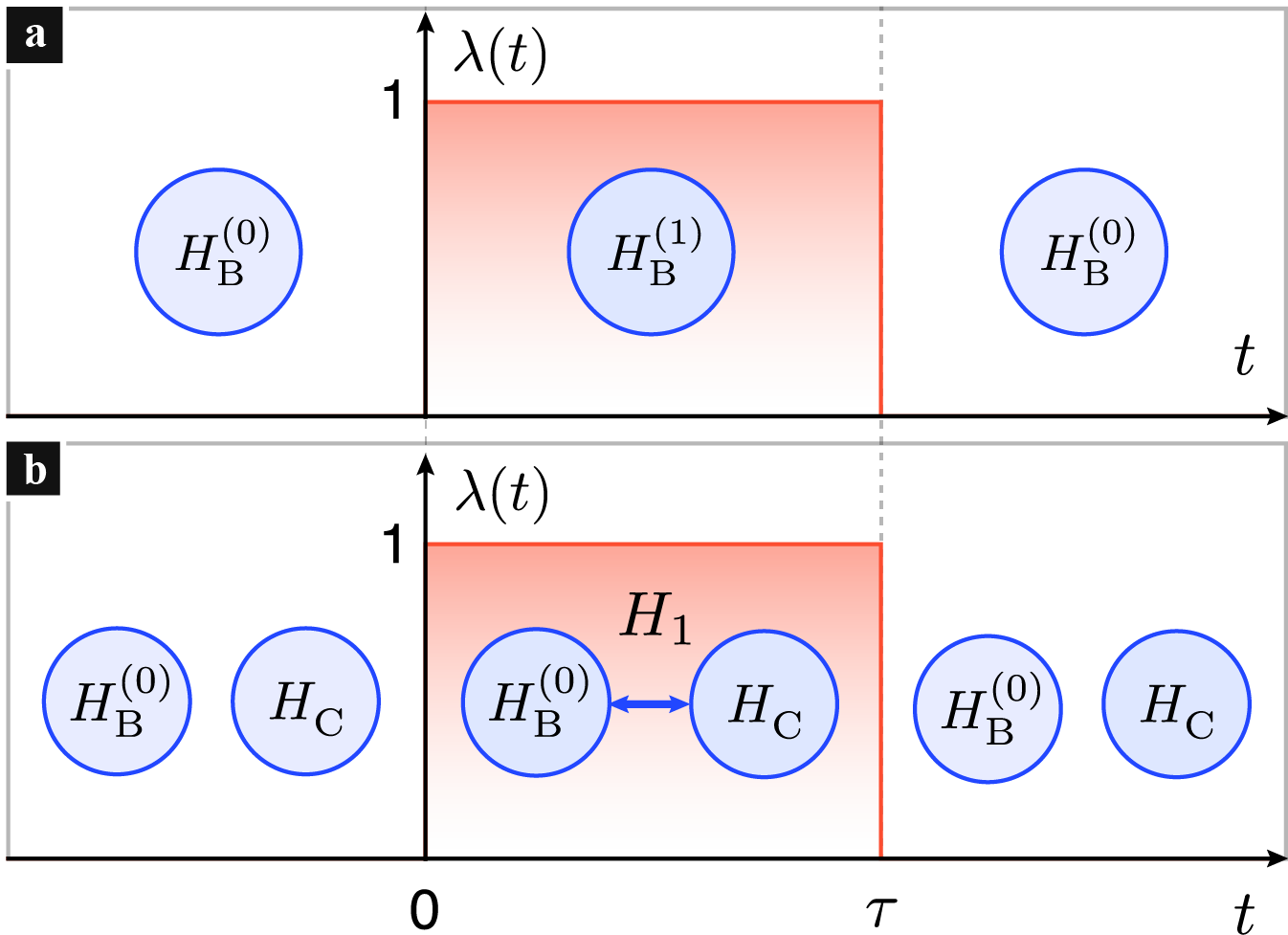}
    \caption{
    (a) In a time-dependent  {\it direct charging protocol}, the Hamiltonian of the battery is externally changed by suddenly quenching on an interaction Hamiltonian $H^{(1)}_{\rm B}$ for a finite amount of time $\tau$, while the battery Hamiltonian $H^{(0)}_{\rm B}$ is turned off during the charging process.
    (b) In a {\it charger-mediated protocol}, a ``charger" ${\rm C}$, initially containing some input energy, transfers energy to the battery $\rm B$. The charging process is enabled by an interaction Hamiltonian $H_{1}$ that couples the charger and the battery for a finite amount of time $\tau$. }
    \label{fig:charging_protocols}
\end{figure}

In the {\it charger-mediated protocol} \cite{Andolina2018,Farina2019,Ferraro2018,Andolina2019,Crescente2020,Delmonte2021}, an auxiliary system, referred to as the "charger" C, is introduced. Initially, the charger contains a certain amount of energy, which is intended to be transferred to the battery B. The charging process occurs due to an interaction between the charger and battery that lasts a finite amount of time $\tau$, as depicted in Fig.~\ref{fig:charging_protocols}~{(b)}.
Thus, the global Hamiltonian of the composite ${\rm CB}$ system is
\begin{align}
\label{eq:protocol1}
&{H}(t)={H}_0+\lambda(t){H}_{1}~, \\
\label{eq:total_bare_Hamiltonian}
&H_0 = {H}_{\rm C}+{H}^{(0)}_{\rm B},
\end{align}
is the bare Hamiltonian (or internal Hamiltonian), ${H}_{1}$ is the interaction Hamiltonian, and $\lambda(t)$ is an external control.
However, in this protocol, the composite system ${\rm CB}$ (described by the composite state $\rho_{\rm CB}(t)$) can exchange work with the environment through the classical control, leading to some energy ambiguity.

The external control $\lambda(t)$ modulates the interaction between the charger and the battery. It may introduce an energy cost $W_{\rm sw}(\tau)$ \cite{Andolina2018} at switching times, specifically when the external control is switched on at $t=0$ and off at $t=\tau$. It is worth noting that, in such a case, the total system energy $E(t) = \tr[H(t)\rho_{\rm CB}(t)]$ remains constant except at the switching times ($0$ and $\tau$). The energy cost can be thus expressed as $W_{\rm sw}(\tau) \equiv {\rm lim}_{\epsilon \to 0} \{[E(\tau+\epsilon)-E(\tau-\epsilon)] + [E(\epsilon)-E(-\epsilon)]\}$. By ensuring that ${H}_{1}$ only exchanges well-defined excitations of the bare Hamiltonian $H_0$, i.e., $[{H}_{0},{H}_{1}]=0$, the energy transfer from the charger to the battery becomes unambiguous as the energy cost vanish, $W{\rm sw}(\tau)=0$.
If the interactions are non-commuting, i.e., $[{H}_{0},{H}_{1}] \neq 0$, the final energy of the quantum battery is partially supplied by the modulation of $\lambda(t)$ instead of being purely provided by the charger.
This introduces an element of arbitrariness in the charging protocol, given that the energy transfer does not exclusively occur between the charger and the battery~\cite{Ferraro2018}.  

\subsection{Charging properties of the Dicke quantum battery}
\label{ss:dicke_quantum_battery}
\subsubsection{The Dicke battery}
In this Section, we discuss the Dicke battery, a quantum battery based on the Dicke model \cite{Dicke54}. 
The Dicke model describes the collective interaction of an ensemble of $N$ TLSs atoms with a single mode of the cavity field:
\begin{equation}
    \label{eq:DickeModel}
    H_{\rm Dicke}=\omega_{\rm c } \hat{a}^{\dagger} \hat{a}+ \omega_{0} \sum_{i=1}^N \hat{\sigma}_i^{z} +  g\sum_{i=1}^N \hat{\sigma}_i^{x} \left(\hat{a}^{\dagger}+\hat{a}\right)~. 
\end{equation}

Here, $\hat{a}$ ($\hat{a}^{\dagger}$) annihilates (creates) a cavity photon with frequency $\omega_{\rm c}$, $\omega_{0} $ is the resonant frequency of a TLs, $\hat{\sigma}_i^{\alpha}$ with $\alpha=x,y,z$ are the components of the Pauli operators $\hat{\sigma}^{\alpha}_i$ of the $i$-th TLS, $g$ is the TLS-cavity coupling parameter. 
Cavities are typically composed of two or more mirrors that reflect light back and forth, creating a standing wave of electromagnetic radiation~\cite{RevModPhys.85.1083}, whose frequencies are determined by the cavity's geometry. In this context, the single mode approximation is typically employed since one of the cavity modes is designed to be resonant with the atomic transition frequency. This resonant mode dominantly interacts with the atoms, while the interaction with off-resonant modes is negligible.
Note that when deriving the Dicke model from a microscopic underlying model, the light-matter coupling $g$ is found to scale as $1/\sqrt{V}$ where $V$ is the volume of the cavity. This scaling has significant implications that will be discussed in Sec.~\ref{sss:TDlimit}.

The Dicke model is a theoretical framework that can describe the phenomenon of \textit{superradiant} emission, which was first predicted by~\citet{Dicke54}. This model also exhibits an {\it equilibrium} quantum phase transition between a normal phase and a superradiant phase~\cite{hepp_lieb,WangHioe1973}, in which the number of photons in the ground state scales extensively with $N$. 

\citet{Ferraro2018} proposed the use of the Dicke model for a quantum battery, due to its experimental feasibility and relation with superradiant emission. In addition, the Dicke model represents a situation where all atoms are collectively coupled to the same cavity mode. Thus, tracing out this cavity mode results in a scenario where all the two-level systems (TLSs) interact with each other, i.e., the so-called all-to-all interaction (corresponding to a participation number $m=N$, see Sec.~\ref{sss:role_of_interaction}). 
These characteristics made the Dicke model an intriguing framework for the study of quantum batteries.

In the Dicke battery, the charging is performed via a \textit{charger-mediated protocol}, where the cavity acts as a charger while the TLSs are seen as the battery system. During the charging, one aims at transferring the energy of the cavity to the TLSs.
The quantum dynamics of this Dicke charging protocol is described by the following Hamiltonian terms
\begin{align}
\label{eq:DickeQB}
H_{\rm C}&= \omega_{\rm c } \hat{a}^{\dagger} \hat{a}~,\\
\label{eq:HQBlocal}
H^{(0)}_{\rm B}&=\frac{\omega_{0}}{2}\sum_{i=1}^N \left(\hat{\sigma}_i^{z}+\mathbb{1}\right)~,\\
\label{eq:HintDicke}
H_{1}&= g \sum_{i=1}^N \hat{\sigma}_i^{x} \left(\hat{a}^{\dagger}+\hat{a}\right)~,
\end{align}
to be compared with the generic {\it charger-mediated protocol} defined by Eq.~\eqref{eq:protocol1}, (${H}(t)={H}_0+\lambda(t){H}_{1}$). During the charging dynamics, where the time $t$ is in the interval $[0,\tau]$ the evolution of the state is dictated by the Dicke Hamiltonian of Eq.~\eqref{eq:DickeModel}.
First, the system is initialized in the state 
\begin{equation}
    \label{eq:initial_dicke}
    \ket{\psi^{(n)}(0)}=\ket{n}\otimes\ket{G},
\end{equation}
where $\ket{n}$ is a Fock state of $n$ photons in the cavity and $\ket{G}=\otimes^N\ket{g}$, $\ket{g}$ being the ground state of each TLS.Since the goal is to fully charge the TLSs, it is convenient to inject into the cavity a number $n$ of photons equal (at least) to $N$. \citet{Ferraro2018} take $n=N$.  Also, in order to favor the exchange of energy between the charger and battery, \citet{Ferraro2018} enforce the resonant condition $\omega_{0}  = \omega_{\rm c}$. Finally, \citet{Ferraro2018} choose the charging time to be the optimal charging time $\tilde{\tau}$, the time that maximize the average power, i.e.~$\langle P \rangle_{\tilde{\tau}}=\textrm{max}_\tau [\langle P\rangle_\tau]$.

\subsubsection{Parallel vs collective charging}

\begin{figure}
    \centering
    \includegraphics[width=0.48\textwidth]{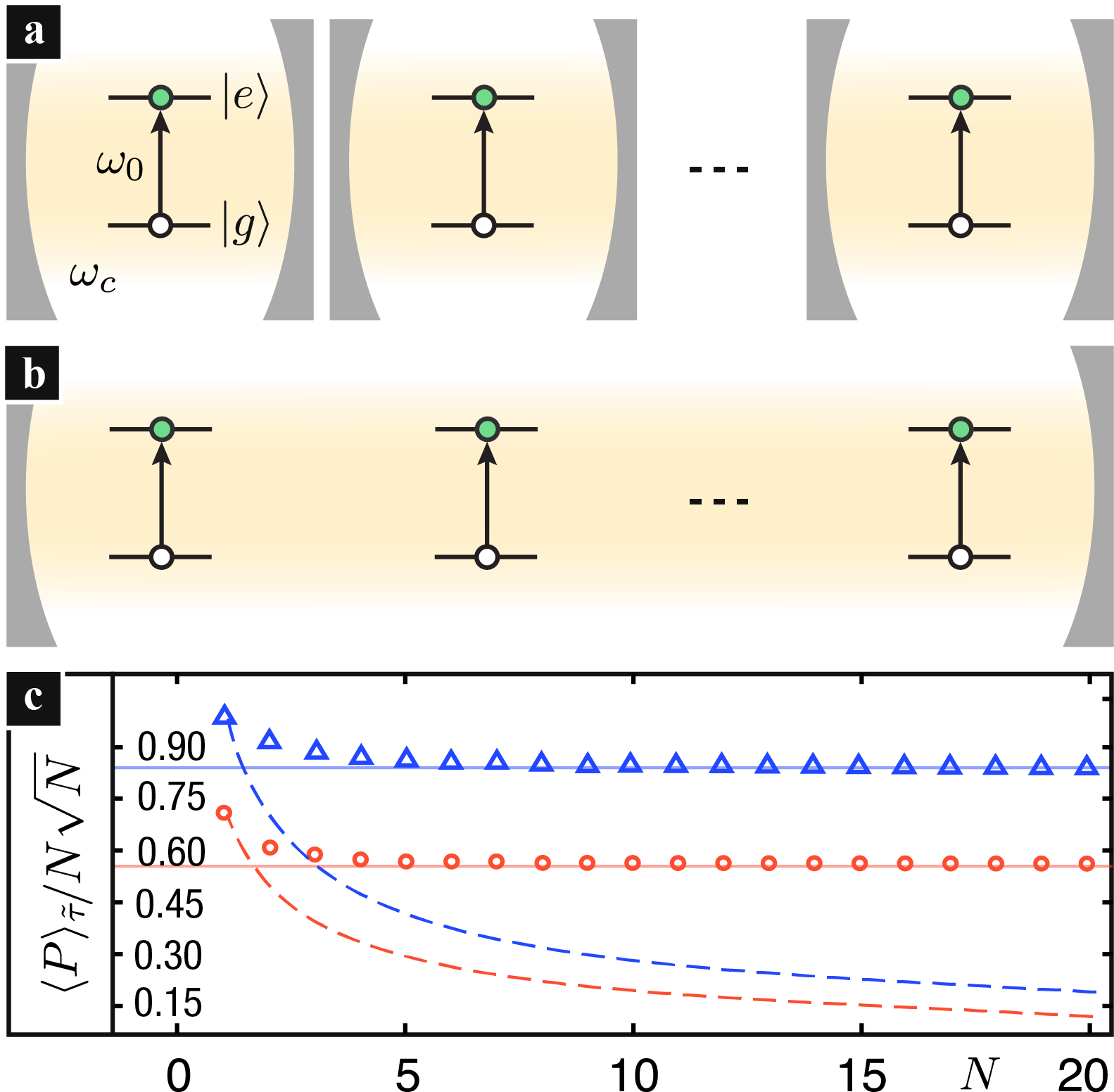}
    \caption{(Color online) (a) A {\it parallel charging protocol} for an array of identical Rabi batteries. The single battery unit consists of a two-level system with transition energy $\omega_0$ between the ground $\ket{g}$ and
excited state $\ket{e}$. Each two-level system is coupled to its own cavity hosting a single mode with frequency $\omega_{\rm c}$.  The corresponding {\it collective charging protocol}: a Dicke quantum battery is composed by $N$ two-level systems coupled to the very same photonic mode.
(c) The maximum average power $\braket{P}_{\tilde{\tau}}$---divided by the factor $N\sqrt{N}$---is plotted as a function of the number $N$ of qubits. Solid (dashed) lines refer to the collective (parallel) protocol. The collective advantage manifests as a saturation to a constant (solid lines) for $N\gg 1$ of the collective-case maximum power divided by $N\sqrt{N}$. Red lines correspond to the weak-coupling regime (i.e.~$g/\omega_0=0.05$), while the blue ones to the strong-coupling regime (i.e.~$g/\omega_0 =0.5$).  Adapted from Ref.~\cite{Ferraro2018}.}
    \label{fig:parallel_collective}
\end{figure}

In order to quantify whether charging the TLSs via the collective coupling $H_{1}$ in Eq.~(\ref{eq:HintDicke}) yields an advantage, \citet{Ferraro2018} compare the maximum charging power $\langle {P_{\sharp}} \rangle_{\tilde{\tau}}$ that can be obtained in the {\it collective charging case}---i.e.~via the Dicke interaction term in Eq.~(\ref{eq:HintDicke})---with the maximum charging power $\langle {P_{\|}} \rangle_{\tilde{\tau}}$ that can be achieved through ``{\it parallel charging}''.
In the latter charging scheme, $N$ identical systems, each comprising a {\it single} TLS interacting with its own cavity containing a single photon ($n=1$), are considered. Each system is a Rabi battery, i.e.~a battery described by the Rabi model \cite{Braak11}, 
\begin{equation}
    \label{eq:RabiModel}
    H_{\rm Rabi}=\omega_{\rm c } \hat{a}^{\dagger} \hat{a}+ \omega_{0} \hat{\sigma}^{z} +  g\hat{\sigma}^{x} \left(\hat{a}^{\dagger}+\hat{a}\right)~. 
\end{equation}
Parallel and collective charging protocols are illustrated in Fig.~\ref{fig:parallel_collective}~{(a)} and~{(b)}, respectively. 

The Authors of Ref.~\citet{Ferraro2018} show that, in the limit $N\gg 1$,
\begin{equation}
\label{eq:power_ratio}
    \frac{ \langle {P_{\sharp}} \rangle_{\tilde{\tau}}}{\langle {P_{\|}}\rangle_{\tilde{\tau}}} \sim \sqrt{N}~.
\end{equation}
This advantage stems from the superextensive scaling of the collective power, $\langle {P_{\sharp}} \rangle_{\tilde{\tau}} \sim N^{3/2}$, and from the extensive scaling of the {\it parallel} charging power $\langle {P_{\|}} \rangle_{\tilde{\tau}} \sim N$, as shown in Fig.~\ref{fig:parallel_collective}~{(c)}. Note that the quantity $\langle {P_{\|}} \rangle_{\tilde{\tau}}$ scales extensively with $N$ because, by construction, the parallel charging scheme is free of collective behavior.
Since in both cases, 
the energy $\langle H^{(0)}_{\rm B}\rangle_{\tilde{\tau}}$ of the battery scales extensively with the number of battery units $N$, this advantage corresponds to a speed-up in the optimal charging time $\tilde{\tau}\sim 1/\sqrt{N}$.

The Dicke battery therefore displays a collective speed-up in the charging time, outperforming the {\it parallel charging} protocol by a $\sqrt{N}$ factor. The Dicke battery model has been widely studied by researchers who have investigated different aspects of it and also proposed generalizations. For instance, the charging process has been optimized in the simplified case where the semi-classical limit is taken~\cite{Crescente2020,Zhang2019a}. Authors have also studied whether this speed-up has a quantum or collective origin ~\cite{Andolina2019a,Julia-Farre2020}  and a two-photon version of the model~\cite{Crescente2020a,Delmonte2021,Crescente2022c}, where an atomic excitation can be converted in two resonant photons and viceversa.
Other aspects of the Dicke model, such as the task of energy extraction, have also been examined~\cite{Andolina2019}. In the remainder of this Section, we will review these issues in more detail.
\subsubsection{The origin of the charging advantage}
\label{sss:originofadvatageTD}

In this Section we provide an in-depth discussion of the microscopic origin of the advantage displayed by a Dicke quantum battery, as reported in Eq.~(\ref{eq:power_ratio}). Using the arguments presented above in Sects.~\ref{sss:role_of_enanglement} and \ref{sss:genuine_quatum_advantage}, we will show that the Dicke battery's collective speed-up has a many-body collective origin rather than a genuine quantum one. Before providing a formal proof of this statement, we offer a series of observations that point towards it.

A first hint of the fact that the Dicke battery's charging speed-up is not related to entanglement can be found in the work by~\citet{Andolina2018}. In this Article, a simplified version of the above mentioned Rabi battery is studied. It consists of a single TLS described by the Hamiltonian $H^{(0)}_{\rm B}=\omega_{0}(\hat{\sigma}^{z}+\mathbb{1})/2$, which is charged by a cavity mode initialized in a Fock state with $n$ photons. The simplification with respect to the Rabi model lies in the fact that the interaction Hamiltonian $H_1$ chosen by~\citet{Andolina2018} contained only rotating terms, i.e.~$H_1= g(\hat{\sigma}^{-}\hat{a}^{\dagger}+\hat{\sigma}^{+}\hat{a})$, where $\hat{\sigma}^{+}$ ($\hat{\sigma}^{-}$) is the Pauli creation (annihilation) operator. The commutation relation $[H^{(0)}_{\rm B},H_1]=0$ is therefore satisfied.
In this elementary case, the time required to reach the maximum energy (denoted as $\bar{\tau}$) can be calculated analytically, and it was found to scale as $\bar{\tau}\sim 1/\sqrt{n}$. This simple example demonstrates that it is possible to speed up the charging process by a factor $1/\sqrt{n}$ for a single TLS by initially placing $n$ photons in the cavity. Only one excitation is transferred from the charger to the quantum battery in this protocol, with the remaining $n-1$ excitations act as a ``catalytic resource'' to increase the speed of the process. Since there is only one battery unit in this protocol, this example also illustrates that entanglement between different units is not necessary to achieve a charging speed-up. Along these lines,~\citet{Zhang} analyze the Dicke battery without assuming $n=N$, showing that the charging times scales as $\bar{\tau}\sim 1/\sqrt{n}$ in the limit $n\gg N$.

Another hint came from the analysis of a Dicke battery in the case in which the cavity is initialized in a coherent state with an average number $N$ of photons. In this case, the charging dynamics leads to very low entanglement between the charger and battery~\cite{Andolina2019}. Despite this low level of entanglement, the Dicke battery still exhibits a collective speed-up $\bar{\tau} \sim 1/\sqrt{N}$ during the charging phase.

Finally, the Authors of Ref.~\cite{Andolina2019a} questioned the quantum origin of the charging speed-up of a series of many-body batteries, including the Dicke one. This study was based on a comparison between quantum mechanical many-body batteries and {\it corresponding} classical models. The correspondence was obtained at the Hamiltonian level and, in the dynamics, by replacing quantum commutators with classical Poisson brackets, as discussed in Sec.~\ref{sss:genuine_quatum_advantage}. Here, we focus specifically on the Dicke battery. Using the fact that the Dicke model has a well established classical analogue~\cite{deAguiar92,Rodriguez18,Carlos19}, one can easily construct a classical Dicke battery, which is described by the following Hamiltonian
\begin{equation}\label{eq:DickeCl}
\mathcal{H}^{(0)}_{\rm B}=N\omega_0\frac{1+\cos(\theta)}{2}~.
\end{equation}
The charger is described by a classical harmonic oscillator
\begin{equation}\label{eq:DickeCl1}
\mathcal{H}_{\rm C}=\frac{\omega_0}{2}\Big(p_a^2+q_a^2\Big)~.
\end{equation}
Finally, charging occurs via the following classical interaction Hamiltonian 
\begin{equation}\label{eq:DickeCl2}
\mathcal{H}_{1}=g\sqrt{2}Nq_a \sin(\theta)\cos(\phi)~.
\end{equation}
Here, $(p_a,q_a)$ and $(\frac{N}{2}\cos(\theta),\phi)$ are conjugate variables~\cite{Carlos19}. The classical protocol in Eq.~(\ref{eq:DickeCl},\ref{eq:DickeCl1},\ref{eq:DickeCl2}) has to be interpreted as a charger-mediated protocol, dictated by a classical Hamiltonian $\mathcal{H}(t)=\mathcal{H}^{(0)}_{\rm B}+\mathcal{H}_{\rm C}+\lambda(t)\mathcal{H}_{1}$.

\citet{Andolina2019a} calculated the charging power of the classical Dicke Hamiltonian, finding that the collective advantage $\Gamma_\mathrm{c}$ introduced in Eq.~\eqref{eq:genuine_quantum_advantage} scales like $\sqrt{N}$. This should be contrasted with the quantum collective advantage $\Gamma_{\rm q}$ of the Dicke battery (see Eq.~\eqref{eq:DickeQB}), which displays the same scaling, i.e.~$\Gamma_{\rm q}\sim \sqrt{N}$.

Hence, for the case of a Dicke battery, the ratio $R$ defined in Eq.~\eqref{eq:genuine_quantum_advantage} was found not to scale with $N$. For Dicke batteries, $R$ depends on the value of the coupling constant $g$ that controls the interaction between the charger and the battery itself. This analysis clarified that no {\it quantum} advantage $R$ scaling  with $N$ can be found in the charging dynamics of the Dicke battery. Similar results were found for other many-body battery models too.

\subsubsection{Charging advantage in the thermodynamic limit}\label{sss:TDlimit}

As discussed in Sec.~\ref{sss:genuine_quatum_advantage},~\citet{Julia-Farre2020} derived a bound for the average charging power, $\langle P\rangle_{\tau}\leq\sqrt{\langle\Delta E_\mathrm{B}^2 \rangle_{\tau}\langle I_E\rangle_{\tau}}$, which allows to distinguish a genuine entanglement-induced speed-up (arising from the first term, the variance of the battery Hamiltonian $\langle\Delta E_\mathrm{B}^2 \rangle_{\tau}$) from a collective many-body speed-up (stemming from the second term, the quantum Fisher information $ \langle I_E\rangle_{\tau}$). The Authors of this work employed the bound on power discussed in Sec.~\ref{sss:genuine_quatum_advantage} to study the charging dynamics of the Dicke battery introduced in Eqs.~\eqref{eq:DickeQB}-\eqref{eq:HintDicke}. In the analysis, $\tau$ was fixed to maximize the energy stored in the battery, i.e.~$\tau=\bar{\tau}$. However, a different normalization with respect to that in Eq.~\eqref{eq:HintDicke} was used for the interaction Hamiltonian $H_1$. Indeed, \citet{Julia-Farre2020} studied the charging dynamics of the Dicke battery by replacing
\begin{equation}
\label{eq:TDscaling}
   g\to g_{\rm TD}\equiv \frac{g}{\sqrt{N}}
\end{equation}
in Eq.~\eqref{eq:HintDicke}.

The normalization mentioned is frequently used when analyzing the phase diagram of the Dicke model, as described by ~\cite{hepp_lieb} and ~\cite{WangHioe1973}. This normalization guarantees that the total energy and energy fluctuations of the Dicke model defined by Eqs.~\eqref{eq:DickeQB}-\eqref{eq:HintDicke} exhibit a well-defined, extensive behavior in the thermodynamic limit. In this limit, one takes $N\to \infty$, $V\to \infty$ while maintaining the ratio $N/V$ constant. Here, $V$ represents the volume of the cavity, which is assumed to scale linearly with the number $N$ of TLSs to ensure a constant density $N/V$ of TLSs. We will come back to this issue in more detail below.

\citet{Julia-Farre2020} derived analytically a bound on the scaling of the quantum Fisher information $I_{E}(t)$, obtaining $\sqrt{I_{E}(t)}\lesssim \sqrt{N}$. Furthermore, they found that, in the weak-coupling $g\ll\omega_0$ limit, the variance of the battery Hamiltonian (see also Sec.~\ref{sss:genuine_quatum_advantage}) $\sqrt{\langle\Delta E_\mathrm{B}^2 \rangle_{\tau}}$ scales as $\sim N^{0.44}$. The bound $\sqrt{\langle\Delta E_\mathrm{B}^2 \rangle_{\tau}\langle I_E\rangle_{\tau}}$ was therefore found to scale sub-linearly with $N$, i.e.~$\sqrt{\langle\Delta E_\mathrm{B}^2 \rangle_{\tau}\langle I_E\rangle_{\tau}}\lesssim N^{0.94}$.
On the contrary, in the strong-coupling $g=\omega_0/2$ regime, the term $\sqrt{\langle\Delta E_\mathrm{B}^2 \rangle_{\tau}}$ was found to scale as $\sim N^{0.92}$, yielding $\sqrt{\langle\Delta E_\mathrm{B}^2 \rangle_{\tau}\langle I_E\rangle_{\tau}} \lesssim N^{1.42}$.
The Authors also found that in the strong-coupling regime, the average power scales extensively as $\langle P \rangle_{\tau} \sim N$, despite the quantum enhancement in $\sqrt{\langle\Delta E_\mathrm{B}^2 \rangle_{\tau}}$. This is because---in this specific case---the bound derived is significantly loose and far from being saturated, with $\langle P \rangle_{\tau}\ll \sqrt{\langle\Delta E_\mathrm{B}^2 \rangle_{\tau} \langle{I_{E}}\rangle_\tau}$. This explains why, in the particular case of the Dicke battery, the bound fails to accurately predict the scaling behavior of the charging power.

We now observe that, if the normalization in Eq.~\eqref{eq:HintDicke} is adopted, as in the work by~\cite{Ferraro2018}, one finds that $\sqrt{\langle{I_{E}}\rangle_\tau}$ scales linearly in $N$, even if $ \sqrt{\langle\Delta E_\mathrm{B}^2 \rangle_{\tau}}\sim \sqrt{N}$ and the bound on power $\langle P \rangle_{\tau}$ scales superextensively as $N\sqrt{N}$, without requiring any quantum superextensive scaling in $\sqrt{\langle\Delta E_\mathrm{B}^2 \rangle_{\tau}}$.
This means that to accelerate the charging process, entanglement generation is not essential in this case. In other words, the battery state does not need to explore highly correlated subspaces, as $ \sqrt{\langle\Delta E_\mathrm{B}^2 \rangle_{\tau}}$ scales as $\sqrt{N}$ even in the absence of system correlations (see Eq.\eqref{eq:BoundHb}). In agreement with the findings of~\citet{Andolina2019a},~\citet{Julia-Farre2020} conclude that the Dicke battery model does not exhibit a genuine quantum advantage. Only recently, a many-body model displaying a quantum advantage has been found and will be discussed in Sec.~\ref{sss:syk}.

We would like to conclude this Section by mentioning that, while the replacement $g\to g/\sqrt{N}$ is necessary when considering the thermodynamic limit with a fixed density $N/V$, there are many instances where $N\gg 1$ is large but finite and there is no need to scale the cavity length to accommodate more battery units. In fact, if we derive the light-matter coupling from fundamental principles, it's found to scale as $g\sim 1/\sqrt{V}$, where $V$ is the cavity volume. In experiments involving a large number of atoms (e.g., $N\sim10^3$ in the experiments by Haroche and collaborators on superradiance \cite{RevModPhys.85.1083}) within a fixed cavity volume, this renormalization does not apply. For instance, cavity quantum electrodynamics experiments that study superradiance are performed by varying the number of atoms $N$ while keeping the cavity volume $V$ constant~\cite{RevModPhys.85.1083} instead of scaling it with the number of atoms $N$. The capacity to accommodate numerous atoms in a single cavity, without altering its volume, stems from the significant size disparity between a typical cavity (approximately a few centimeters, or around one centimeter for microwave cavities) and the effective 'size' of an atom (a few micrometers for Rydberg atoms). Hence, to describe these experiments the normalization of Eq.~\eqref{eq:TDscaling} is not used. As discussed in Ref.~\cite{Andolina2019a}, whether one uses Eq.~\eqref{eq:DickeQB} with or without the renormalization $g\to g/\sqrt{N}$ ultimately depends on the specific experimental setup.
In scenarios where the thermodynamic limit is taken, namely the limit $N\to \infty$ and $V\to \infty$ with a constant ratio $N/V$, the correct approach is to apply the renormalization $g\to g/\sqrt{N}$.  

\subsubsection{Variations of the Dicke batteries}

Due to the success of the Dicke battery paradigm, several variations of the Dicke model have been subsequently proposed. Recent studies on trapped-ion~\cite{Felicetti15} and superconducting-flux-qubit~\cite{Felicetti18} setups have shown the potential to suppress the dipole contribution, which is linear in the photon coupling (Eq.~\eqref{eq:HintDicke}), thereby enhancing the two-photon coupling. If the TLSs of a Dicke battery are set to be resonant with twice the cavity frequency, $\omega_0 = 2\omega_\mathrm{c}$, their dynamics is dominated by two-photon processes, describing by the so-called {\it two-photon Dicke model}. This model presents an intriguing phase diagram with two quantum criticalities: i) the superradiant phase transition~\cite{Felicetti17} and ii) a spectral collapse~\cite{Garbe20}. 
In this context,~\citet{Crescente2020a} focuses on two-photon Dicke quantum batteries, in which the interaction Hamiltonian in \eqref{eq:DickeQB} is replaced by the following one,
\begin{equation}
H_{1}= g_{2{\rm ph} } \sum_{i=1}^N \hat{\sigma}_i^{x} \left[\big(\hat{a}^{\dagger}\big)^2+\big(\hat{a}\big)^2\right]~,
\end{equation}
where $g_{2{\rm ph}}$ is coupling strength for two-photon processes. The Authors found that the maximum charging power scales quadratically in $N$, $\langle P\rangle_{\tilde{\tau}}\sim N^2$, which is $\sqrt{N}$ times faster than the conventional Dicke battery. However, it was noted by very same Authors that if consistency with the thermodynamic limit (discussed in Sec.~\ref{sss:originofadvatageTD}) is enforced, the two-photon coupling needs to be rescaled as $ g_{2{\rm ph} }\to  g_{2{\rm ph} }/N$, which exactly cancels the superextensive scaling of the power, yielding $\langle P\rangle_{\tilde{\tau}}\sim 1$. Additionally, the scaling of energy fluctuations~\cite{Crescente2020a}, the extractable work, and the dependence upon the initial photonic state~\cite{Delmonte2021} have also been studied for this model. Recently,~\citet{Gemme2023a} have also studied the case of driving a Dicke quantum battery with off-resonant pulses, via exchange of virtual photons.

The Dicke model is a widely used tool to describe atoms interacting with a cavity. In some physical situations, however, it may require extensions. For example, in a Bose-Einstein condensate coupled to an optical cavity, direct dipole-dipole interactions between different atoms need to be considered~\cite{Rodriguez18,Baumann2010}.

\citet{Dou2022} considered an extended Dicke model that includes an interatomic interaction term in addition to the atom-photon Dicke coupling Hamiltonian in Eq.~\eqref{eq:HintDicke}. This model was found to exhibit faster battery charging in the strong coupling regime, with a maximum power scaling of $\langle P\rangle_{\tilde{\tau}}\sim N^{1.88}$. Not surprisingly, in the weak-coupling regime, the power was found to scale as $\sim N^{1.5}$ like for the conventional Dicke model.
Another variation of the Dicke model was studied by~\citet{Dou2022b} who examined the charging dynamics of a Heisenberg spin-chain battery (see Sec.~\ref{sss:spin-chain}) and found that the maximum power scaling was at most $\langle P\rangle_{\tilde{\tau}}\sim N^{0.75}$. A superconducting implementation of an extend Dicke battery has been recently proposed by~\citet{Dou2023transmon}.
However, by coupling the spin-chain battery to a common cavity field, it was possible to enhance the maximum power scaling to $\sim N^{2}$~\cite{Dou2022b}. The extended Dicke model has also been studied by~\citet{Zhao2022}, while~\citet{Yang2023threelevel} recently studied a three-level version of the Dicke battery.

\subsection{Charging properties of other many-body batteries}

\subsubsection{Spin-chain and spin-networks batteries}
\label{sss:spin-chain}

A spin chain is a one-dimensional array of TLSs that interact with each another through specific exchange interactions.
These theoretical models have been widely studied in the field of condensed matter physics where are closely related to the study of magnetism \cite{Auerbach1994} and have also been applied to quantum devices, communication, and computation~\cite{Zueco2009}.
The first spin-chain model of a many-body quantum battery was proposed by~\citet{Le2018}. In such a model, energy is injected via the direct charging protocol introduced in Eq.~\eqref{eq:protocol}, where the battery Hamiltonian is
\begin{equation}
    \label{eq:spin-chain_Hamiltonian}
 H_{\rm B}^{(0)}=B\sum_{i=1}^N\hat{\sigma}_i^z-\sum_{i<j}g_{ij}\big[\hat{\sigma}_i^z\hat{\sigma}_j^z+\alpha(\hat{\sigma}_i^x\hat{\sigma}_j^x+\hat{\sigma}_i^y\hat{\sigma}_j^y ) \big]~,
\end{equation}
and the charging Hamiltonian is $
 H_{\rm B}^{(1)}=\omega \sum_{i=1}^N\hat{\sigma}_i^x$. Here, $B$ is the strength of an external Zeeman field and $g_{ij}$ is the interaction strength. The anisotropy parameter $\alpha$ can be tuned to recover Ising ($\alpha=0)$, XXZ ($0<\alpha<1$), and XXX ($\alpha=1)$ Heisenberg models, respectively~\cite{Le2018}. Additionally, a transverse magnetic field, parameterised by $\omega$, is used to charge the system. Either nearest neighbor, $g_{ij}=g\delta_{i,j}$, or long-range interactions $g_{ij}=g/|i-j|^p$, with $p>0$, are considered.

Differently from customary battery models, energy can be stored in the interactions between the different spins, as the battery Hamiltonian is not a sum of local independent terms.
This implies that the eigenstates of the internal Hamiltonian $H_{\rm B}^{(0)}$ can be entangled, if the coupling strength $g_{ij}$ is non-vanishing, due to the presence of $2$-body interactions between the spins.
With reference to the discussion in Sec.~\ref{sss:role_of_interaction}, the $N$ battery units are interacting via a $k=2$-body interaction term with arbitrarily long range, corresponding to a participation number $m$ that scales with the system size as $m=N$. This comparison, in light of Eq.~\eqref{eq:quantum_advantage_order_participation}, implies that the power is bounded to scale quadratically with the size, i.e., $\langle P\rangle_{\tilde{\tau}}\sim N^2$ in the long-range scenario. This is indeed what it is observed in this model, as will be discussed momentarily.

\citet{Le2018} study the role of the anisotropy $\alpha$ and interaction range. They found that the maximum charging power obtainable in the charging process increases with increasing $\alpha$. 

Interestingly, the isotropic coupling of the XXX Heisenberg model ($\alpha=1$) resulted in the independent charging of each spin. This phenomenon occurs even in the presence of interactions because for $\alpha=1$, the battery Hamiltonian $H_{\rm B}^{(1)}$ commutes with these interactions due to rotational invariance of the model. Consequently, the XXX model exhibits an extensive maximum power $\langle P\rangle_{\tilde{\tau}} \sim N$.
Conversely, the full anisotropy of the XXZ model, \emph{i.e.} $\alpha=-1$, leads to much higher power compared to the independent case. 

In the weak-coupling $\sum_{i<j}g_{ij}\ll N \omega$  regime, the scaling of the maximum power is analyzed.
In particular, in the case of nearest-neighbor interactions, the power is extensive in $N$. 
When the coupling strength decays algebraically as $1/|i-j|$ ($p=1$), where $i$ and $j$ denotes the lattice size along the chain, the charging power grows superextensively as $\langle P\rangle_{\tilde{\tau}}\sim N\log(N)$. 
Finally, for a ``uniform'' $p=0$ interaction strength, the Authors find a superextensive power $\langle P\rangle_{\tilde{\tau}}\sim N^2$.
These results are in full agreement with Eq.~\eqref{eq:quantum_advantage_order_participation}, provided that one considers the case $k=2$ of two-body interactions and takes into account the fact that the participation number $m$ is limited by the specific range of the coupling ($m$ scale with the system size, $m=N$, in the long-range case). This reflects the ``symmetry'' between the roles of internal and charging Hamiltonians. Indeed, in the present case, the charging Hamiltonian is local and does not increase correlations between different spins. The advantage in the power is rather given by the energy stored in the $2$-body interactions, which are present in the battery Hamiltonian. 

Finally, the Authors discuss a way to implement the global entangling Hamiltonian in Eq.~\eqref{eq:global_charging_hamiltonian} proposed by~\citet{Binder2015a} and~\citet{Campaioli2017}.
Indeed, in the limit of strong nearest-neighbor attractive interactions ($g\gg\omega$), it is possible to write an effective global entangling Hamiltonian with a $k=N$-body interaction. However, this effective interaction scales as $\omega(\omega/g)^N$, and since it has been obtained in the perturbative regime $g\gg \omega$, exponentially vanishes in the limit of large $N$. 
Due to this fact, the power deposited in the battery is actually worse when the spins traverse the correlated shortcut suggested by~\citet{Binder2015}, and vanishes in the $N\to\infty$ limit.
The study of spin-chain batteries has been further extended to disordered models, see e.g.~\citet{Rossini19} and~\citet{Ghosh2020}. Additionally, separate investigations have focused on charging spin chains through the use of a cavity mode~\cite{Dou2022b} or other spin systems~\cite{Gao2022,Yang2020,Qi2021,Lu2021,Zhao2021,Liu2021,Konar2021,Peng2021,Arjmandi2022,Yao2022a,Barra2022,Ghosh2022,Crescente2023a,Rodriguez2023a,Guo2023analytically}.

\subsubsection{SYK quantum batteries}
\label{sss:syk}

Motivated by the discussions reported in Sects.~\ref{sss:genuine_quatum_advantage} and~\ref{sss:originofadvatageTD},~\citet{Rossini2020} proposed a model of a many-body battery that unequivocally presents a genuine quantum advantage---certified by using the bound of~\citet{Julia-Farre2020}.

This implementation relies on the so-called Sachdev-Ye-Kitaev (SYK) model~\cite{Sachdev1993,Kitaev15}. The SYK model describes quantum matter defying the Landau paradigm of normal Fermi liquids in that it displays no quasiparticles. It has garnered a great deal of attention in recent years due to its unique properties, which include fast scrambling~\cite{Maldacena2016}, nonzero entropy density at vanishing temperature~\cite{Georges2001}, and volume-law entanglement entropy in all its eigenstates~\cite{Liu2018SYK}. Moreover, this model is holographically connected to the dynamics of the ${\rm AdS}_2$ horizon of a quantum black hole~\cite{Kitaev15}.

\begin{figure}
    \centering
    \includegraphics[width=0.48\textwidth]{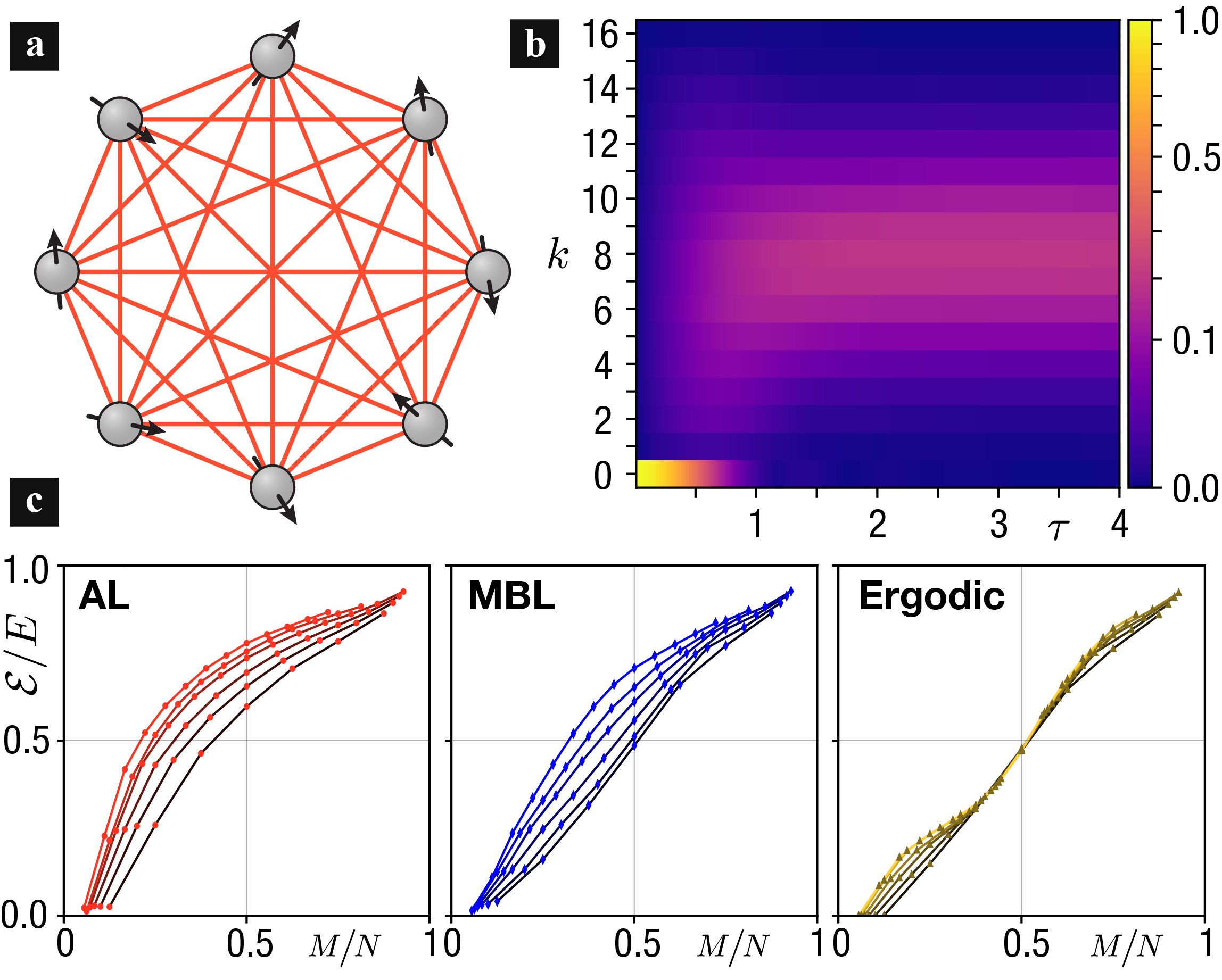}
    \caption{(Color online) (a) A sketch of the SYK model \cite{Sachdev1993,Kitaev15}, which is a Hamiltonian for randomly interacting fermions with all-to-all interactions. (b) The dynamics of the populations $p_k$ of the $k-$th energy level associated to the battery Hamiltonian $H^{(0)}_{\rm B}$ to be compared with Fig.~\ref{fig:gqa}(b). The charging Hamiltonian consists in the c-SYK Hamiltonian $H^{(1)}_{\rm B}$ of Eq.~\eqref{eq:SYK}. This global charging scheme takes a \textit{shortcut} in the Hilbert space by driving the battery through maximally entangled states.  This can be clearly seen in the short-time behavior of the populations. In this regime, the populations do not propagate locally in the Hilbert space, similarly to the global charging (see Fig.~\ref{fig:gqa}(b)).
    (c) The fraction of useful energy $\mathcal{E}/E$ as a function of the fraction $M/N$, at fixed time $\tau$. This quantity can be used to discriminate if the system is in an Anderson localized (AL), many-body localized (MBL) or ergodic phase. 
    }
    \label{fig:syk_ergodicity}
\end{figure}

The SYK battery introduced by~\citet{Rossini2020} is charged via a direct charging protocol, as in Eq.~\eqref{eq:protocol}. The internal battery Hamiltonian $H_{\rm B}^{(0)}$ is a sum of local terms, given by 
\begin{equation}\label{eq:Ham0SYK}
H_{\rm B}^{(0)} = \sum_{i=0}^N \omega_0\sigma_i^{y}~,
\end{equation}
while charging is performed via the complex SYK (c-SYK) Hamiltonian,
\begin{equation}\label{eq:HamB}
H_{\rm B}^{(1)} = \sum_{i,j,k,l=1}^N J_{i,j,k,l}\hat{c}^\dagger_i \hat{c}^\dagger_j \hat{c}_k \hat{c}_l~.
\end{equation}
Here, $\hat{c}^\dagger_j$ ($\hat{c}_j$) is a spinless fermionic creation (annihilation) operator. The SYK model is illustrated in Fig.~\ref{fig:syk_ergodicity}(a). This has to be understood in its spin-$1/2$ representation, which is obtained by the Jordan-Wigner (JW) transformation, i.e.~$\hat{c}^\dagger_j= \hat \sigma^{+}_j \big( \Pi_{m=1}^{j-1}  \hat \sigma_m^{z} \big)$, where $\hat{\sigma}^{\pm}_j\equiv( \hat{\sigma}^{x}_j \pm i \hat{\sigma}^{y}_j)/2$.
The couplings $J_{i,j,k,l}$ are zero-mean Gaussian-distributed complex random variables,
with variance $\langle \! \langle J^2_{i,j,k,l} \rangle \! \rangle = J^2/N^3$, where $\langle \! \langle J^2_{i,j,k,l} \rangle \! \rangle$ denotes an average over different disorder realizations. Crucially, this scaling ensures that the SYK battery model of~\citet{Rossini2020} has a well-defined thermodynamic limit. Notably, this precludes any potential collective advantage that might have been present if a different normalization was chosen.
Note that Eq.~\eqref{eq:Ham0SYK} differs from the battery Hamiltonian $H_{\rm B}^{(0)}$ in Eq.~\eqref{eq:HQBlocal}, since $H_{\rm B}^{(1)}$ commutes with the battery Hamiltonian in Eq.~\eqref{eq:HQBlocal} and, hence, cannot injects energy in the system.
This model displays a superextensive scaling of the optimal charging power, $\langle P\rangle_{\tau}\sim N^{3/2}$.
Furthermore, the Authors provided a rigorous certification of the quantum origin of the charging advantage of the c-SYK model by considering the following bound,
\begin{equation}
 \label{eq:averaged_power_bound_loose}
\langle P \rangle_{\tau} \leq \sqrt{\langle\Delta E_\mathrm{B}^2 \rangle_{\tau} \langle\Delta E_{1}^2 \rangle_{\tau}}~,
\end{equation}
where $\langle\Delta E_\mathrm{B}^2 \rangle_{\tau}$ ($\langle\Delta E_{1}^2 \rangle_{\tau}$) is the time-average variance of $H_{\rm B}^{(0)}$ ($H_{\rm B}^{(1)}$). Eq.~\eqref{eq:averaged_power_bound_loose} is a loose version 
of the bound on the average power derived by~\citet{Julia-Farre2020}.
In the SYK battery model introduced above, $\langle\Delta E_{1}^2 \rangle_{\tau}$ scales linearly in $N$, while the power enhancement is linked to a quadratic scaling of $\langle\Delta E_{1}^2 \rangle_{\tau}$.
This fact hints at a genuine quantum advantage (this concept has been discussed in Sec.~\ref{sss:genuine_quatum_advantage}) displayed by the c-SYK model with respect to the charging task.
The Author also examined a bosonic version of the SYK battery and a parallel-charging scheme, showing that both models do not display any quantum advantage. 
The poor performance of the bosonic SYK battery compared to the c-SYK battery suggests that non-local JW strings for fermions are crucial for maximizing entanglement production during the time evolution and therefore correlations between the battery units. This result is in accordance with the bound derived by \citet{Gyhm2022} discussed in Sec.~\ref{sss:role_of_interaction}, $P(t)\leq \gamma k N$, being $k$ the interaction order and $\gamma$ a constant. When the Hamiltonian in Eq.~\eqref{eq:HamB} is represented in the spin basis, via the JW transformation, a Jordan string, $\Pi_{m=i}^{i+k} \hat \sigma_m^{z}$, emerges in the Hamiltonian with an interaction order of $k\sim N$. Conversely, when considering the bosonic SYK model, Jordan strings are absent, resulting in an interaction order of $k=4$.

The work by~\citet{Rossini2020} provided the first quantum many-body battery model where fast charging occurs due to the maximally-entangling underlying quantum dynamics envisioned by Refs.~\cite{Binder2015a, Campaioli2017}. 
Still, the non-local interactions peculiar to the SYK model may be extremely challenging to be realized in practice and the feasibility of such a many-body battery remains disputed (see, however, the discussion in Sec.~\ref{s:architectures} below).

\subsection{Work extraction}
\label{sss:workextractionDicke}

Most of the previous discussion was focused on the scaling of the charging dynamics with the number $N$ of battery units. However, the performance of a battery cannot be captured by a single figure of merit, such as the charging power $\braket{P}_\tau$.
For example, a ``good'' battery, should not only be charged in a small amount of time, but also have the capability to fully deliver such energy in a useful way, i.e., to perform work. This ability is measured by the ergotropy $\mathcal{E}$, which was introduced in Sec.~\ref{sss:ergotropy}. In the context of a many-body battery, the presence of correlations and entanglement between different constituents may induce limitations on the task of energy extraction~\cite{Oppenheim2002}.
\citet{Andolina2019} studied the maximum work ${\mathcal E}(\tilde{\tau})$ that can be extracted from $N$ TLSs in a Dicke battery (see Sec.~\ref{ss:dicke_quantum_battery}), where $\tilde{\tau}$ is the optimal charging time (i.e. the time that maximizes the charging power).
It was observed that, for finite-size batteries, the extractable energy ${\mathcal E}(\tilde{\tau})$ constitutes a small fraction of the total energy ${E(\tilde{\tau})}$ stored in the battery. This reduction is due the presence of correlations (entanglement) between the charger and battery, proving that quantum effects can be detrimental for work extraction.
However, this issue can be mitigated in the limit of a large number of battery units, since the fraction of energy locked by correlations becomes negligible, i.e.~$\lim_{N \rightarrow \infty} {\mathcal{E}(\tilde{\tau})}/{E(\tilde{\tau})} = 1$, regardless of the initial state of the charger.

As further demonstrated by~\citet{Rossini19}, this is a general property of quantum charging processes of closed Hamiltonian systems, which is ultimately linked to the integrability of the dynamics and does not depend on the details of the underlying microscopic model.
 \citet{Rossini19} considered a disordered quantum Ising chain Hamiltonian charged via a direct charging protocol. The quantum Ising Hamiltonian that was studied has a rich equilibrium phase diagram presenting many-body localized (MBL), Anderson localized (AL), and ergodic phases.
In this work, the ergotropy of a subsystem of $M\leq N$ batteries units, ${\mathcal E}_M(\tilde{\tau})$, properly normalized by the energy of the same subsystem $E_M(\tilde{\tau})$, can be used to discriminate different thermodynamic properties of the eigenstates of the system, as shown in Fig.~\ref{fig:syk_ergodicity}(b). Indeed, considering half of the chain ($M=N/2$), the ratio ${\mathcal E}_{N/2}(\tilde{\tau})/E_{N/2}(\tilde{\tau})$ saturates to a finite constant when the thermodynamic $N\to \infty$ limit is taken if the ergodic phase is considered. 
In contrast, in the MBL and AL phases, the energetic cost of creating correlations becomes negligible in the thermodynamic limit, i.e.~${\mathcal E}_{N/2}(\tilde{\tau})/E_{N/2}(\tilde{\tau})\to 1$.
This stems for the fact that in these phases the dynamics is restricted to a sub-portion of the Hilbert space. 
The findings of this study demonstrate that ergotropy can effectively distinguish between different thermodynamic characteristics of a quantum system and reveal insights of its underlying dynamics. 

It is worth to mention that the issue of energy extraction has been investigated in waveguide quantum electrodynamics  setup by \citet{MonselPRL2020}, \citet{Maffei2021}, and in random quantum systems by~\citet{Caravelli2021}.


\section{Charging precision}
\label{s:charging_precision}

In the previous sections we have focused on the power of many-body quantum batteries. While powerful charging and work extraction seem always desirable, speed can come at the expense of precision, e.g., in terms of work fluctuations. In this section, we look at some results on the precision of unitary charging and work extraction~\cite{Friis2018,Santos2019a,Santos2020,Rosa2020,Crescente2020a,Delmonte2021,Moraes2021,Abah2022,Imai2022,Dou2022,Dou2022a,Hu2022}, which aim to mitigate fluctuations in the final energy of the battery, as well as work fluctuations during the charging procedure.

\subsection{Bosonic batteries and Gaussian unitaries}
\label{ss:gaussian_precision}

\citet{Friis2018} focus for the first time on the charging precision and study a bosonic quantum battery given by an ensemble of harmonic oscillators, i.e.~$H_0 = \sum_j \omega_j \hat{a}_j^\dagger \hat{a}_j^\phdagger$. Since no charger subsystem is involved, $H_0 = H_\mathrm{B}^{(0)}$ of Eq.~\eqref{eq:total_bare_Hamiltonian}. Thus, we will refer to $H_0$ as the bare Hamiltonian of the battery to lighten the notation.
The Authors consider a quantum battery in an initial, completely passive state $\rho_0 = G_\beta[H_0] = \bigotimes_j G_\beta[H_{0,j}]$, where $H_{0,j} = \omega_j \hat{a}_j^\dagger \hat{a}_j^\phdagger$ is the Hamiltonian of the $j$-th mode and
\begin{equation}
    \label{eq:gibbs_harmonic}
    G_\beta[H_{0,j}] = \Big(1-e^{-\beta\omega_j}\Big)\sum_n e^{-n\beta\omega_j}\ketbra{n_j}{n_j}~.
\end{equation}
Here, $\ket{n_j}$ is the (Fock) state with $n$ particles in the $j$-th mode.
The system is charged by some unitary $U$ to a final state $\rho=U \rho_0 U^\dagger$, such that $W = \tr[H_0\rho] - \tr[H_0\rho_0]$ is the energy deposited by $U$. To quantify the \textit{charging precision}, they consider two quantities. First, they study fluctuations in the final energy of the battery, quantified by the variance of the internal Hamiltonian with respect to the final state
\begin{equation}
    \label{eq:variance_energy_deposited}
    \Delta E^2(\rho) = \tr[H_0^2\rho]-\tr[H_0\rho]^2~.
\end{equation}
This quantity is also used to calculate the increase in the standard deviation of the Hamiltonian, $\Delta\sigma = \Delta E(\rho) - \Delta E(\rho_0)$.

Second, they quantify the energy fluctuations $\Delta W$ during the charging process, using
\begin{equation}
    \label{eq:energy_fluctuations}
    \Delta W^2 = \sum_{m,n} p_{m\to n}(W_{m\to n}-W)^2,
\end{equation}
where $W_{m\to n} = \tr[H_0\ketbra{n}{n}] - \tr[H_0\ketbra{m}{m}]$ is the energy difference between two energy levels\footnote{Here the energy levels are with respect to $H_0$. If $H_0$ is a single mode battery then $\ket{n}$ is also the Fock state with $n$ particles.} $m$ and $n$, and $p_{m\to n} = p_m |\braket{n|U|m}|^2$ is the transition probability from an initial state $\ket{n}$ with population $p_n = \tr[\rho_0 \ketbra{n}{n}]$ (calculated over the energy basis of the Hamiltonian $H_0$) to a final state $\ket{m}$ with population $p_m$.

While it is possible to construct optimal protocols to maximize charging precision by minimizing either $\Delta E$ or $\Delta W$, the operations necessary for powerful and precise charging may be hard to implement in practice~\cite{Friis2018}. For example, global entangling operations like $\ketbra{G}{E}+{\rm h.c.}$ are notoriously difficult to implement for spin chains~\cite{Mathew2020}, while generating coherence between energy levels with arbitrarily large separation is hard when considering harmonic oscillators.~\citet{Friis2018} therefore focus on Gaussian unitaries, a special class of unitary operations that is obtained from driving Hamiltonians that are at most quadratic in the creation and annihilation operators $a_j^\dagger$ and $a_{j}$, such as \textit{squeezing} and \textit{displacements}~\cite{Gardiner2000}. In this case, the Authors show that neither $\Delta E$ nor $\Delta W$ are bounded from above and, in fact, increase with $W$ for infinite dimensional systems\footnote{The Authors also show that, for finite dimensional systems, upper bounds on $\Delta W$ and $\Delta E$ are finite.}. They then provide lower bounds on $\Delta E$ and $\Delta W$ for single-mode and multi-mode systems, presenting the respective optimal protocols for maximizing charging precision as a function of $W$ and temperature $T$. Importantly,~\citet{Friis2018} show that the best Gaussian operations produce energy fluctuations that vanish asymptotically when compared to $W$ for large energy supply.
It turns out that that best Gaussian charging protocols for single-mode batteries can be obtained by using a combination of squeezing and displacement operations, while pure squeezing is the worst performing Gaussian charging. Such clear-cut interpretation is harder to obtain for the case of multi-mode batteries, for which more work is needed to further clarify the role of coherences and correlations. 

Further work in this direction could leverage on the resource-theoretic approach laid out by \citet{Friis2018} for the class of Gaussian operations and \textit{Gaussian passivity}, which introduces a set of free states that are not generally passive, but passive with respect to Gaussian operations~\cite{Ein-Eli2016}.
Another interesting outlook is to study the role of higher-order interactions (i.e., interactions mediated by $\hat{a}^n$ and $(\hat{a}^\dagger)^n$ with $n>1$) on the performance of bosonic batteries, 
as done by~\citet{Delmonte2021} for the case of two-photon couplings. 

\subsection{Adiabatic quantum charging}
\label{ss:transitionless}
\citet{Santos2019a} propose an approach to mitigate work fluctuations based on the use of stimulated Raman adiabatic passage (STIRAP)~\cite{Gaubatz1990,Vitanov2017}, akin to transitionless quantum driving~\cite{Berry2009}. The protocol is based on the idea of slowly varying the interaction Hamiltonian in order to prevent any undesired transition between its eigenstates.
To achieve adiabatic driving, the Authors consider a time-dependent interaction Hamiltonian $H_1(t)$ such that $[H_1(0),\widetilde{\rho}_0]=0$, where $\widetilde{\rho}_0 = e^{iH_0t}\rho_0 e^{-iH_0t}$ is the initial state of the battery in the interaction picture. The interaction Hamiltonian is then changed sufficiently slowly to prevent any transition between its eigenstates, until some target state $\rho^\star = \rho(\tau)$ is reached at time $\tau$.

The Authors study adiabatic charging for the case of a three-level system, well-known in the field of STIRAP~\cite{Vitanov2017}, using the interaction Hamiltonian
\begin{equation}
    \label{eq:adiabatic_driving}
    H_1(t) = \Omega_{12}(t)\ketbra{1}{2}+\Omega_{23}(t)\ketbra{2}{3} + {\rm h.c.}~,
\end{equation}
where the time-dependent transition frequencies $\Omega_{12}(t) = \Omega_0 f(t)$ and $\Omega_{23}(t)=\Omega_0[1-f(t)]$ are chosen such that $f(0)=0$ and $f(\tau) = 1$, $\Omega_0$ being the maximal frequency of the interaction. 
Santos \etal compare the the energy deposited onto the battery $W(\tau)$ with the bandwidth of the Hamiltonian $W_\mathrm{max}:=w[H_0]=\epsilon_3-\epsilon_1$, i.e., the maximal amount of energy that can be deposited onto the battery\footnote{This is also the \textit{ergotropy} of the active state $\ket{3}$, since the battery is initialized in the completely passive state $\ket{1}$.}. 
They also compare the average charging power $\braket{P} = W(\tau)/\tau$ 
with the maximal achievable power $\braket{P}_\mathrm{max} := W(\tau)/\tau_\mathrm{QSL}$, which they calculate using a Margolus-Levitin form of the QSL in Eq.~\eqref{eq:qsl}, i.e.~$\tau_\mathrm{QSL} = \pi/2 W_\mathrm{max}$.
\begin{figure}
    \centering
    \includegraphics[width=0.48\textwidth]{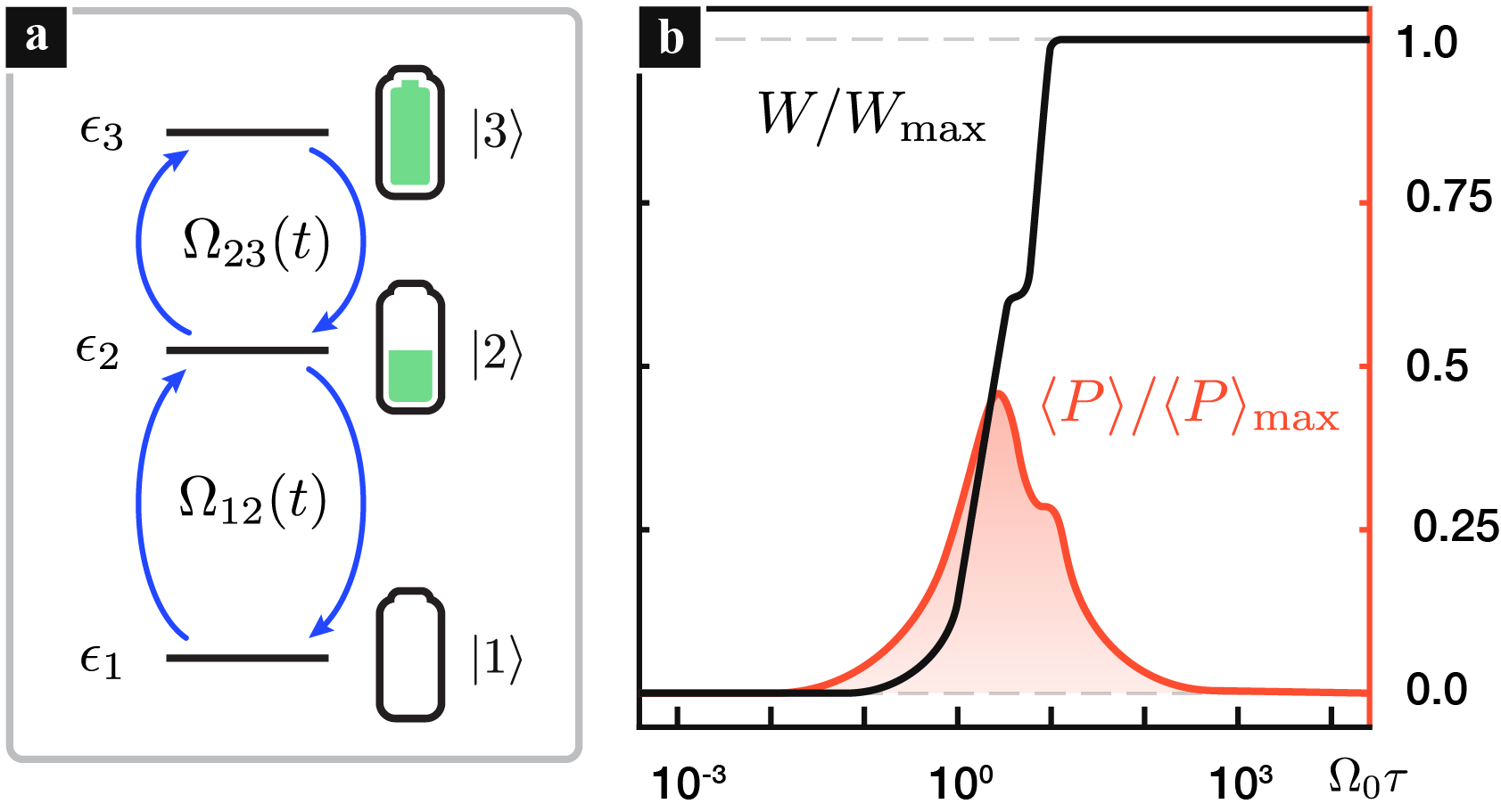}
    \caption{(Color online) Adiabatic charging of a three-level quantum battery, adapted from Ref.~\cite{Santos2019a}. (a) Energy diagram and couplings of the three-level quantum battery considered by~\citet{Santos2019a}, with $H_0 = \sum_{i=1}^3 \epsilon_i \ketbra{i}{i}$ and $H_1(t)$ given in Eq.~\eqref{eq:adiabatic_driving}.~\citet{Santos2019a} suggest that superconducting transmons would be particularly suitable candidates to implement this protocol. (b)  Adiabatic driving charges the system completely for large $\Omega_0\tau$, i.e., when the charging is slow, revealing a trade-off between precision and power.}
    \label{fig:adiabatic}
\end{figure}
These quantities are studied as a function of the  dimensionless parameter $\Omega_0\tau$. As expected, far from adiabaticity, i.e.~for $\Omega_0\tau\ll1$, the battery performs poorly and $W/\mathcal{E}, \braket{P}/\braket{P}_\mathrm{max} \approx 0$. As $\Omega_0\tau$ grows, the energy deposited grows until the battery can be fully charged, $W/\mathcal{E} =1$, as shown in Fig.~\ref{fig:adiabatic}. On the other hand, the charging power grows until it reaches a maximum value at $\Omega_0\tau\sim 1$, and then decreases for larger values of $\Omega_0\tau$. Crucially, this result suggests a trade-off relation between charging power and precision, dictated by the competition between adiabaticity and the saturation of QSLs. 

As discussed by the Authors, maximal power $\braket{P}/\braket{P}_\mathrm{max}\approx 1$ may be achieved with adiabatic driving by means of \textit{shortcuts to adiabaticity} (STA), i.e., by engineering a fast process that leads to the same final state and work fluctuations as of an
indefinitely slow processes. While STA can lead to the joint optimization of work, power and precision, it must come at the cost of a larger energy expenditure on the control interaction, as demonstrated by~\citet{Campbell2017}, thus affecting the efficiency of the protocol~\cite{Uzdin2012,Campaioli2019}. \citet{Santos2019a} also consider the effect of dephasing and relaxation, showing that the competition between decoherence and adiabaticity leads to an optimum for the charging time $\tau$ such that the evolution is fast enough to beat decoherence but not too fast to avoid deviations from adiabaticity. 

This adiabatic charging approach has been experimentally realized for three-level transmons using stimulated Raman adiabatic passage (STIRAP)~\cite{GaubatzJCP1990} by~\citet{Hu2022}, and is applicable to any other architecture, as discussed in Sec.~\ref{s:architectures} 
This work has also inspired similar protocols for powerful and precise charging, such as that by~\citet{Dou2020},~\citet{Dou2022a}, and~\citet{dou2022Lipkin}, as well as the work by~\citet{Santos2020} and~\citet{Moraes2021} for the transitionless quantum driving of 
quantum batteries consisting of two spin-$1/2$ particles coupled to a spin-$1/2$ charger.
These results shown the promising role of adiabatic driving for work injection and extraction. Outlooks should focus on the trade-off between charging power, precision and energetic cost of the charging protocol. As mentioned above, the relation between the speed of evolution and the cost of STA has been studied by~\citet{Campbell2017}. Extending this work by explicitly accounting for finite deviations from adiabaticity could clarify the feasibility of powerful and precise charging of quantum batteries.
Another interesting direction could be to jointly optimize charging power and precision by choosing a suitable target function $\mathcal{F}(\braket{P},W,\delta W)$, along the lines suggested by~\citet{Binder2015a}. 

\subsection{Entanglement and work fluctuations}
\label{ss:entanglement_precision}

Fabrication defects and other sources of noise introduce disorder in the energies of each subsystem (diagonal disorder) and in the couplings between them (off-diagonal disorder). In the presence of disorder, unitary dynamics is often characterized by temporal fluctuations, that manifest at different time scales. These fluctuations affect the energy deposited on the battery $W(t)$ and need to be mitigated to improve charging precision. \citet{Rosa2020} focus on these temporal fluctuations, and show that they can be suppressed by preparing many-body quantum batteries in highly entangled states. Note that, in practice, entanglement in many-body system is usually rather fragile, due to exponentially fast suppression of correlations as an effect of decoherence. For this reason, the feasibility of the protocol proposed by~\citet{Rosa2020} depends on the ability to sustain highly entangled states for sufficiently long timescales.

\citet{Rosa2020} consider the standard, local $N$-spin battery model in Eq.~\eqref{eq:local_hamiltonian}, with $H_j = h \hat{\sigma}_j^{z}$, where $h$ represents the energy scale of each subsystem.
The battery is charged via the  direct charging protocol in Eq.~\eqref{eq:protocol}. The Authors consider two different models for the interaction $H^{(N)}_1$, 
\begin{align}
    \label{eq:MBL}
    & H_\MBL^{(N)} =  \sum_{j=1}^{N}\Big( J_j\hat{\sigma}_j^x \hat{\sigma}_{j+1}^x +J_2\hat{\sigma}_j^x\hat{\sigma}_{j+2}^x\Big)~,\\
    \label{eq:SYK}
    & H_\SYK^{(N)} = \sum_{i<j<k<l} J_{i,j,k,l} \hat{\gamma}_i\hat{\gamma}_j\hat{\gamma}_k\hat{\gamma}_l~,
\end{align}
where $J_j$ and $J_2$ are the nearest-neighbor and next-to-nearest-neighbor Ising couplings, and $\hat{\gamma}_i$ is the Majorana fermion operator\footnote{It holds that $\hat{\gamma}_i = \hat{\gamma}_j^\dagger$ and $\{\hat{\gamma}_i,\hat{\gamma}_j\}=\delta_{ij}$.}. Together with the internal Hamiltonian $H^{(N)}_0$, the many-body localized Hamiltonian $H_\MBL^{(N)}$ is a model for many-body quantum systems that exhibits different quantum phases (eigenstate thermalisation hypothesis, Anderson localized, many-body localized, spin-glass) depending on the value of its parameters~\cite{Kjall2014,Nandkishore2015}. 
The SYK Hamiltonian $H_\SYK^{(N)}$, discussed extensively in Sec.~\ref{sss:syk}, is characterized by non-local interactions that promote the formation of highly entangled states~\cite{Rosa2020}.
\begin{figure}
    \centering
    \includegraphics[width=0.46\textwidth]{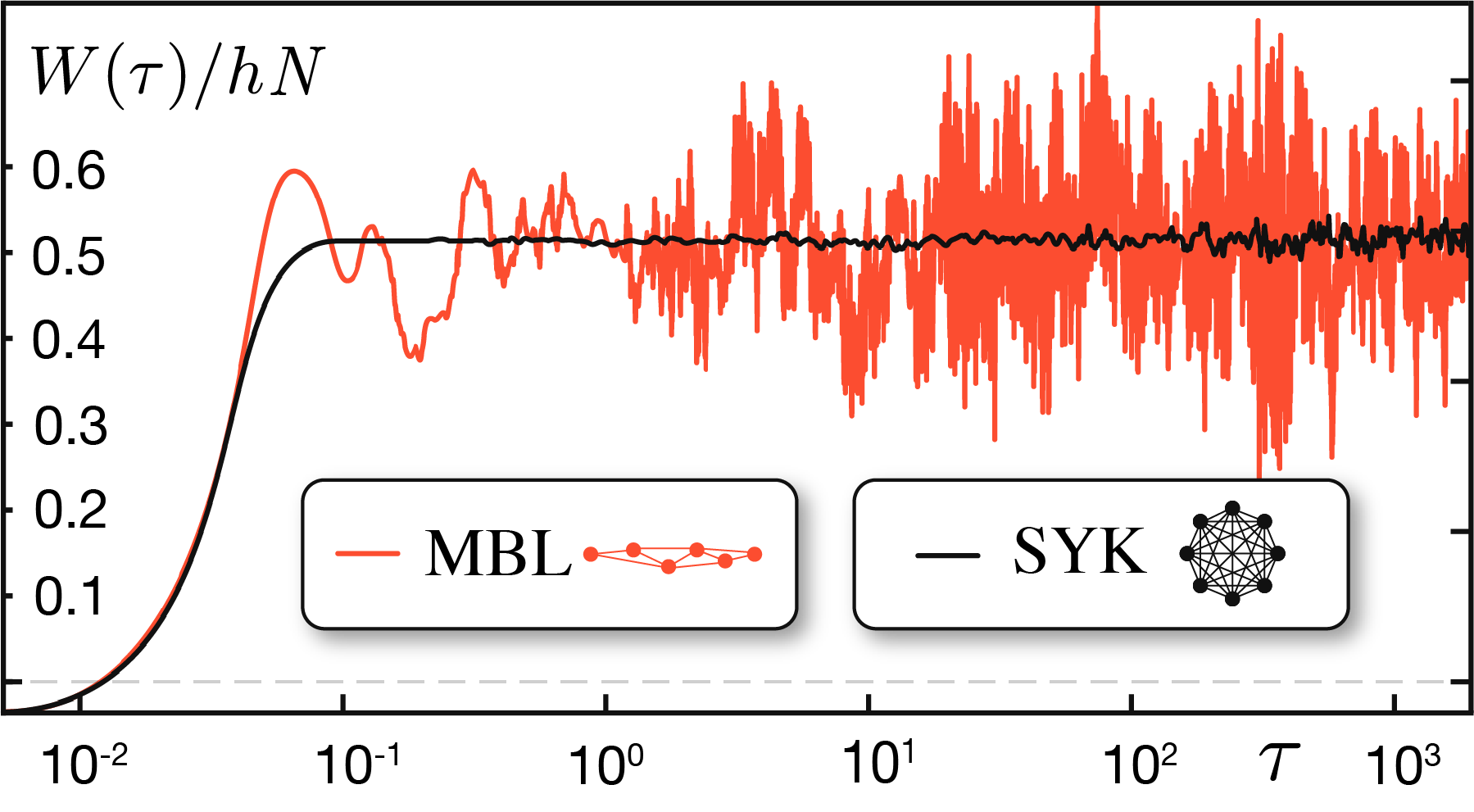}
    \caption{(Color online) Temporal fluctuations affecting the charging precision can be suppressed in many-body quantum batteries, by promoting the formation of volume-law entanglement. Work fluctuations in SYK batteries (black line) are exponentially suppressed over all time scales as localized states transition to highly-entangled states. Instead, MBL quantum batteries are characterized by localized states and  exhibit large work fluctuations at all time scales. Adapted from Ref.~\cite{Rosa2020}.}
    \label{fig:syk_precision}
\end{figure}

The Authors study the fluctuations in $W(\tau)$ relative to the Hamiltonian \textit{bandwidth}, $Nh$, using the parameter $R(\tau) = W(\tau)/(Nh)$. They show that fluctuations of $R(\tau)$ are suppressed in the presence of non-local correlations, as illustrated in Fig.~\ref{fig:syk_precision}. Furthermore they show that the amplitude of the fluctuations decreases with the number $N$ of subsystems involved. 
Rosa \etal provide extensive mathematical evidence to support the thesis that energy fluctuations in $W(\tau)$ are mitigated by the rapid and homogeneous formation of highly entangled states over many energy levels (volume-law entanglement), which is typical of the SYK Hamiltonian and other models for quantum chaos~\cite{Bianchi2021}. 

Temporal fluctuations in the stored  energy are exponentially suppressed at all time scales, a property that the Authors link to the collective and non-local thermalisation of chaotic systems~\cite{Rosa2020}. Finally, the Authors conjecture that the SYK model sets an upper bound on the precision of quantum batteries. Testing this conjecture is therefore an interesting outlook. Since SYK models can provide both powerful and precise charging~\cite{Rossini2020,Rosa2020}, further work in this direction should focus on experimentally viable SYK charging architectures, as discussed in Sec.~\ref{s:architectures}. The relation between entanglement and work fluctuations has also been considered by~\citet{Caravelli2021} and~\citet{Imai2022}, and would be of interest to further understand charging precision. Another interesting outlook is to exploit quantum optimal control~\cite{Mazzoncini2022,Rodriguez2022a} to generate fluctuation-free active states in many-body systems based on more experimentally feasible models, as done for the case of \textit{entanglement storage} in Ref.~\cite{Caneva2012}.

\section{Open quantum batteries}
\label{s:open_quantum_batteries}

In this Section we introduce the concept of \emph{open quantum battery} (OQB). An OQB is an open quantum system that may be charged by an auxiliary system, i.e., the \emph{charger}, denoted with $\mathrm{C}$, and/or may interact with the external environment, denoted with $\mathrm{E}$. 
As described in the previous sections, if the battery system was isolated, closed or perfectly protected from the outside, then it would always evolve unitarily as in Eq.~\eqref{eq:unitary_work_extraction}. 
Conversely, the dynamics of an OQB is generated by the Hamiltonian of the universe $H$, whose microscopic details may not be fully accessible,
\begin{equation}
    \label{eq:open_quantum_battery}
    H = H_\mathrm{B} + H_\mathrm{C} + H_\mathrm{E} + H_\mathrm{int},
\end{equation}
where the interaction terms $H_\mathrm{int}$ may couple battery, charger and environment to each other. In this regard, we are going to propose the wording \emph{dissipative quantum battery} (DQB) to denote, more specifically, an OQB subject to an external environment $\mathrm{E}$, that is usually unknown, uncontrollable, and whose degrees of freedom far outnumber those of the battery and charger subsystem~\cite{CarregaNJP2020,Carrasco2022}. By this notion, we will refer to DQBs regardless of the presence of an auxiliary charger subsystem.

The dynamics of an open battery, generated by the Hamiltonian of Eq.~\eqref{eq:open_quantum_battery}, can be described using density operator master equations~\cite{Breuer2002}, such as the general Liouville-von Neumann equation $\dot{\rho}_\mathrm{B}(t) = \mathcal{L}_{t}[\rho_\mathrm{B}(t)]$ where $\mathcal{L}_t$ is a time-dependent superoperator. The latter prescribes a reduced-state master equation for the dynamics of the battery's density operator $\rho_\mathrm{B}(t) = \Lambda_{t;0}[\rho_\mathrm{B}(0)]$, which approximates the underlying process
\begin{equation}
    \label{eq:general_evolution}
    \rho_\mathrm{B}(t) = \tr_\mathrm{CE}[\rho(t)] = \tr_{\mathrm{CE}}\Big[\mathcal{U}(t;0)[\rho_0]\Big],
\end{equation}
where $\rho(t)$ is the state of the universe at time $t$, starting from an initial state $\rho_0$. The unitary map $\mathcal{U}_{t;0}[\cdot] = U(t;0) \cdot U(t;0)^\dagger$, is generated by the total Hamiltonian $H$, such that $U(t;0) = \mathcal{T}\{-i\int_0^t ds H(s)\}$. The dynamics of OQBs has been studied extensively in the Markovian regime~\cite{Farina2019,Gherardini2020,Quach2020}, using the Gorini-Kossakowski-Sudarshan-Lindblad (GKSL) quantum master equation~\cite{Dariusz2017},
\begin{equation}
    \label{eq:Lindblad_open_dynamics}
\dot{\rho}_\mathrm{B}(t) = -i[H_\mathrm{B},\rho_\mathrm{B}(t)]+\mathcal{D}[\rho_\mathrm{B}(t)],
\end{equation}
where $\mathcal{D}[\rho] = \sum_{k} \gamma_k (\hat{L}_k^\phdagger\rho \hat{L}_k^\dagger - \{\hat{L}^\dagger_k \hat{L}_k^\phdagger,\rho\}/2) $ is superoperator modeling the interaction between the battery and the rest of the universe, via a set of operators $\hat{L}_{k}$ generating incoherent transitions at rates $\gamma_k$~\cite{Breuer2002}. Some Authors have also considered open quantum batteries beyond the Markovian approximation~\cite{Kamin2020,Ghosh2021}. In non-Markovian OQBs, memory effects due to the interaction with the environment (and the charger) become important for a complete description of the battery time-evolution. 

The coupling between environment and battery is crucial in Eq.~\eqref{eq:open_quantum_battery}. Dissipative quantum batteries are subject to decoherence and energetic relaxation induced by their interaction with the external environment. These processes tend to deteriorate resourceful active states, leading them towards passive (via dephasing, depolarization) or even completely passive states (via thermalisation). If these processes are not properly prevented, their timescale puts a limit on the lifetime of active states, and thus, on how long the OQB can keep its charge. Furthermore, the performance of some OQB models can decrease over time when many charging-discharging cycles are performed, in an \emph{aging} processes analogue to that of electrochemical batteries~\cite{Pirmoradian2019}. 

Any practical application of a quantum battery has to face the presence of the environment. Hence, protection mechanisms from energy leakages and decoherence are essential for device implementations~\cite{Bai2020,Xu2022}, and will be reviewed in Sec.~\ref{ss:oqb_stabilisation} and~\ref{ss:protection_from_energy_loss}. In the current literature, this issue was first addressed in Refs.~\cite{Liu2019,Gherardini2020,Quach2020} and soon after in~\cite{Mitchison2021,Yao2021,YaoPRE2022}. It has been shown that stabilizing a DQB can be recast into the thermodynamic problem of \emph{rectifying} the \emph{irreversible entropy production}~\cite{Camati2016,BatalhaoBook2019,LandiRMP2021} due to the interaction with {\rm E}. To this end, \emph{open-} and \emph{close-loop} control procedures have been thus proposed, and will be revised in Sec.~\ref{ss:oqb_stabilisation}. 

\subsection{Charging a dissipative quantum battery}
\label{ss:dissipative_charging}

The presence of an auxiliary charger subsystem may not be sufficient to enable the charging of a DQB. Depending on our ability to control the interaction between the battery, the charger and the environment, several charging protocols can be considered, each of them accompanied with different performance.   

\subsubsection{Dissipative charging}

The concept of dissipative charging was firstly discussed by~\citet{Barra2019}, and then extended by~\citet{Hovhannisyan20,Chang2021}. These Authors introduce the idea of charging a quantum battery via the interaction with a dissipative mechanism. 
\citet{Barra2019} propose to engineer a dissipative process that involves an ensemble of auxiliary quantum systems $\mathrm{A}$, all initialized in a thermal (Gibbs) state $G_{\beta}[H_\mathrm{A}]$ at some temperature $(k_{B}\beta)^{-1}$. A sequence of interactions between the battery and individual elements of this ensemble is then used to drive the battery's state to an \emph{active} equilibrium $\pi$, i.e., an equilibrium state with positive ergotropy $\mathcal{E}[\pi] > 0$. This scheme is akin to that of collisional models, also applied to the charging and stabilization of dissipative quantum batteries~\cite{Seah2021,Landi2021,Ciccarello2022,Barra2022a,Salvia2022,Chen2022c,Shaghaghi2022,Shaghaghi2023a}.

For this to occur, the battery is coupled with an auxiliary subsystem $\mathrm{A}$ via a time-independent interaction $V$ for a time interval $\tau$. For the duration of this interaction, the battery's states evolves as $\rho \to \rho' = \tr_\mathrm{A}[U \rho\otimes G_\beta[H_\mathrm{A}] U^\dagger] =:\Phi[\rho]$, with $U = \exp[-i(H_\mathrm{B}+H_\mathrm{A}+V)]$. This implicitly defines a dissipative dynamical map $\Phi$ on the battery, which is there assumed to have a unique fixed point $\pi$, such that $\Phi[\pi] = \pi$ and $\lim_{n\to\infty}\Phi^n[\rho_0] = \pi$ for any initial state $\rho_0$ of the battery. The Authors then recall that any unitary operator $U$ associated with a map with equilibrium satisfies $[U,H'_\mathrm{B}+H_\mathrm{A}] = 0$, for some Hamiltonian $H'_\mathrm{B}$ on the battery's space, such that $\pi = G_\beta[H'_\mathrm{B}]$~\cite{Barra2017}. 

Now, if $H'_\mathrm{B} = H_\mathrm{B}$ then $\pi$ is a completely passive state and $\mathcal{E}[\pi] = 0$. Instead, if $H'_\mathrm{B} \neq H_\mathrm{B}$, the equilibrium $\pi$ may be active with respect to the bare Hamiltonian of the battery $H_\mathrm{B}$, leading to $\mathcal{E}[\pi] > 0$.~\citet{Barra2019} obtain a general condition for reaching an active equilibrium, which stems from the fact that $[U,H'_\mathrm{B}+H_\mathrm{A}]=0$ if $[H_\mathrm{B},H'_\mathrm{B}] = 0$ and $[H'_\mathrm{B}+H_\mathrm{A},V] = 0$. They conclude that $\pi$ is an active equilibrium if there exists a pair $(i,j)$ such that $(\epsilon_i-\epsilon_j)(\epsilon'_i - \epsilon'_j)\leq 0$, with $\epsilon_i$, $\epsilon'_i$ being the eigenenergies of $H_\mathrm{B}$ and $H'_\mathrm{B}$, respectively. 

While it is clear what $H_\mathrm{B}'$ should be used for $\pi$ to be active in $H_\mathrm{B}$, such as $H'_\mathrm{B} = -H_\mathrm{B}$, the power of the result by~\citet{Barra2019} is that it provides a prescription for the interaction $V$ to obtain an effective Hamiltonian $H'_\mathrm{B}$ with associated active equilibrium $\pi$. The Authors show that a generic interaction $V = \sum_\alpha \hat{B}_\alpha\otimes \hat{A}_\alpha$ leads to an active states if the battery (auxiliary) interaction operators $\hat{B}_\alpha$ ($\hat{A}_\alpha$) satisfy $[H_\mathrm{B},\hat{B}_\alpha] = \lambda_\alpha \hat{B}_\alpha$ ($[H_\mathrm{A},\hat{A}_\alpha] = \lambda_\alpha \hat{A}_\alpha$), for some $\{\lambda_\alpha\}$, leading to $H'_\mathrm{B} = - H_\mathrm{B}$.

\citet{Barra2019} also examine the energetic cost of charging $W_\mathrm{ch}$, associated with performing such repeated interactions $V$ leading to $\pi$. They obtain $W_\mathrm{ch} = \tr[(H_\mathrm{B}-H'_\mathrm{B})(\pi-\rho_0)]$, which vanishes if $\pi$ is a thermal equilibrium state of $H_\mathrm{B}$. They are then able to quantify the efficiency $\eta$ of the associated charging process, quantified as the ratio between the ergotropy of $\pi$ and the cost of the charging process,
\begin{equation}
    \label{eq:dissipative_efficiency}
    \eta := \frac{\mathcal{E}[\pi]}{W_\mathrm{ch}} = 1 - \frac{|Q_\mathrm{ch}|}{W_\mathrm{ch}},
\end{equation}
in agreement with the 2nd law of thermodynamics and analogy with the Carnot limit,
where $Q_\mathrm{ch} = \tr[H'_\mathrm{B}(\pi-\sigma_\pi)]$ is the total heat exchanged, with $\sigma_\pi$ begin the passive state of $\pi$, as in Eq.~\eqref{eq:passive_of_rho}. It follows that, for $H_\mathrm{B} = -H'_\mathrm{B}$ and $\rho_0 = \sigma_\pi$ the efficiency is $\eta = 1/2$. The Authors also use an example involving a 2-qubit battery to show that unit efficiency $\eta\to1$ can be achieved in the limit of of zero temperature $\beta\to\infty$, while for finite temperatures, unit efficiency leads to vanishing ergotropy $\mathcal{E}\to0$.

These results show that for a dissipative charging process, there exists a trade-off between the charging efficiency $\eta$, and the maximum amount of extractable work, in the form of the ergotropy $\mathcal{E}$. Hence, one cannot maximize both at the same time, and this is a general property of thermalisation-assisted charging~\cite{Barra2019,Hovhannisyan20}. To overcome this issue, other solutions have been proposed, as discussed in~\ref{ss:environment-assisted_charging}. These make use of an environment that is able to bring a DQB to an active equilibrium state which is not necessarily thermal.

\subsubsection{Environment-assisted charging}
\label{ss:environment-assisted_charging}

This charging mechanism described in the previous section can be generalized by relaxing the condition on $\pi$ to be thermal, which may be attained for a more structured environment, where, in the general case, also a charger can be included. Thus, in the paradigm of \emph{environment-assisted charging}, the effect of such structured environment can assist the charging process. Among the proposed approaches, we can make a clear distinction between two complementary topics that we are going to present: First, the case of charging assisted by non-Markovian effects on the battery dynamics~\cite{Kamin2020,Ghosh2021}, and second, the beneficial consequences for charging performance that are provided by engineering the interactions between the battery and the external environment \cite{TabeshPRA2020,Xu2021a,XuFP2023,Xu2023c}. In both cases the environment is known at least partly, so that some modeling of the experimental data can be built. 

Let us start from the case of \textit{charging assisted by non-Markovian effects}. In the work by~\citet{Kamin2020}, both the battery and the charger are given by TLSs resonantly coupled and interacting with an independent environment. The latter is given by an amplitude damping reservoir, modeled as an ensemble of quantum harmonic oscillators.~\citet{Kamin2020} determine that in the under-damped regime~\footnote{
The interaction between the battery and the charger is stronger than the coupling among such single quantum systems and the common environment. 
} the presence of the environment leading to non-Markovian effects allows for the optimal energy transfer from the charger to the battery. Thus, albeit the environment {\rm E} generally degrades the charging of the battery, there are parameter regimes whereby the charging performance are kept close to the ideal case, represented by the absence of {\rm E}. The Authors also show that the environment is naturally responsible for the \emph{discharging} of the battery when its coupling with the charger is turned off. However, as before, this energy loss is less prominent if the battery dynamics exhibits non-Markovianity. 

In another work,~\citet{Ghosh2021} consider a case-study where the presence of an external noise source leading to non-Markovian effects effectively improves the performance of the charging process with respect to the noiseless case. 
By extending the findings of~\citet{Kamin2020}, this work lays the foundations for a paradigm of charging assisted by non-Markovianity. The key difference with Ref.~\cite{Kamin2020} is that both the battery and the charger are quantum many-body systems. Specifically, in \cite{Ghosh2021} the battery is initially prepared as the ground state of an one-dimensional XY model with a transverse magnetic field in open boundary conditions, and is charged/discharged via interactions with \emph{local} bosonic reservoirs. It is thus shown that, in the transient regime, the DQB can store energy faster and it can return a higher amount of extractable work if the DQB is affected by non-Markovian dephasing noise sources. The enhancement per spin of the transverse XY model is independent of the size of the quantum many-body battery.

A complementary scenario is that of \textit{engineered  battery-environment interactions}. 
In the work by~\citet{Xu2021a}, the Authors consider a battery and the charger given by TLS, similarly to~\cite{Kamin2020}.~\citet{Xu2021a} show that, by tuning the spectral density function of the environment, the performance of the charging process increases when a stronger coupling strength with the environment is enabled in the so-called \emph{band-gap} configuration. This means to require that the spectral density function of the environment is constituted of both a positive- and a negative-weighted Lorentzian spectra, both centered around the same frequency. In such band-gap configuration, one can fully charge the DQB and extract the total stored energy as useful work. The Authors also consider other configurations for the DQB-environment interaction different from the band-gap one. Another approach proposed by~\citet{TabeshPRA2020}, consists of a battery and charger that are not directly coupled, but that are allowed to interact through a common environment, similarly to Ref.~\cite{Quach2020}. In this scenario the battery, can be charged more efficiently in the strong coupling regime of the battery-environment interaction. 

\subsection{Active charging and stabilization methods}
\label{ss:oqb_stabilisation}

To see how charging and stabilisation are achieved in DQBs, it is helpful to frame them as a control task, whereby some control operations are used to increases the energy of the battery with respect to its internal Hamiltonian $H_\mathrm{B}$, and then protect it from leaking back into the environment.
With reference to the charging protocols described in Secs.~\ref{s:fundamental_theory} and~\ref{s:quantum_battery_models}, a battery can be fully charged if the transformation $\ket{g} \rightarrow \ket{e}$ can be carried out, with $\ket{g}$ and $\ket{e}$ being the ground and maximally excited states of the battery, respectively. Through the lens of quantum optimal control theory, 
a quantum battery is thus \emph{completely chargeable} if there exist a dynamical map $\Lambda_{0;\tau}$ such that $\Lambda_{0;\tau}[\ketbra{g}{g}] = \ketbra{e}{e}$ within some target time $\tau$. These conditions, as well as less stringent ones, outline the control task to be solved, which falls within the more general concept of \emph{controllability}~\cite{Altafini2012,Dalessandro2021}. 

\subsubsection{Controllability of closed and open quantum batteries}
For isolated and closed quantum batteries, we are asking for the existence of a unitary operator $U(\tau;0)$ so that $\rho(\tau) = \mathcal{U}(\tau;0)[\rho_0] = \ketbra{e}{e}$. 
If such unitary operator $U(\tau;0)$ does not exist, one may relax the control requirements of the charging process. For example, one could vary the target time $\tau$ or chose a target state $\rho^\star$ different from the maximally-charged state associated with $H_\mathrm{B}$. 
Alternatively, we may still seek full charging $\ket{e}\to\ket{g}$, by \textit{opening} the quantum battery to external resources. In the first instance, this can be achieved via the charger~\cite{Ferraro2018,Farina2019}, as seen in Sec.~\ref{ss:dicke_quantum_battery} for the case of the Dicke quantum battery. A charger can be employed regardless of whether the battery is interacting with the external environment. However, in the presence of a dissipative environment~\cite{CarregaNJP2020,Carrasco2022}, one would seek for conditions to guarantee that the charging process is energetically efficient, despite the effect of dissipation. Some of these conditions are reviewed in Sec.~\ref{ss:dissipative_charging}, such as the \emph{dissipative charging} mechanism~\cite{Barra2019,Hovhannisyan20} in which charging is assisted by a thermal dissipative process able to bring the battery to an \emph{active} equilibrium state. Other requirements and methods will be discussed in Sec.~\ref{ss:oqb_stabilisation}.

In the current literature, several solutions have been proposed to realize a charger. Among the most meaningful, it is worth mentioning a thermal heat engine~\cite{LevyPRA2016}, an external quantised light field~\cite{Ferraro2018,Andolina2019,MonselPRL2020}, an ancilla system acting as a controllable switch~\cite{Farina2019}, and a stream of coherently-prepared quantum units~\cite{Seah2021,Landi2021,Salvia2022}. Concerning the latter, a \emph{collisional charging} protocol has been suggested, in which a quantum battery is charged via repeated interactions with a sequence of subsystems, such as qubits~\cite{Seah2021}. There, each subsystem is allowed to interact with the battery for a finite time. In collisional models charging speed-up can be achieved by exploiting coherences in the state of the qubits~\cite{Seah2021}.

\subsubsection{Stabilization power and its energy cost}

Let us now consider a scenario where we aim to charge a DQB onto the maximally charged state $\rho^\star = \ketbra{e}{e}$ from an arbitrary initial quantum state 
at the desired time $\tau$, and stabilize $\rho^\star$ for an arbitrarily long time until the external user requests for the stored energy. For such a purpose, unitary operations are no longer sufficient, since they are \emph{isentropic} (thus, adiabatic and reversible) processes and cannot decrease the entropy changes induced by ${\rm E}$. Hence, a non-unitary operation is needed, clearly requiring an extra energy cost, dubbed as \emph{stabilization cost}, due to the use of auxiliary systems. In the following, we are going to discuss both \textit{open}- and \textit{closed-loop} control strategies, and corresponding stabilization costs.

For these strategies to be energetically favorable, the stabilization cost $W_\mathrm{stab}$ must be smaller than the energy storage capacity $E_\mathrm{max} = \tr[H_\mathrm{B}(\rho^\star-\ketbra{g}{g})]$. 
To this end,~\citet{Gherardini2020} introduce two figures of merit: The instantaneous \emph{relative stabilization cost},
\begin{equation}
\eta_{\rm stab}(t) := \frac{ W_{ \rm stab }(t) }{ E_{\mathrm{max}} },
\end{equation}
and the instantaneous \emph{relative excess stabilization cost},
\begin{equation}\label{eq:excess_stab_cost}
\zeta_{\rm stab}(t) := \frac{ W_{ \rm stab }(t) - \mathfrak{L}(t) }{ E_{\mathrm{max}} },
\end{equation}
that identifies the excess cost besides the total amount of energy spent to compensate the energy leakage $\mathfrak{L}(t)$ due to decoherence and thermalisation. For open dynamics described by the GKSL Eq.~(\ref{eq:Lindblad_open_dynamics}), $\mathfrak{L}(t) := \tr[H_\mathrm{B}\mathcal{D}[\rho_\mathrm{B}(t)]]$. Ideally, one should allow for $W_{ \rm stab }(t) = \mathfrak{L}(t)$ such that $\eta_{\rm stab}(t) = \mathfrak{L}(t)/E_{ \mathrm{max} }$ and $\zeta_{\rm stab}(t)=0$ for any time $t$, so as the energy needed to stabilize the DQB is minimized. This intuitive result can be analytically demonstrated with thermodynamics arguments~\cite{Gherardini2020}. 

The power $P_{\rm stab}(t) := \dot{W}_{\rm stab}(t)$ in stabilizing a DQB has a \emph{lower bound}~\cite{Gherardini2020}, which determines the minimum amount of energy that has to be supplied from the outside. 
The lower bound derives from the 1st law of thermodynamics~\cite{HorowitzPRL2015}:
\begin{equation}
\dot{E}(t) = \dot{W}_{ \rm stab }(t) + \mathcal{J}_{\mathrm{E}\to\mathrm{B}}(t) \,,
\end{equation}
where $E(t) = \tr[H_\mathrm{B}(t)\rho_\mathrm{B}(t)]$ is the instantaneous total battery energy\footnote{Assuming the ground state to have zero energy.}, while $\mathcal{J}_{\mathrm{E}\to\mathrm{B}}(t) := \dot{E}_{\mathrm{EB}}(t)$ denotes the instantaneous \emph{energy current} driven into the battery by the external environment.
It can be thus shown~\cite{Gherardini2020} that the stabilization power $P_{\rm stab}(t)$ follows the inequality
\begin{equation}\label{eq:stabilization_power}
P_{\rm stab}(t) \geq \dot{E}(t) - T_{\pi}\dot{S}[\rho_\mathrm{B}(t)] \,,
\end{equation}
where $T_{\pi}$ is an effective temperature able to parameterize $\pi$, invariant state of the DQB under the influence of the environment~\footnote{
If the battery is a TLS or the invariant state $\pi$ is thermal, then the parameterization is exact. Otherwise, the value of $T_{\pi}$ can be determined by setting further constraints on the energy or the entropy associated to $\pi$~\cite{Salvia2020energyupperbound}. Another possibility to set-up the value of $T_{\pi}$ is to invoke the concept of \emph{virtual temperatures}~\cite{BrunnerPRE2012}. 
}. 
In case the open dynamics of the battery is governed by Eq.~(\ref{eq:Lindblad_open_dynamics}), then $\dot{S}[\rho_\mathrm{B}(t)] = - \tr[\mathcal{D}[\rho_\mathrm{B}(t)]\log\rho_\mathrm{B}(t)]$. 

The lower bound of Eq.~(\ref{eq:stabilization_power}) also entails an inequality on the time-derivative of the non-equilibrium free-energy $F(t) := E_\mathrm{EB}(t) - T_{\pi}S[\rho_\mathrm{B}(t)]$ for the battery without the charger and the controller. Such inequality reads 
$\dot{F}(t) \leq 0$, whereby the non-equilibrium free-energy $F(t)$ reduces over time due to the increasing of the battery von Neumann entropy. Hence, the action of the control procedure assisting the charger is to invert this time-behavior of $F(t)$. As a consequence, the \emph{minimum stabilization power} is the one allowing for $\dot{F}(t) = 0$ (i.e., constant non-equilibrium free-energy), such that the entropy due to the environment is perfectly compensated by the control. Therefore, implicitly from Eq.~(\ref{eq:stabilization_power}), the \emph{minimum energy cost} to stabilize a DQB shall be proportional to $E(t) - T_{\pi}S[\rho_\mathrm{B}(t)]$.

\subsubsection{Charging and stabilization via sequential measurements}

\citet{Gherardini2020} introduce a first control strategy for the stabilization of a DQB, given by a non-unitary open-loop control that requires to perform a sequence of projective energy measurements at discrete consecutive times~\cite{GherardiniPRA2019,GherardiniPRE2021}. The energy measurements have to be close enough in time, so as to freeze the dynamics of the DQB in the corresponding energy basis, in agreement with the \emph{quantum Zeno regime}~\cite{SmerziPRL2012,MuellerPRA2016}. In this way, the energy basis becomes a decoherence-free subspace, and the DQB is stabilized in the maximally-charged state\footnote{The DQB is repeatedly brought back to $\ketbra{e}{e}$ with probability as much greater as $\rho_\mathrm{B}(t)$ and $\ketbra{e}{e}$ are statistically \emph{indistinguishable}, i.e., their difference is detectable by no measurement devices~\cite{WoottersPRD1981}.} $\rho_e :=\ketbra{e}{e}$.
Albeit maintaining the quantum Zeno regime imposes quite stringent constraints~\cite{SmerziPRL2012,GherardiniNJP2016,MuellerPRA2016}], the time intervals among projective measurements can be optimized in order to minimize the stabilization power $P_{\rm stab}$.

\begin{figure}
    \centering
    \includegraphics[width=0.45\textwidth]{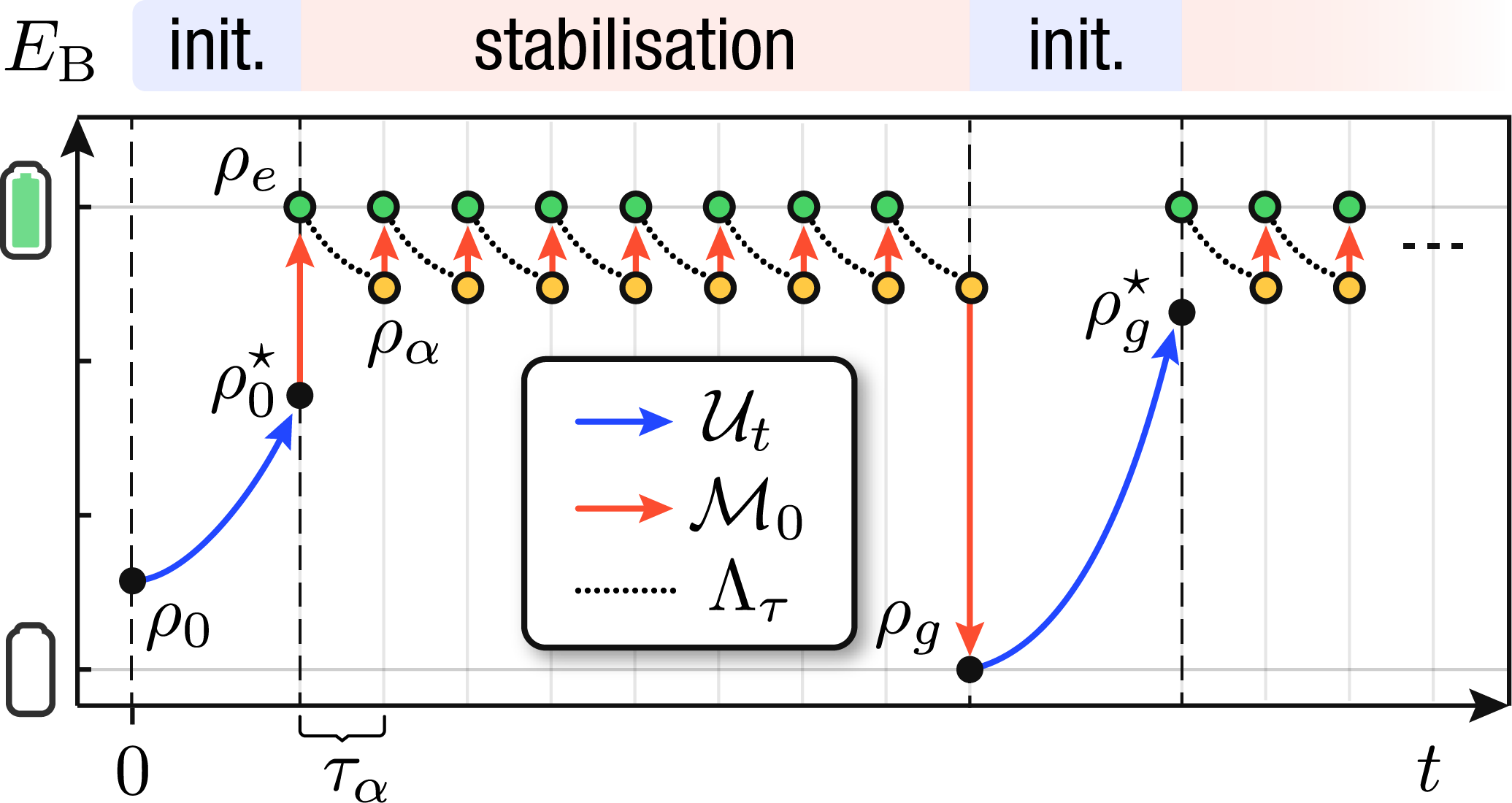}
    \caption{(Color online) Single-run pictorial representation of sequential measurements protocol for a TLS battery. After the initialization step from the initial state $\rho_0$, the stabilization protocol consists of intermittent free evolutions, fast unitary controlled dynamics $\mathcal{U}_t$ (blue solid lines) and projective measurements $\mathcal{M}_0$ (solid red lines) in time intervals of duration $\tau$. 
    The green dots denote the maximally-charged state $\rho_e$, while the yellow dots represent the battery state $\rho_{\alpha}$
    to which the DQB decays as an effect of the environmental action $\Lambda_{\tau}$, here exaggerated for illustrative purposed. Finally, $\rho_{0}^{\star}$ and $\rho_{g}^{\star}$ are, respectively, the nearest states to $\rho_e$ on the unitary orbit of $\rho_0$ and $\rho_g = |g\rangle\!\langle g|$. 
    Adapted from \cite{Gherardini2020}.}
    \label{fig:stab_scheme}
\end{figure}

Let us now outline how the sequential measurements protocol works for the case of a TLS battery.
As shown in Fig.~\ref{fig:stab_scheme}, three steps of the protocol can be identified. {\it (i) Initialization}. Any initial state $\rho_0 := \rho_\mathrm{B}(0)$ of the battery is driven to the unitarily-connected state that is closest to $\rho_e$, here denoted with $\rho_{0}^{\star}$. This transformation is allowed by controlling the battery through a time-dependent Hamiltonian as in Eq.~\eqref{eq:unitary_work_extraction}. The initialization step has to be performed sufficiently fast for the dynamics to be approximately unitary despite the presence of the external environment. 

{\it (ii) Full charging \& stabilization via energy projective measurements}. Once the battery is in $\rho_{0}^{\star}$, a projective energy measurement (defined in the eigenbasis of $H_\mathrm{B}$) is performed. The battery's state collapses into the maximally-charged state $\ket{e}$ with probability $p_{e}[\rho_{0}^{\star}]=\tr[\rho_{0}^{\star}\rho_e]$. Subsequently, the interaction with the environment leads the battery to lower-energy state $\rho_\alpha = \Lambda_{\tau_\alpha}[\rho_e]$.
A sequence of projective measurements $\mathcal{M}_0$ on the basis of the battery Hamiltonian $H_\mathrm{B}$ repeatedly brings the battery to state $\rho_\alpha\xrightarrow{\mathcal{M}_0}\rho_e$ at consecutive times, with probability $p_e[\rho_\alpha]$. In order for the protocol to be successful, the time $\tau_\alpha$ between subsequent measurements must be much smaller than the energy timescale of both the battery free-evolution and the decoherence time dictated by {\rm E}. In such way, the states $\rho_{\alpha}$ are sufficiently close to $\rho_e$ such that $p_e[\rho_{\alpha}]\gg 1/2$. 

{\it (iii) $\to$ (i) Re-initialization}. When a projective measurement collapse $\rho_\alpha$ onto the ground state $\ketbra{g}{g}$, a new initialization is performed, by applying unitary driving as in step {\it (i)}.

For an unstable DQB subject to the external environment (thus not fully chargeable without applying non-unitary control actions), the attainment of the quantum Zeno regime in the energy eigenbasis of $H_\mathrm{B}$ allows for full charging and stabilization. However, this is assured at the price of extra energetic and entropic costs. On average, no \emph{energetic cost} is associated with the jumps of the energy projective measurements, independently on which quantum state is measured. 
However, there is a cost associated with storing and erasing the information associated with these measurements, connected to the \emph{entropic cost} of the protocol. Let $p_k[\sigma]$ be the probability for some $d$-level battery's state $\sigma$ to collapse on one of its energy eigenstates $\epsilon_k$. These probabilities carry the \emph{informational content} of the outcomes obtained by measuring the state $\sigma$. The \emph{storing} and \emph{erasure} of each measurement outcome requires an entropy production, whose rate of change over time is proportional to the time-derivative of the Shannon entropy $S_\mathrm{sh}[\sigma] := - \sum_{k}p_{k}[\sigma]\log p_{k}[\sigma]$. 

Hence, implicitly, as an effect of the \emph{Landauer's principle}~\cite{Landauer2010a,BerutNature2012,Zhen2021}, the irreversible erasure of the measurement information content is responsible for an energetic cost $W_{\rm Zeno}$. 
Formally,
\begin{equation}
W_{\rm Zeno} = \overline{m}\beta_{\rm eras}^{-1}S_\mathrm{sh}[\rho_{\alpha}],
\end{equation}
where $\overline{m}$ denotes the average number of projections in the quantum Zeno regime (step {\it (ii)}), and $\beta_{\rm eras}$ is the inverse temperature of the macroscopic system that erases the memory containing the measurement outcomes. 
We conclude by observing that the time $\tau_\alpha$ between measurements has to obey a \emph{trade-off} condition: $\tau_\alpha$ cannot be too large to guarantee high stabilization performance, but neither too small so as the entropic cost and energy consumption, both proportional to $\overline{m}$, become unsustainable.

\subsubsection{Charging and stabilization with feedback-control}

Another approach to charging and stabilization consists in using \emph{linear feedback}, as proposed by~\citet{Mitchison2021}.
In their work, the Authors consider a finite-dimensional battery and show that feedback (or closed-loop) control is able to fully charge the battery by increasing its average total energy as a monotonic function. They also show that this protocol can be used to keep the battery in a charged state $\rho_\mathrm{B}(t)=\rho^\star$ despite the presence of the environment {\rm E}. Moreover, the feedback asymptotically makes the target charged state, chosen by the user, a non-equilibrium steady-state that is \emph{robust} to small additional (i.e., not explicitly modeled) thermal noise.

\begin{figure}
    \centering
    \includegraphics[width=0.42\textwidth]{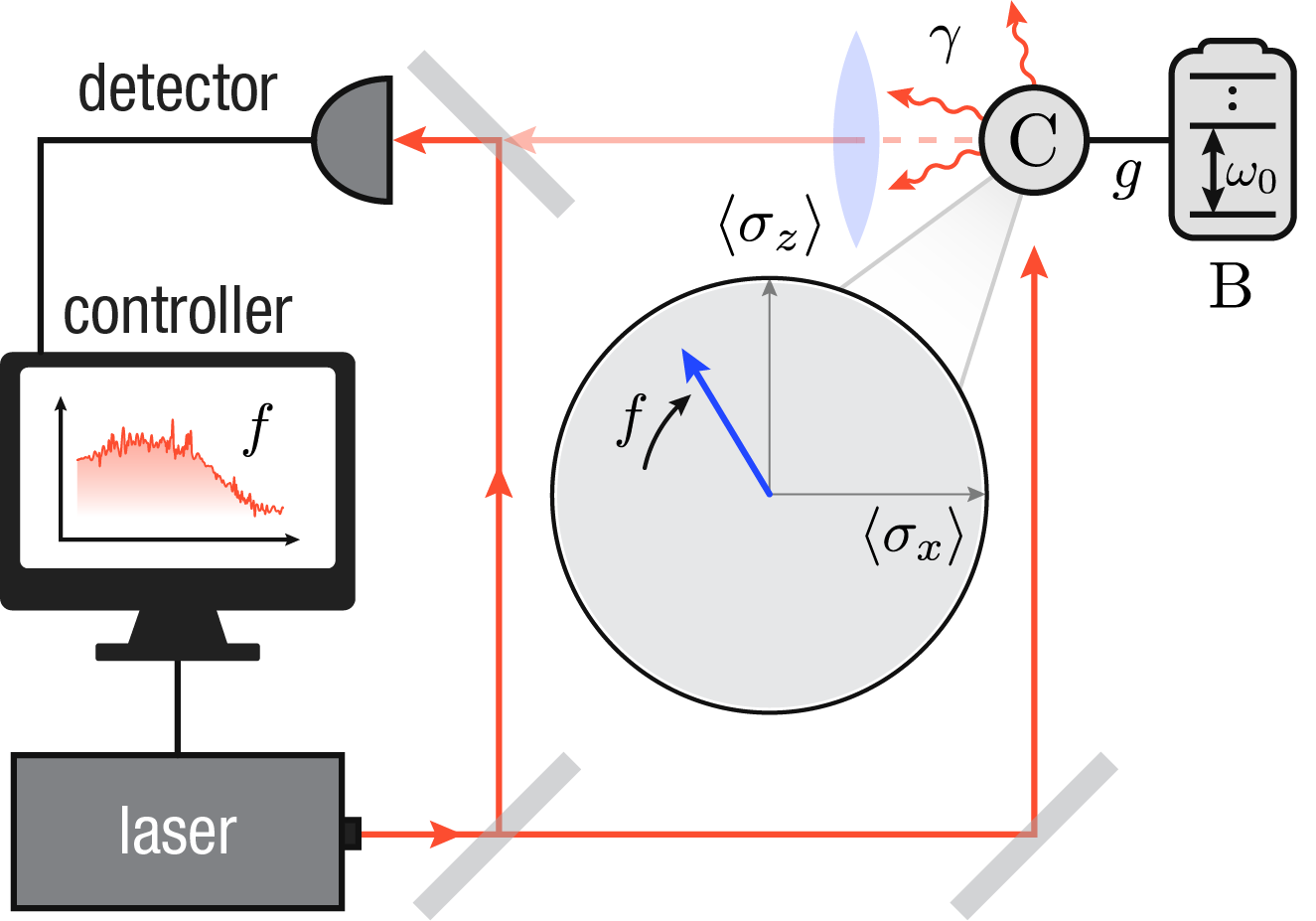}
    \caption{(Color online) Charging and stabilisation via feedback control. 
    The quantum battery $\mathrm{B}$ is coupled to a charger $\mathrm{C}$, which is affected by the external environment. Due to this interaction, the charger spontaneously emits photons at some rate $\gamma$, which are collected by a detector. The resulting measurement records trigger a feedback loop that modulates the intensity of a control laser pulse driving $\mathrm{C}$. The driving laser pulse is made interfere with the photons emitted by the charger thanks via interferometric homodyne measurement. 
    In the figure we also depict a feedback action operating on the charger, where $f$ is the feedback strength and the blue arrow is related to the stabilization of the Bloch vector as close as possible to the vertical axis $z$. Figure and caption adapted from~\cite{Mitchison2021}.}
    \label{fig:feedback}
\end{figure}

The procedure proposed by~\citet{Mitchison2021}, illustrated in Fig.~\ref{fig:feedback}, is based on homodyne-like continuous measurements and linear feedback control~\cite{Belavkin1987,WisemanPRL1993,Wiseman1994}. A finite-dimensional quantum battery is
coupled to a two-level quantum charger that pumps energy into the battery. In the following, $\sigma_{\rm C}^{i}$, with $i=x,y,z$, will denote the standard Pauli operators acting on the charger. The battery resonantly exchanges energy with the charger via the interaction Hamiltonian $H_{1}=g(\hat{s}_\mathrm{B}\hat{\sigma}^{+}_{\rm C} + \hat{s}^{\dagger}_\mathrm{B}\hat{\sigma}^{-}_{\rm C})$, where $g$ is the battery-charger coupling strength, $\hat{\sigma}^{\pm}_{\rm C}:=(\hat{\sigma}^x_{\rm C} \pm i\hat{\sigma}^y_{\rm C})/2$ and $\hat{s}_\mathrm{B} := \sum_{k=1}^{d-1}|k-1\rangle\!\langle k|$ denotes the lowering operator for the battery. 

The bare Hamiltonian $H_\mathrm{B} + H_\mathrm{C}$, with  $H_\mathrm{C} = \hat{\sigma}^z_{\rm C}/2$, commutes with $H_{1}$: $[H_\mathrm{B} + H_\mathrm{C},H_{1}]=0$. Thus, the local energy of both the battery and the charger does not change on average. Instead, $\omega_0 \propto \|H_\mathrm{B} + H_\mathrm{C}\|_\mathrm{op}$ is taken as the largest energy scale of the composite battery-charger system. While the battery is considered a well-isolated system, the charger is coherently driven via the Rabi Hamiltonian $H_{\rm drive}(t) = \Omega(t)\hat{\sigma}_{\rm C}^{y}$, with $\Omega(t) \gg \omega_0$ $\forall t$, in the interaction picture with respect to $H_{\rm B}+H_{\rm C}$ and in the rotating wave-approximation. In this way, fast energy transfer to the battery is enabled. However, the charger is affected by the external environment, represented by a multi-mode field that induces spontaneous emission from the charger. Therefore, in the absence of feedback control, the composite battery-charger system evolves following a driven-dissipative dynamics. 

The linear feedback control in~\cite{Mitchison2021} consists in measuring the photons that are spontaneously emitted by the charger via a homodyne interferometer. The latter returns the signal 
\begin{equation}
r(t)dt = \tr[\rho_\mathrm{BC}(t|r)\hat{\sigma}_{\rm C}^{x}]dt + \frac{d\omega(t)}{\sqrt{\varsigma\gamma}} \,,
\end{equation}
with $\gamma$ spontaneous emission rate, $\varsigma$ measurement efficiency and $\rho_{\rm BC}(t|r)$ the instantaneous state of the battery-charger system conditioned on the measurement outcome $r$. The Wiener increment $d\omega(t)$ models the noise in the detector output that makes the measurement signal a stochastic process. The feedback loop, then, is closed by applying a driving field directly proportional to the measurement record to the charger:
$\Omega(t) = \Omega_0 - fr(t-\tau)$, where $\tau > 0$ is the lag in the feedback loop, $f$ the feedback strength and $\Omega_0$ is a constant function. 

In the nearly-ideal case of negligible feedback lag $\tau\approx 0$, the dynamics of the battery subject to feedback is well described by a Markovian GKSL master equation for the ensemble-averaged density operator $\overline{\varrho}_{\rm BC}(t) := \mathbb{E}_{r}[\rho_{\rm BC}(t|r)]$, with $\mathbb{E}_{r}[\cdot]$ denoting the average over the detector noise.
For $\tau > 0$ the Markovian description 
is no longer valid. Hence, in such a case, one needs to solve a stochastic Ito equation, and then average over several realizations of the corresponding quantum dynamics~\cite{Mitchison2021}. 

In the ideal case of noiseless measurement signals (i.e., maximum measurement efficiency $\varsigma$) and instantaneous feedback, a battery can be fully charged.
Nevertheless, in \cite{Mitchison2021}, a good performance of the charging process and battery stabilization are observed even under more realistic constraints of inefficient measurements 
and time delay $\tau$ in the feedback loop. 
However, for sufficiently large $\tau$ the control action stops counteracting both the measurement back-action and the detrimental presence of the environment on the charger. As a result, the feedback loop breaks and behaves as a random sequence of quantum measurements \cite{GiachettiPRE2021}, so that the charger relaxes to the maximally mixed-state, which is equivalent to a thermal state at infinite temperature.     

Feedback control is certainly an up-and-coming solution. However, further in-depth studies need to be carried out: Evaluation of the energetic and entropic costs of the feedback control mechanisms in \cite{Mitchison2021,Yao2021,YaoPRE2022}, as well as the investigation of the corresponding charging precision. Furthermore, the energy cost of the work extraction will have to be taken into account, as well as a fully-autonomous approach~\cite{MitchisonCP2019,HernandezPRXQuantum2022} to enable thermodynamic process including feedback, certainly deserve further investigations. 

\subsection{Protection from energy losses}
\label{ss:protection_from_energy_loss}

Let us now discuss some approaches based on the preparation of charged states that are then protected from the underlying decoherence process. These protection methods can be thought of as \textit{passive}, in the sense that they do not require energy expenditure following the state preparation protocol, in contrast with the \textit{active} stabilisation methods discussed in Sec.~\ref{ss:oqb_stabilisation}. 
Some passive protection methods have been introduced in
Refs.~\cite{Liu2019,Quach2020,Liu2021b}.

\subsubsection{Protection via decoherence-free subspaces}
\label{sss:decoherence-free_subspaces}

\citet{Liu2019} propose an approach to engineer loss-free open quantum batteries, i.e., OQBs that are protected from losing energy to the environment. Their scheme consists in preparing the battery in some energetically favorable state of a decoherence-free subspace (DFS). The latter is a subset of the states space that is protected from decoherence, in which the dynamics is perfectly unitary and unaffected by the environment~\cite{Lidar2003}. DFSs have been extensively studied in the theory of open quantum systems and quantum information~\cite{Zanardi1997}, as a promising mean to shield quantum states from noise sources that limit the performance of quantum computing~\cite{Lidar1998,Beige2000,Bacon2000} and quantum key distribution~\cite{Walton2003}. This strategy can also be used to prevent energy loss, by engineering a relaxation-free subspace~\cite{Deponte2007}, i.e., a DFS unaffected by thermal relaxation.

The working principle of \citet{Liu2019}'s loss-free battery is based on the decomposition of the battery's Hilbert space into subspaces that are invariant with respect to the dynamics induced by the total Hamiltonian. Following~\citet{Liu2019}, let us consider a system (battery) described by some quantum network Hamiltonian 
\begin{equation}
\label{eq:network_Hamiltonian}
    H_\mathrm{B} = \sum_{j} E_j \ketbra{j}{j} + \sum_{(j,k)} (J_{jk} \ketbra{j}{k} + {\rm h.c.}),
\end{equation}
where $E_j$ is the energy of each site in the network and $J_{jk}$ the coupling between pairs of sites $(j,k)$. The Authors consider an environment given by a bath of uncorrelated, harmonic vibrational modes, coupled to the system only via a subset $\mathcal{S}$ of site operators $\ketbra{j}{j}$, here associated with \textit{surface} sites. If there exists a unitary symmetry operator $\hat{\Pi}$ such that
\begin{equation}
    \label{eq:symmetery_DFS}
    [\hat{\Pi},H_\mathrm{B}] = 0, \quad [\hat{\Pi},\ketbra{j}{j}] = 0 \quad \forall \; \ketbra{j}{j} \in \mathcal{S},
\end{equation}
then $\hat{\Pi}$ and $H_\mathrm{B}$ admit a common eigenbasis $\{\ket{\psi_\alpha^{(k)}}\}$, where the index $\alpha$ is associated with eigenvalues $\lambda_\alpha$, whose degeneracy $d_\alpha$ is spanned by index $k=1,\cdots,d_\alpha$. Under these conditions one can decompose the system's Hilbert space $\mathcal{H}_\mathrm{B} = \bigoplus_{\alpha} \mathcal{H}_\mathrm{B,(\alpha)}$,
such that the dynamics generated by the total system-environment Hamiltonian $H_\mathrm{B}+H_\mathrm{int}+H_\mathrm{E}$ leaves these subspaces $ \mathcal{H}_\mathrm{B,(\alpha)}$ invariant. 
See~\cite{Lidar2003} for a review of different DFS conditions and formulations. 

\begin{figure}
    \centering
    \includegraphics[width=0.46\textwidth]{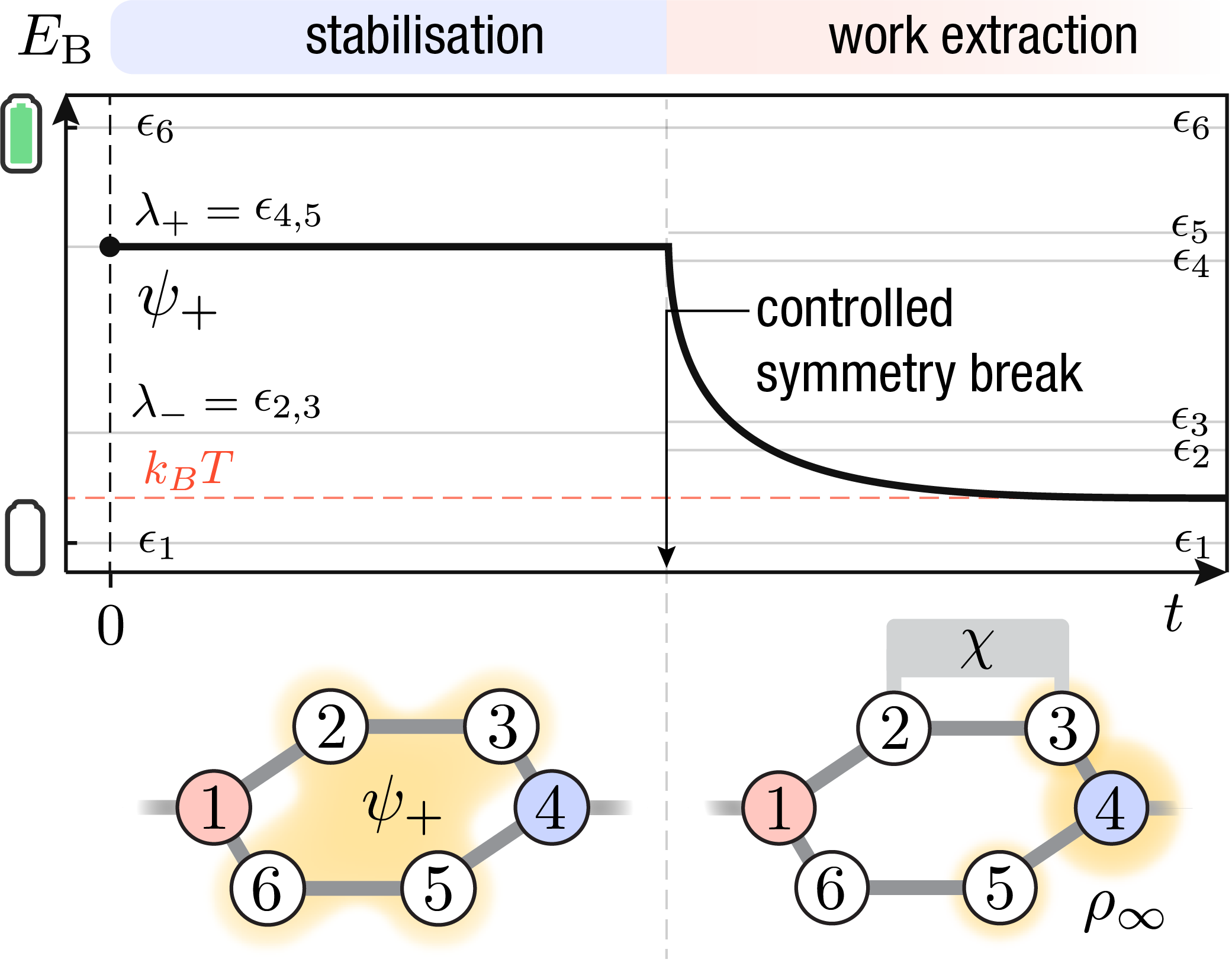}
    \caption{~\citet{Liu2019} consider Eq.~\eqref{eq:network_Hamiltonian} with nearest-neighboring couplings $J_{jk}$ with cyclic boundary conditions. Bulk sites $j=2,3,5,6$ are set to have the same energy $E_j = E_\mathrm{bulk}$, different from $E_1, E_4$. All couplings are chosen to be $J_{jk} = J_0$. The environment is coupled to the battery only via surface site operators $\mathcal{S} = \{\ketbra{1}{1}$, $\ketbra{4}{4}\}$. Under these conditions, the unitary operator 
        $\hat{\Pi} = \exp[i(\ketbra{2}{6}+{\rm h.c.})]\exp[i(\ketbra{3}{5}+{\rm h.c.})]$,
    corresponding the independent and simultaneous exchange of sites $2\leftrightarrow6$ and $3\leftrightarrow5$, commutes with $\mathcal{S}$. This system has two DFSs associated with the eigenvalues $\lambda_\pm = E_\mathrm{bulk}\pm J_0$ of $\Pi$, and its eigenstates $\ket{\psi_{\pm}} = \frac{1}{2} (\ket{6}-\ket{2}\pm \ket{5}\mp\ket{3})$. An initial charged state $\ket{\psi_+}$ can be preserved indefinitely. Work can be extracted by funneling energy to a target site ($j=4$) by breaking the symmetry, here via a small perturbation $\chi$ of the energy of bulk sites $2,3$.}
    \label{fig:symmetry_protection}
\end{figure}

To illustrate this approach, Liu \etal consider an quantum battery consisting of six chromophores (or color sites)~\cite{Mikhnenko2015}, forming a network with ring topology~\cite{Trudeau1994}, illustrated in Fig.~\ref{fig:symmetry_protection}. Such system, drawn from similarity with the tight-binding model of a benzine molecule~\cite{Bancewicz1989}, can be realized with excitonic~\cite{Alicki2019,Jang2018} and superconducting architectures~\cite{Blais2021}, as well as on other quantum technology platform. This system can be engineered to have two \textit{dark} states\footnote{In analogy with the optically-dark states that arise, under similar conditions, in coupled TLSs interacting with light. Here the states may be considered acoustically-dark, i.e., prevented from exchanging energy with the environment via emission or absorption of a phonon.} $\ket{\psi_\pm}$, not subject to decoherence or relaxation. Specifically, if the composite system is initialized in a state $\rho_0 = \ketbra{\psi_\pm}{\psi_\pm}\otimes\rho_\mathrm{E}$, then the battery will remain in the state $\ketbra{\psi_\pm}{\psi_\pm}$ for all times.

Let us make a few consideration. First, the robustness of these symmetry-protected dark states depends on how well the conditions of Eq.~\eqref{eq:symmetery_DFS} are met. The lifetime of dark states depends on the magnitude of the disorder affecting energies and couplings in the sites basis, as well as on the temperature and the strength of the coupling with the environment~\cite{Liu2019}. 
Second, charging the battery onto such dark states is a challenging task in its own right. Sudden quenches, such as optical pumping with a pulsed laser, can bring both system and environment out-of-equilibrium and generate correlations between 
strongly coupled exciton and phonons~\cite{Wells2008}, and may not provide a direct path to DFSs. However, an established approach to prepare dark states is to adiabatically vary the battery Hamiltonian, with techniques such as the STIRAP, recently demonstrated in the experiment of~\cite{Hu2022} following the proposal of Ref.~\cite{Santos2019a}. 

It is worth mentioning that in excitonic architectures, bulk states are typically weakly-coupled to vibrational modes that drive the system toward thermal relaxation~\cite{Barford2014,Lemaistre1999}. These systems also suffer from radiative and non-radiative losses, such as exciton recombination, internal conversion and inter-system crossing. It is therefore important to develop approaches to prevent these losses when working with excitonic architectures, such as those suggested by~\citet{Davidson2020,Davidson2022} for transport, and by~\citet{Quach2020} for energy storage.

Further work is needed to generalise loss-free quantum batteries to the case of many-body systems. A systematic approach to finding robust and energetically favorable relaxation-free subspaces for scalable many-body system would be invaluable. Another observation is that dark states like $\ket{\psi_{\pm}}$ are not the most energetic states of the battery Hamiltonian and, instead, sit roughly in the middle of its spectrum. This is akin to the subradiant states of Dicke models, which sit around the midpoint of the Dicke ladder~\cite{Glicenstein2022}. Therefore, it is important to understand if there is a trade-off relation between the robustness of dark states and their energetic content, and whether they are compatible with superextensive power scaling. Future work should explore the possibility of using high-energy ratchet states\footnote{Excited states that can absorb but not emit light~\cite{Higgins2017}.} while retaining the absorption enhancement~\cite{Quach2022}.

\subsubsection{Engineering spontaneous dark-state preparation}
\label{sss:dark_states}

Loss-free quantum battery rely on our ability to prepare dark states. However, adiabatic passage and quantum optimal control may not always be viable options, due to limitations on the set of control operations and their bandwidth. Interestingly, protected states can also form spontaneously due to the interaction with an auxiliary charger and the environment. This approach, first proposed by~\citet{Quach2020}, can be adopted to protect many-body excitonic quantum batteries from radiative (and vibrational) losses, while retaining the superextensive scaling in the charging power discussed in Secs.~\ref{s:fundamental_theory} and~\ref{s:quantum_battery_models}.

The system considered in Ref.~\cite{Quach2020} consists of an $N_\mathrm{B}$-body battery and an $N_\mathrm{C}$-body charger, both given by ensembles of uncoupled quantum TLSs with energy gap $\omega$. The battery and the charger are also coupled with a thermal reservoir, given by a bath of optical harmonic modes at temperature $T$. The total system-reservoir Hamiltonian is
\begin{equation}
    \label{eq:dark_state_total_Hamiltonian}
    \begin{split}
        H = \: &\omega(\hat{J}_\mathrm{B}^{z}+\hat{J}_\mathrm{C}^{z}) + \int d\bm{k} E_{\bm{k}}^{\phdagger} \hat{r}^\dagger_{\bm{k}} \hat{r}^{\phdagger}_{\bm{k}}  \\
        &+\frac{g}{2}\Big[ (\hat{J}_\mathrm{B}^{+}+\hat{J}_\mathrm{C}^{+}) \hat{R} + (\hat{J}_\mathrm{B}^{-}+\hat{J}_\mathrm{C}^{-}) \hat{R}^\dagger\Big],
    \end{split}
\end{equation}
where $\hat{J}_i^{x,y,z}$ denotes the usual collective spin operator of ensembles $i=\mathrm{B},\mathrm{C}$ and $\hat{J}_i^{\pm} := \hat{J}_i^{x}\pm i\hat{J}_i^{y}$. The reservoir is here represented via the linear dispersion relation $E_{\bm{k}}$ as a function of the wave vector $\bm{k}$, with creation and annihilation operators $\hat{r}_{\bm{k}}^{\phdagger}$ and $\hat{r}_{\bm{k}}^\dagger$, respectively. The parameter $g$ represents the coupling strength between the battery-charger system and the reservoir, while the reservoir operator $\hat{R} = \int d\bm{k} \, \kappa_{\bm{k}} \hat{r}_{\bm{k}}$ is expressed as a function of a general continuous function $\kappa_{\bm{k}}$, whose form depends on the considered system. The reservoir is assumed to be weakly coupled ($g\ll\omega$) with the system and approximately in the thermal state $G_\beta[H_\mathrm{R}] = \exp[-\beta H_\mathrm{R}]/\mathcal{Z}$, with $H_\mathrm{R} = \int d\bm{k} E_{\bm{k}}^{\phdagger} \hat{r}^\dagger_{\bm{k}} \hat{r}^{\phdagger}_{\bm{k}}$.
Under these conditions, the dynamics of the system's density operator $\rho(t)$ is well-approximated by the Markovian GKSL master equation
\begin{equation}
    \label{eq:dark_state_QME}
            \dot{\rho}(t) =  - i \omega [\hat{J}_\mathrm{B}^z+\hat{J}_\mathrm{C}^z,\rho(t)] \\
             + \gamma (\overline{n}+1)\mathcal{D}_-[\rho]+\gamma\overline{n}\mathcal{D}_{+}[\rho],
\end{equation}
where $\mathcal{D}_{\pm}[\rho] := 2\hat{O}\rho \hat{O}^\dagger - \{\hat{O}^\dagger \hat{O},\rho\}, \quad \text{with} \quad 
    \hat{O} =\hat{J}_\mathrm{B}^{\pm}+\hat{J}_\mathrm{C}^{\pm}$
with $\gamma$ being a function of $g$, and $\overline{n} = 1/[\exp(\beta\omega)-1]$ being the mean thermal population. Interestingly, although battery and charger are not directly coupled, the superoperators $\mathcal{D}_\pm$ (associated with the system-reservoir interaction) can generate correlations between the spins of the two subsystems. 
The Authors consider an initial product state $\ket{\psi_0} = \ket{G}_\mathrm{B}\otimes\ket{E}_\mathrm{C}$, given by the ground state of the battery and excited state of the charger.
The dynamical map generated by Eq.~\eqref{eq:dark_state_QME}, leads $\ket{\psi_0}$ to a steady state $\rho_\mathrm{\infty}$ that partially overlaps with a dark state forbidden from further decaying towards a completely passive thermal equilibrium state. Even at zero temperature, this approach allows to prepare a battery state with finite energy.

With a numerical study conducted for $N_\mathrm{B}\leq 10$, the Authors shows that this charging protocol is characterized by a superextensive scaling in the charging power. The observed scaling of the power density is linear $\braket{P}_\tau/N_\mathrm{B} \propto N_\mathrm{B}$, exceeding that of the Dicke model in the regime discussed in Sec.~\ref{ss:dicke_quantum_battery}. While this power scaling is accompanied by a growth of the rate of entanglement generation, the nature of this scaling is likely to be related to the \textit{pressure} generated by the number of excitations available in the charger, in analogy with the Dicke model. Notably, if the charger is removed, the battery would lose its energy to the bath at a rate that is also superextensive in $N_\mathrm{B}$. By destroying the coherences between the spins via some dephasing operation, one can slow this process down and restore the single-spin relaxation rate~\cite{Quach2020}.

A key finding of Ref.~\cite{Quach2020} is that for low $N_\mathrm{B}$, the capacity of this battery, defined as the energy density in its steady state, also displays superextensive scaling. While this scaling improves upon previous models, usually exhibiting extensive capacity, it relies on the fact that the steady state of the battery $\rho_{\infty,\mathrm{B}}=\tr_\mathrm{B}[\rho_\infty]$ has a rather low energetic content, when compared to the maximally charged $\ket{G}_\mathrm{B}$, whose energy is proportional to $N_\mathrm{B}\omega$. 
As for the effect of temperature, it is shown that there exists a trade-off relation between the energy flow from the bath to the battery and the stability of the dark-states, hindered by thermal fluctuations. While at low temperature the battery gains energy primarily from the charger, as $T$ increases the energy flows from the bath to the battery. However, 
in the large-temperature limit, the ergotropy of the battery vanishes, i.e., $\lim_{T\to\infty}\mathcal{E} = 0$.

Besides the charger-mediated stabilization approach proposed by~\citet{Quach2020}, quantum \textit{metastability}~\cite{Macieszczak2016} may offer another path to the spontaneous formation of protected states. In this phenomenon, related to that of \textit{prethermalisation}~\cite{Berges2004}, relaxation rapidly leads the system towards long-lived non-equilibrium states, from which thermal equilibration is reached over exponentially longer time scales. Recent studies have shed light on the nature of metastability and its relation decoherence-free subspaces~\cite{Macieszczak2021}. These results may help in developing a path towards engineered metastability in quantum batteries, starting from a Hamiltonian prescription of battery, charger and interactions with the environment~\cite{Arjmandi2022b,Song2022a}.

\section{Experimental implementations}
\label{s:architectures}

This Section is devoted to the presentation of the most promising platforms for the realization of quantum batteries. To this end, we can broadly divide suitable platforms in two macro-categories: (i) platforms that operate at ultra-low temperatures, and (ii) platforms that work at room temperature. 

Regarding the first macro-category, any platform for quantum computing is suitable for the realization of quantum batteries. Obvious examples for the realization of scalable solid-state Dicke batteries operating in the quantum coherent regime, which were discussed by~\citet{Ferraro2018}, include semiconductor quantum dots~\cite{Burkard2021} and superconducting materials~\cite{Blais2021}. Other alternatives include NV-centers in diamond~\cite{Schirhagl2014,Casola2018,Barry2020},
neutral atoms, Rydberg atoms~\cite{Adams2020}, and trapped ions~\cite{Bruzewicz2019}. The realization of SYK quantum batteries is much more challenging in that no experimental system where the SYK model has been realized is to date available. As discussed by \citet{Rossini2020}, however, potential candidates include graphene quantum dots with disordered edges subject to quantizing magnetic fields~\cite{Chen2018,Brzezinska2022}
and strange metals~\cite{Patel2019,Hartnoll2022}, i.e, metals whose properties are not well described by Landau's Fermi liquid theory.

Battery platforms in the second, room-temperature macro-category should---in the first place---contain quantum systems with level spacing's $\Delta E$ greatly exceeding the room-temperature thermal energy $k_B T \approx 25~{\rm meV}$, in order to be sufficiently robust to energy relaxation. Secondly, they are expected to be dominated by a collective behavior rather than being based on entanglement generation~\cite{Cruz2021}, simply because the decoherence rate is faster as temperature increases. In Sec.~\ref{sss:organic} we will discuss a particularly promising platform for the realization of room-temperature Dicke batteries.

\subsection{Superconductors}
\label{sss:superconductor}
It is well known that circuit QED setups~\cite{Blais2021} 
offer the opportunity to simulate---in the sense of {\it quantum simulation}~\cite{Daley2022}---some of the most important models of quantum optics. For example, the Tavis-Cummings (TC) model, which, similarly to the Dicke model, also describes a system of $N$ TLSs coupled to a single-mode cavity, was simulated in a circuit QED setup by~\citet{Fink09}. Note that the TC model differs from the Dicke model in that it does not contain counter-rotating terms. Circuit QED therefore offers us a natural platform to fabricate Dicke quantum batteries, discussed in Sec~\ref{ss:dicke_quantum_battery}. In such a platform, TLSs can be realized by using superconducting qubits dubbed ``transmons''~\cite{Koch2007} while the cavity is realized by using a coplanar waveguide resonator. The qubits are positioned at the antinodes of the first-harmonic standing wave electric field and have nearly identical couplings to the cavity mode. The resonators are typically~\cite{Fink09} realized by employing optical lithography and metal evaporation techniques on suitable substrates, while the transmons are fabricated by combining e-beam lithography and shadow evaporation techniques for the metal of choice---typically Aluminum and Niobium, and, more recently, Tantalum~\cite{Place2021}---and its oxide. Recent developments on materials science aspects can be found in recent review articles, e.g.~\citet{Polini2022}. Circuit QED setups are operated in the microwave regime (typical level spacing being on the order of a few GHz) and display timescales on the order of nanoseconds.

A preliminary prototype of a superconducting quantum battery has been realized experimentally by~\citet{Hu2022}. The device contained a {\it single} superconducting qutrit (i.e.~a three-level system) coupled to a single-mode cavity. This system cannot display neither a quantum nor a collective advantage since it contains a single quantum unit. The work of Hu \textit{et al.}, however, constitutes a very important first step towards the realization of highly-controllable superconducting quantum devices capable of storing energy. Indeed, by using two microwave pulses with time-dependent Rabi frequencies, the Authors were able to resonantly drive the qutrit and implement a number of controlled adiabatic charging processes (such as those discussed above in Sec.~\ref{ss:transitionless}). Moreover, by reconstructing the full density matrix of the qutrit via quantum state tomography, \citet{Hu2022} were able to accurately measure the stored energy at each time, during the charging and discharging dynamics.

The physics of quantum batteries based on superconducting qubits can also be studied on platforms like IBM Quantum Experience, as recently done by~\cite{Gemme2022}. Despite their limitations, these controllable experimental setups offer a great starting point to test the working principles of quantum energy storage. However, custom experiments are likely required to explore charging and protection from energy loss in many-body quantum batteries.

\subsection{Quantum dots}
\label{sss:quantum dot}

\citet{Wenniger2022} conducted an experimental study of energy transfer between a two-level system and a reservoir of electromagnetic modes. In this work, the energy transfer occurred via spontaneous emission from a InGaAs quantum dot (the charger) to a micropillar optical cavity (the battery). The quantum dot was resonantly excited by a pulsed Ti:Sapphire laser in a cryostat at 5-20 K, to be brought into a superposition of the ground and excited state. 
\citet{Wenniger2022} considered also a work extraction phase, during which the energy stored in the battery flows into another system (a laser field) through homodyne-type interference. After spontaneous emission of the quantum dot into cavity, the work transferred to the battery corresponds to the coherent part of the emitted field, while the incoherent component corresponds to the heat exchanged. The latter is an energy loss mechanism and does not play a part during work extraction. This work is an important step towards the implementation of a fully operative quantum battery, as it involves a battery system with finite energy bandwidth that can be charged and from which energy can be extracted. Further efforts could extend this setup to the case of many-body quantum batteries, and study approaches to protect charged states from energy relaxation.

\subsection{Organic microcavities}
\label{sss:organic}

In Ref.~\cite{Quach2022}, the Authors present an experimental implementation of a many-body charging protocol, realized using an organic semiconductor coupled to a confined optical mode in a microcavity at room temperature. The Authors provide some evidence that the system displays superextensive scaling of energy absorption, interpreting the experimental results in terms of the Dicke quantum battery model discussed in Sec.~\ref{ss:dicke_quantum_battery}.
The set-up consists of a microcavity, formed by two dielectric mirrors~\cite{Vahala2003}, that contains a thin layer of a low-mass molecular semiconductor dispersed in a polymer matrix. The organic semiconductor used in this study was the dye Lumogen-F Orange (LFO)~\cite{Russo2022a}. Operating around the 0-0 transition\footnote{In the presence of vibrational sub-levels, 0-0 transitions are those between the lowest vibrational states.}, the LFO molecules can be considered as a two-level system. As these two-level systems interact with a common cavity mode, the experiment has been interpreted using the Dicke Hamiltonian of Eq.~\eqref{eq:DickeModel}. The number of two-level systems was varied by controlling the concentrations of the dye molecules.

In \cite{Quach2022}, charging and energy storage dynamics were measured using ultrafast transient-absorption spectroscopy, allowing femtosecond charging times to be measured~\cite{Cerullo2007}. In this technique, the microcavity was excited with a pump pulse, and the stored energy (i.e., the number of excited molecules) as a function of time was measured with a second probe pulse. The experimental data suggested superextensive energy storage capacity and charging. 
\citet{Quach2022} also observed that the retention of energy in the system benefits of a finely-tuned balance between cavity coupling and decoherence channels, allowing the device to charge quickly and yet discharge much more slowly, relative to the charging rate. This effect, analogue to that discussed in Sec.~\ref{ss:protection_from_energy_loss}, provides an example of a realistic noisy environment that can aid the implementation and application of useful quantum batteries.

A practical challenge discussed in the study~\cite{Quach2022} was that high concentrations of dye molecules led to quenching, i.e., the formation of optically dark electronic ground states that suppress light absorption. Overcoming this limitation would require careful choice of materials to mitigate or eliminate quenching. The study noted that there are classes of materials where such quenching is suppressed. For example, fluorescent molecules can be surrounded by a ``molecular cage'' or ``bumper'', e.g., consisting of protein $\beta$ sheets~\cite{Dietrich2016}. These cages force a minimum distance between the chromophores, reducing the intermolecular electronic couplings, and thus, allowing for the suppression of exciton-exciton quenching at high concentrations~\cite{Dietrich2016}. Such materials are a promising route to study this type of quantum battery at higher molecular concentrations, and thus at higher energy densities. 

\subsection{Nuclear spins}
\label{sss:nuclear spin}

\citet{Joshi2022} have recently used nuclear magnetic resonance to study energy injection and extraction in nuclear spin systems. The Authors consider molecular structures with star-topology, consisting of a so-called \textit{battery-spin} surrounded by $\mathcal{N}=3$ to 36 \textit{charger-spins}. In this work the battery is a single spin-$1/2$ and the charger is a collection of $\mathcal{N}$ spin-$1/2$ systems. Although these systems resemble the spin-network models discussed in Sec.~\ref{sss:spin-chain}, here the ``battery'' is a single-body system ($N=1$), while the ``charger'' is composed of $\mathcal{N}>1$ sub-units. Note that we used $\mathcal{N}$ to indicate the number of charger subsystems, not to be confused with the number $N$ of battery subsystems used in Secs.~\ref{s:fundamental_theory} and~\ref{s:quantum_battery_models}. The experiments were conducted at an ambient temperature of 298~K. The initial thermal equilibrium state of the charger-spins were driven out of equilibrium by inverting their populations with a $\pi$ pulse. Following this, the energy of the battery-spin was measured as a function of time and of the number $\mathcal{N}$ of charger spins.~\citet{Joshi2022} suggest that their results are an indication of the charging advantage discussed in Secs.~\ref{s:fundamental_theory} and~\ref{s:quantum_battery_models}. Although similar in nature, the charging speed-up measured by~\citet{Joshi2022} is not exactly the one described in Sec.~\ref{s:fundamental_theory}, due to the fact that the battery is composed of a single constituent ($N=1$). Instead, we argue that it is rather a case of the \textit{supertransfer} mechanism described in Ref.~\cite{Taylor2018}. 

Quantum state tomography was performed to find a $\sqrt{\mathcal{N}}$ advantage in ergotropy as well. To keep the battery spin in the charged state, the charger spins were continually charged (i.e., driven out of equilibrium) after an experimentally determined delay. Interestingly, the Authors also considered a \textit{load-spin} subsystem, to which the battery-spin could deposit its energy after a suitable storage time. In the full 38-spin system, the battery-spin could store energy for up to 2 minutes. This result is a promising step towards the implementation of room-temperature quantum batteries based on nuclear spins. Further work could extend this model to the case of many-body quantum batteries (i.e., $N>1$), and clarify the nature of the charging speed-up scaling with $\mathcal{N}$.

\section{Conclusions and outlooks}
\label{s:conclusions_and_outlooks}

In the past decade, quantum batteries have been a framework for studying energy injection, storage, and extraction processes in the quantum regime. Despite the significant progress made by this rapidly growing field, we are only beginning to understand the potential and the applications of quantum energy storage. A complication is posed by the fact that batteries are multifaceted devices, whose performance is evaluated using a variety of figures of merit. Therefore, we could argue that a quantum battery is well defined only with a clear goal in mind. Hence, the following question is of paramount importance: ``\textit{What is the use of a quantum battery?}'' Here, we outline a few possible answers, elaborating on how they set a path for future developments.

One option is to focus on the advantages that quantum batteries may offer over classical ones. At the beginning of this Colloquium we set out to determine if genuine quantum effects can improve the performance of energy storage devices. The results that we reviewed suggest that the answer is affirmative, as discussed in Secs.~\ref{s:fundamental_theory} and~\ref{s:quantum_battery_models}, at least in some systems and under conditions intended to preserve such key quantum features. The example discussed in Sec.~\ref{sss:role_of_enanglement} explicitly shows the importance of entanglement generation for the joint optimization of power and work. Furthermore, the emblematic case of the SYK model in Sec.~\ref{sss:syk} shows how entanglement leads to enhancements in charging power and precision that cannot be attained classically. 

Another avenue for quantum batteries is their integration with emerging quantum technologies, which are expected to earn a prominent role in a variety of practical tasks~\cite{Acin2018,Deutsch2020}, like optimizations~\cite{Preskill2018,Atzori2019}, simulations~\cite{Altman2021}, and measurements~\cite{Degen2017,Albarelli2020}.
As a major example, quantum computation necessitates examining its energy consumption for practical implementation and optimization \cite{Auffeves2021}. \citet{Bennett1973} connected logical reversibility in computation to thermodynamic reversibility, suggesting no heat generation in logically reversible computations that prevent information loss, adhering to Landauer's principle~\citet{Landauer1961}. Quantum computation is inherently reversible because it uses unitary operations, which are reversible linear transformations that preserve the inner product between quantum states and enable the system to return to its initial state without information loss. This feature allows, at least in theory, quantum computation to be performed without generating heat. However, due to the quantum speed limit \cite{Deffner2017}, finite-time operations require energy input, necessitating an energy source for a reversible quantum computer, even without heat production.

In their recent work,~\citet{Chiribella2021} demonstrate the necessity of a quantum energy storage component for performing reversible quantum computing. Their proposed approach involves implementing a reversible quantum gate $G$, which might not conserve energy, on the system used for the computation through a free unitary $U_G$. The latter acts on both the system and the battery by preserving their combined energy. Interestingly, the battery only needs to supply a finite energy for $U_G$ to approximate the desired gate $G$ with arbitrary precision.~\citet{Chiribella2021} discuss the energetic requirements for reversible quantum computing in relation to those imposed by Landauer's principle for irreversible processes. These results speak clearly about the importance that quantum batteries have for energy-efficient computing, encouraging the investigation of several aspects, such as the relation to the quantum speed limit~\cite{Deffner2017} and with the finite-time Landauer principle~\cite{Proesmans2020}.

Beyond computation, quantum batteries could prove resourceful for other quantum devices that rely on coherences and entanglement. The scheme proposed by~\citet{Chiribella2021} for the case of computing also finds application in quantum metrology~\cite{Chiribella2017}, and has general validity for reversible quantum operations. In this context, it is likely that the energy storage components would require some unique conditions to correctly interface with some quantum device. For example, batteries might need to operate at energy and time scales that can only be achieved with, say, transmons, just like SQUIDs and NV-centers are among the few available platforms to offer nanotesla sensitivity for magnetometry~\cite{Buchner2018,Barry2020}.
Based on these perspectives, we can outline a roadmap for the development of quantum batteries. In what follows, we start from theory, proceeding through experiments to arrive at the future possible applications.

From a theoretical standpoint, there are several interesting outlooks that take the research on quantum batteries beyond the scaling of charging and extraction power. A key question is to understand if the amount of extractable energy affects the relaxation timescale. To this end, decoherence-free subspaces~\cite{Kwiat2000,Lidar2003b,Cappellaro2006}, metastability~\cite{Lan2018,Macieszczak2021} (see Sec.~\ref{sss:decoherence-free_subspaces}), and other anomalous relaxation phenomena~\cite{Lu2017,Baity-Jesi2019,Wildeboer2022,Carollo2021} could offer a viable way to overcome the trade-off apparently imposed by the relaxation rate $\gamma_R \propto \exp(\Delta E/k_B T)$, which is exponentially faster with an increasing energy gap $\Delta E$. It is then crucial to find models that permit these phenomena, and experimental platforms suitable for their realization. Another important question is to determine the limits of energy density (the useful energy per component) of these devices, which ultimately depend on the details of the considered physical architecture.

Proof-of-principle experiments are the next step to demonstrate advantages and uncover new aspects of quantum energy storage. While recent experiments (see Sec.~\ref{s:architectures}) have made some significant steps forward, a fully \emph{operational} quantum battery is yet to be demonstrated. To this end, we argue that any practical implementation should involve (1) a system with finite-energy bandwidth, (2) the extraction of some work to carry out a task, and (3) charging and extraction time scales that are much shorter than the relaxation time scale. With the experimental work on quantum batteries still in its infancy, it is desirable to consider all the existing platforms for quantum technologies~\cite{DeLeon2021}, such as superconducting qubits and cavities, neutral atoms, trapped ions, color centers, quantum dots, cavity and waveguide QED, and other systems~\cite{Smyser2020,Huang2023a}. 

Finally, we should make some considerations on the relation between the physical architectures and the applications discussed above. In Secs.~\ref{s:fundamental_theory} and~\ref{s:quantum_battery_models} we extensively discussed the difference between quantum and collectives speed-ups, originating from quantum correlations and collective interactions, respectively. This distinction suggests that ``cold'' platforms would be ideal to explore quantum speed-ups. This is because, currently, most of the experimental set-ups that can reliably sustain entanglement are characterized by low energies, low densities and low temperatures. Since these conditions are achieved using cryostats and other dedicated and energy-consuming controls, care must be taken in the accounting of the expended energetic resources, as argued by~\citet{Auffeves2021} and~\citet{Fellous-asiani2022}.
On the other hand, room temperature set-ups offer a path towards energy storage devices that exploit collective speed-ups. Architectures based on quantum dots and organic molecules are characterized by energy densities that are favorable for optoelectronic applications~\cite{Ostroverkhova2016}, and could find application for light-harvesting devices~\cite{Calvin1983,Curutchet2017,Kundu2017,Jang2018}. Nevertheless, as discussed above, there are some important issues that have to be addressed before quantum batteries can make a direct technological impact, specifically in terms of charging and stabilization efficiency, and of energy density.

The management of energetic resources is at the center of a social and political discussion. As mentioned in the introduction, scientific revolutions often stem from fundamental research. Thus, investigating not only immediate solutions but also basic science is essential. Given that, it may be surprising that the quantum technology research communities have only recently started to address this urgent issue~\cite{Auffeves2021}. Interestingly, recent findings are pointing at the fact that quantum technologies can deeply change the way we harvest~\cite{Curutchet2017,Wang2018}, store~\cite{Ho2018,Gao2021}, deliver~\cite{Zhong2016,Mattioni2021,Davidson2022}, and convert energy~\cite{Lindenfels2019,Ono2020}. This cross-disciplinary effort is effectively leading to the emergence of a quantum energy sector~\cite{Metzler2023}. Quantum batteries are an important part of this effort, and offer a platform to further deepen our fundamental understating of energy storage. 

\begin{acknowledgments}
\noindent
The Authors thank Prof.~Kavan Modi for insightful discussions. FC acknowledges that results incorporated in this standard have received funding from the European Union Horizon Europe research and innovation programme under the Marie Sklodowska-Curie Action for the project SpinSC. SG acknowledges the European Commission under GA n.~101070546--MUQUABIS, the PNRR MUR project PE0000023-NQSTI, and the Royal Society IEC\textbackslash R2\textbackslash 222003. 
M.P. is a cofounder and shareholder of Planckian. All other authors declare no competing interests. 
G.M.A. acknwoledges funding from the European Research Council (ERC) under the European Union's Horizon 2020 research and innovation
programme (Grant agreement No. 101002955 -- CONQUER).
\end{acknowledgments}

\bibliography{main.bbl}

\vfill
\clearpage

\end{document}